\documentclass{article}%
\usepackage{amsmath}
\usepackage{amsfonts}
\usepackage{amssymb}
\usepackage{graphicx}%
\newtheorem{theorem}{Theorem}
\newtheorem{acknowledgement}[theorem]{Acknowledgement}

\newtheorem{corollary}[theorem]{Corollary}

\newtheorem{definition}[theorem]{Definition}

\newtheorem{proposition}[theorem]{Proposition}
\newtheorem{remark}[theorem]{Remark}

\newenvironment{proof}[1][Proof]{\noindent\textbf{#1.} }{\ \rule{0.5em}{0.5em}}
\begin{document}

\title{Clifford Valued Differential Forms, Algebraic Spinor Fields, Gravitation,
Electromagnetism and "Unified" \ Theories\thanks{This paper is an expanded
version of the material containing in \cite{caro2004}(math-ph/0407024) and
\cite{roca2004}(math-ph/0407025), which are published in \textit{Int. J. Mod.
Phys. D }\textbf{13}(8), 1637-1659\ (2004) and \textit{Int. J. Mod. Phys. D
}\textbf{13}(9), 1879-1915\ (2004). }}
\author{E. Capelas de Oliveira and W. A. Rodrigues Jr.\\Institute of Mathematics, Statistics and Scientific Computation\\IMECC-UNICAMP CP 6065\\13083-970 Campinas-SP, Brazil\\walrod@ime.unicamp.br\hspace{0.15in}capelas@ime.unicamp.br}
\date{revised July 18 2004}
\maketitle
\tableofcontents

\begin{abstract}
In this paper, we show how to describe the general theory of a linear metric
compatible connection with the theory of Clifford valued differential forms.
This is done by realizing that for each spacetime point the Lie algebra of
Clifford bivectors is isomorphic to the Lie algebra of $Sl(2,\mathbb{C)}$. In
that way, the pullback of the linear connection under a trivialization of the
bundle is represented by a Clifford valued 1-form. That observation makes it
possible to realize that Einstein's gravitational theory can be formulated in
a way which is similar to a $Sl(2,\mathbb{C)}$ gauge theory. Some aspects of
such approach are discussed. Also, the theory of \ the covariant spinor
derivative of spinor fields is introduced in a novel way, allowing for a
physical interpretation of some rules postulated for that covariant spinor
derivative in the standard theory of these objects. We use our methods to
investigate some \textit{polemical} issues in gravitational theories and in
particular we scrutinize a supposedly "unified" field theory of gravitation
and electromagnetism proposed by M. Sachs and recently used in a series of
papers by other authors. Our results show that Sachs did not attain his
objective and that recent papers based on that theory are ill conceived and
completely invalid both as Mathematics and Physics.

\end{abstract}

\section{Introduction}

In this paper we introduce the concept of Clifford valued differential
forms\footnote{Analogous, but non equivalent concepts have been introduced in
\cite{dimakis,vatorr,tucker}. In particular \cite{dimakis} is a very complete
paper using clifforms, i.e., forms with values in a abstract Clifford
algebra.}, mathematical entities which are sections of $\mathcal{C\ell
}(TM)\otimes%
%TCIMACRO{\dbigwedge }%
%BeginExpansion
{\displaystyle\bigwedge}
%EndExpansion
T^{\ast}M$. We show how with the aid of this concept we can produce a very
beautiful \textit{description} of the theory of linear connections, where the
representative of a given linear connection in a given gauge is represented by
a \textit{bivector} valued 1-form. The notion of an exterior covariant
differential and exterior covariant derivative of sections of $\mathcal{C\ell
}(TM)\otimes%
%TCIMACRO{\dbigwedge }%
%BeginExpansion
{\displaystyle\bigwedge}
%EndExpansion
T^{\ast}M$ \ is crucial for our program and is thus discussed in details. Our
\textit{natural} definitions (to be compared with other approaches on related
subjects, as described, e.g., in
\cite{baezm,beentucker,frankel,goeshu,nasen,palais,sternberg}) parallel in a
noticeable way the formalism of the theory of connections in principal bundles
and their associated covariant derivative operators acting on associated
vector bundles. We identify Cartan \textit{curvature} 2-\textit{forms} and
\ \textit{curvature bivectors}. The curvature 2-forms satisfy Cartan's second
structure equation and the curvature bivectors satisfy equations in complete
analogy with equations of gauge theories. This immediately suggests to write
Einstein's theory in that formalism, something that has already been done and
extensively studied in the past (see e.g., \cite{carmeli,caleini}). Our
methodology suggest new ways of taking advantage of such a formulation, but
this is postpone for a later paper. Here, our investigation of the
$Sl(2,\mathbb{C})$ \ nonhomogeneous\ gauge equation for the curvature bivector
is restricted to the relationship between that equation and Sachs theory
\cite{s1,s2,s3} and the problem of the energy-momentum `conservation' in
General Relativity.

We recall also the concept of covariant derivatives of (\textit{algebraic})
spinor fields in our formalism, where these objects are represented as
sections of real spinor bundles\footnote{Real Spinor fields have been
introduced by Hestenes in \cite{hestenesspinors}, but a rigorous theory of
that objects in a Lorentzian spacetime has only recently been achieved
\cite{moro,28}.} and study how this theory has as matrix representative the
standard two components spinor fields (\textit{dotted} and undotted) already
introduced long ago, see, e.g., \cite{carmeli,penrose, penrindler, pirani}.
What is \textit{new} here is that we identify that in the theory of algebraic
spinor fields the \textit{realization} of some rules which are used in the
standard formulation of the matrix spinor fields, e.g., why the covariant
derivative of the Pauli matrices must be null, imply some constraints, with
admit a very interesting geometrical interpretation. Indeed, \ a possible
realization of that rules in the Clifford bundle formalism is one where the
vector fields defining a global tetrad $\{\mathbf{e}_{\mathbf{a}}\}$ must be
such that $D_{\mathbf{e}_{\mathbf{0}}}\mathbf{e}_{\mathbf{0}}=0$, i.e.,
$\mathbf{e}_{\mathbf{0}}$ is a geodesic reference frame and along each one of
its integral lines, say $\sigma$, and the $\mathbf{e}_{\mathbf{d}}$
$(\mathbf{d}=1,2,3)$ are Fermi transported, i.e., they are not rotating
relative to the local gyroscope axes. For the best of our knowledge this
important fact is here disclosed for the first time.

We use the Clifford bundle formalism and the theory of Clifford valued
differential forms to analyze some polemic\ issues in presentations of
gravitational theory and some other theories. In particular, we scrutinized
Sachs "unified" theory as described recently in \cite{s3} and originally
introduced in \cite{s1}. We show that unfortunately there are some
\textit{serious} mathematical errors in Sachs theory. To start, he identified
erroneously his basic variables $q_{\mu}$ as being quaternion fields over a
Lorentzian spacetime. Well, they are \textit{not}. The real mathematical
structure of these objects is that they are matrix representations of
particular sections of the even Clifford bundle of multivectors
$\mathcal{C\ell}(TM)$ (called paravector fields in mathematical literature) as
we proved in section 2. Next we show that the identification of a `new'
antisymmetric field in his theory is indeed nothing more than the
identification of some combinations of the curvature bivectors\footnote{The
curvature bivectors are physically and mathematically equivalent to the Cartan
curvature 2-forms, since they carry the same information. This statement will
become obvious from our study in section 4.}, an object that appears naturally
when we try to formulate Einstein's gravitational theory as a
$Sl(2,\mathbb{C)}$ gauge theory. \ In that way, any tentative of identifying
such an object with any kind of \textit{electromagnetic field} as did by Sachs
in \cite{s1,s2,s3} is clearly \textit{wrong}. \ We note that recently in a
series of papers, Evans\&AIAS group (\cite{a1}-\cite{a20},\cite{ea4}%
-\cite{ea8},\cite{cleevans}\cite{e0}-\cite{e4}) \ uses Sachs theory in order
to justify some very \textit{odd} facts, which must be denounced. Indeed, we
recall that:

(\textbf{i}) On March 26 2002, the United States Patent and Trademark Office
(USPTO) in Washington issued US Patent no. 6,362,718 for a Motionless
Electromagnetic Generator (MEG). This \ would be \ `remarkable'\ device has
been projected by retired lieutenant colonel Tom E. Bearden of Alabama and
collaborators. They claimed MEG produces more output energy than the input
energy used for its functioning!

Of course, nobody could think that the officers at the US Patent office do
\textit{not} know the law of energy-momentum conservation, which in general
prevents all Patent offices to veto all free energy machines, and indeed that
energy momentum conservation law has been used since a long time ago as a
\textit{golden rule}.

So, affording a patent to that device must \ have a reason. A possible one is
that the patent officers must somehow been convinced that there are
theoretical reasons for the functioning of MEG. How, did the patent officers
get convinced?

We think that the answer can be identified in a long list of papers published
in respectable (?) Physics journals signed by Evan\&AIAS group and quoted
above\footnote{Note that Bearden is one of the members of the AIAS group. We
mention also that in the AIAS website the following people among others are
listed as emeritus fellows of the Foundation: Prof. Alwyn van der Merwe, Univ.
of Denver, Colorado, USA, Prof. Mendel Sachs, SUNY, Buffalo, USA, Prof. Jean
Pierre Vigier, Institut Henri Poincare, France. Well, van der Merwe is editor
of \textit{Foundations of Physics} and \textit{Foundations of Physics
Letters}, Sachs is one of the authors we criticize here and Vigier is on the
editorial board of \textit{Physics Letters A} for decades and is one of the
AIAS authors. This eventually could explain how AIAS got their papers
published...}. There, they claimed that using Sachs theory there is a
`natural' justification for an entity that they called the $\mathbf{B}_{3}$
field and that appears (according to them) in their \ `new' $O(3)$
electrodynamics and \ `unified' field theory. According to them, the
$\mathbf{B}_{3}$ field is to be identified with $\mathbf{F}_{12}$,
where\textbf{ }$\mathbf{F}_{\mu\nu}=-\mathbf{F}_{\nu\mu}$ (see Eq.(\ref{37})
below) is a mathematical object \ that Sachs identified in \cite{s1,s2,s3}
with an electromagnetic field after \ `taking the trace in the spinor
indices'. Evans\&AIAS group claim to explain the operation of MEG. It simply
pumps energy from the $\mathbf{F}_{12}$ existing in spacetime. However, the
Mathematics and Physics of Evans\&AIAS used in their papers are unfortunately
only a pot pourri of nonsense as we already demonstrated
elsewhere\footnote{For more details on the absurdities propagated by
Evans\&AIAS in ISI indexed journals and books see \cite{carrod,warxaias}. The
second citation is a reply to Evans' paper \cite{e0}.} and more below. This,
of course invalidate any theoretical justification for the patent.

It would be great if the officers of USPTO would know enough Mathematics and
Physics in order to reject immediately the \textit{theoretical} explanations
offered by the MEG inventors. But that unfortunately was not the case, because
it seems that the knowledge of Mathematics and Physics of that officers was no
great than the knowledge of these disciplines by the referees of the
Evans\&AIAS papers.

Of course, theoretical explanations apart and the authors prejudices it can
happen that MEG \textit{works}. However, having followed with interest in the
internet\footnote{See http://groups.yahoo.com/group/free\_energy/.} the work
of supposedly MEG builders, we arrived at the conclusion that MEG did not work
until now, and all claims of its inventors and associates are simply due to
\textit{wrong} experimental measurements. And, of course, that must also been
the case with the USPTO officers, if they did realize any single experiment on
the MEG device. And indeed, this may be really the case, for in a recent
article \cite{mac} we are informed that in August last year the Commissioner
of Patents, Nicholas P. Godici informed that it was a planned a re-examination
of the MEG patent. We do not know what happened since then.

(\textbf{ii}) Now, \ is energy-momentum conservation a \textit{trustworthy}
law of the physical world? To answer that question we discuss in this paper
the shameful problem of the energy-momentum \ `conservation' in General
Relativity. \ 

Yes, in General Relativity there are \textit{no} conservation laws of energy,
momentum and angular momentum \textit{in general}, and this fact must be clear
once and for ever for all (even for school boys, that are in general fooled in
reading science books for laymen).

To show this result in an economic and transparent way a presentation of
Einstein's gravitational theory is given in terms of tetrads fields, which has
a very elegant description in terms of the calculus in the Clifford bundle
$\mathcal{C\ell}(T^{\ast}M)$ described in Appendix. Using that toll, we recall
also the \textit{correct} wave like equations solved by the tetrad
fields\footnote{The set $\{\theta^{\mathbf{a}}\}$ is the dual basis of
$\{e_{\mathbf{a}}\}$.} $\theta^{\mathbf{a}}$ in General Relativity. This has
been done here in order to complete the debunking of recent Evans\&AIAS papers
(\cite{cleevans},\cite{e0}-\cite{e4}) claiming to have achieved \ (yet)
another `unified' field theory. Indeed, we show that, as it is the case with
almost all other papers written by those authors, these new ones are again a
compendium of very bad Mathematics and Physics.

\section{Spacetime, Pauli and Quaternion Algebras}

In this section we recall very well known facts concerning three special
\textit{real} Clifford algebras, namely, the \textit{spacetime} algebra
$\mathbb{R}_{1,3}$, the \textit{Pauli} algebra $\mathbb{R}_{3,0}$ and the
\textit{quaternion} algebra $\mathbb{R}_{0,2}=\mathbb{H}$ and the relation
between them.\footnote{This material is treated in details e.g, in the books
\cite{hestenessob,lounesto,porteous1,porteous2}. See also
\cite{femoro101,femoro201,femoro301,femoro401,femoro501,femoro601,femoro701}.}

\subsection{Spacetime Algebra}

We define the spacetime algebra $\mathbb{R}_{1,3}$ as being the Clifford
algebra associated \ with Minkowski vector space $\mathbb{R}^{1,3}$, which is
a four dimensional real vector space, equipped with a Lorentzian bilinear
form
\begin{equation}%
%TCIMACRO{\TeXButton{eta}{\mbox{\boldmath{$\eta$}}}}%
%BeginExpansion
\mbox{\boldmath{$\eta$}}%
%EndExpansion
:\mathbb{R}^{1,3}\times\mathbb{R}^{1,3}\rightarrow\mathbb{R}. \label{1}%
\end{equation}

Let $\{\mathbf{m}_{0}\mathbf{,m}_{1}\mathbf{,m}_{2}\mathbf{,m}_{3}\}$ be an
arbitrary orthonormal basis of $\mathbb{R}^{1,3}$, i.e.,%
\begin{equation}%
%TCIMACRO{\TeXButton{eta}{\mbox{\boldmath{$\eta$}}}}%
%BeginExpansion
\mbox{\boldmath{$\eta$}}%
%EndExpansion
(\mathbf{m}_{\mu}\mathbf{,m}_{\nu})=\eta_{\mu\nu}=\left\{
\begin{array}
[c]{ccc}%
1 & \text{if} & \mu=\nu=0\\
-1 & \text{if} & \mu=\nu=1,2,3\\
0 & \text{if} & \mu\neq\nu
\end{array}
\right.  \label{2}%
\end{equation}
As usual we resume Eq.(\ref{2}) writing $\mathbf{\eta}_{\mu\nu}=\mathrm{diag}%
(1,-1,-1,-1)$. We denote by $\{\mathbf{m}^{0}\mathbf{,m}^{1}\mathbf{,m}%
^{2}\mathbf{,m}^{3}\}$ the \textit{reciprocal} basis of $\{\mathbf{m}%
_{0}\mathbf{,m}_{1}\mathbf{,m}_{2}\mathbf{,m}_{3}\}$, i.e., $%
%TCIMACRO{\TeXButton{eta}{\mbox{\boldmath{$\eta$}}}}%
%BeginExpansion
\mbox{\boldmath{$\eta$}}%
%EndExpansion
\mathbf{(m}^{\mu}\mathbf{,m}_{\nu})=\delta_{\nu}^{\mu}$. We have in obvious
notation $\mathbf{\eta}(\mathbf{m}^{\mu}\mathbf{,m}^{\nu})=\eta^{\mu\nu}=$
$\mathrm{diag}(1,-1,-1,-1)$.

The spacetime algebra $\mathbb{R}_{1,3}$ is generate by the following
algebraic fundamental relation%
\begin{equation}
\mathbf{m}^{\mu}\mathbf{m}^{\nu}\mathbf{+m}^{\nu}\mathbf{m}^{\mu}=2\eta
^{\mu\nu}. \label{3}%
\end{equation}

We observe that (as with the conventions fixed in the Appendix) in the above
formula and in all the text the Clifford product is denoted by
\textit{juxtaposition} of symbols. $\mathbb{R}_{1,3}$ as a vector space over
the real field is isomorphic to the exterior algebra $%
%TCIMACRO{\dbigwedge }%
%BeginExpansion
{\displaystyle\bigwedge}
%EndExpansion
\mathbb{R}^{1,3}=%
%TCIMACRO{\dsum \limits_{j=0}^{4}}%
%BeginExpansion
{\displaystyle\sum\limits_{j=0}^{4}}
%EndExpansion
$ $%
%TCIMACRO{\dbigwedge \nolimits^{j}}%
%BeginExpansion
{\displaystyle\bigwedge\nolimits^{j}}
%EndExpansion
\mathbb{R}^{1,3}$ of $\mathbb{R}^{1,3}$. We code that information writing $%
%TCIMACRO{\dbigwedge }%
%BeginExpansion
{\displaystyle\bigwedge}
%EndExpansion
\mathbb{R}^{1,3}\hookrightarrow\mathbb{R}_{1,3}$. \ Also, $%
%TCIMACRO{\dbigwedge \nolimits^{0}}%
%BeginExpansion
{\displaystyle\bigwedge\nolimits^{0}}
%EndExpansion
\mathbb{R}^{1,3}\equiv\mathbb{R}$ and $%
%TCIMACRO{\dbigwedge \nolimits^{1}}%
%BeginExpansion
{\displaystyle\bigwedge\nolimits^{1}}
%EndExpansion
\mathbb{R}^{1,3}\equiv\mathbb{R}^{1,3}$ . We identify the exterior product of
vectors by
\begin{equation}
\mathbf{m}^{\mu}\mathbf{\wedge m}^{\nu}\mathbf{=}\frac{1}{2}\left(
\mathbf{m}^{\mu}\mathbf{m}^{\nu}\mathbf{-m}^{\nu}\mathbf{m}^{\mu}\right)  ,
\label{4}%
\end{equation}
and also, we identify the scalar product by
\begin{equation}%
%TCIMACRO{\TeXButton{eta}{\mbox{\boldmath{$\eta$}}}}%
%BeginExpansion
\mbox{\boldmath{$\eta$}}%
%EndExpansion
\mathbf{(m}^{\mu}\mathbf{,m}^{\nu}\mathbf{)=}\frac{1}{2}\left(  \mathbf{m}%
^{\mu}\mathbf{m}^{\nu}\mathbf{+m}^{\nu}\mathbf{m}^{\mu}\right)  . \label{5}%
\end{equation}
Then we can write
\begin{equation}
\mathbf{m}^{\mu}\mathbf{m}^{\nu}=%
%TCIMACRO{\TeXButton{eta}{\mbox{\boldmath{$\eta$}}}}%
%BeginExpansion
\mbox{\boldmath{$\eta$}}%
%EndExpansion
\mathbf{(m}^{\mu}\mathbf{,m}^{\nu}\mathbf{)+m}^{\mu}\mathbf{\wedge m}^{\nu}.
\label{6}%
\end{equation}
\ From the observations given in the Appendix it follows that an arbitrary
element $\mathbf{C}\in\mathbb{R}_{1,3}$ can be written as sum of
\textit{nonhomogeneous multivectors}, i.e.,%
\begin{equation}
\mathbf{C}=s+c_{\mu}\mathbf{m}^{\mu}\mathbf{+}\frac{1}{2}c_{\mu\nu}%
\mathbf{m}^{\mu}\mathbf{m}^{\nu}+\frac{1}{3!}c_{\mu\nu\rho}\mathbf{m}^{\mu
}\mathbf{m}^{\nu}\mathbf{m}^{\rho}+p\mathbf{m}^{5} \label{7}%
\end{equation}
where $s,c_{\mu},c_{\mu\nu},c_{\mu\nu\rho},p\in\mathbb{R}$ and $c_{\mu\nu
},c_{\mu\nu\rho}$ are completely antisymmetric in all indices. Also
$\mathbf{m}^{5}\mathbf{=m}^{0}\mathbf{m}^{1}\mathbf{m}^{2}\mathbf{m}^{3}$ is
the generator of the pseudo scalars. As matrix algebra we have that
$\mathbb{R}_{1,3}\simeq\mathbb{H(}2)$, the algebra of the $2\times2$
quaternionic matrices.

\subsection{Pauli Algebra}

Now, the Pauli algebra $\mathbb{R}_{3,0}$ is the Clifford algebra associated
with the Euclidean vector space $\mathbb{R}^{3,0}$, equipped as usual, with a
positive definite bilinear form. \ As a matrix algebra we have that
$\mathbb{R}_{3,0}\simeq\mathbb{C}\left(  2\right)  $, the algebra of
$2\times2$ complex matrices. Moreover, we recall that $\mathbb{R}_{3,0}$ is
isomorphic to the even subalgebra of the spacetime algebra, i.e., writing
$\mathbb{R}_{1,3}=$ $\mathbb{R}_{1,3}^{(0)}\oplus$ $\mathbb{R}_{1,3}^{(1)}$ we
have,
\begin{equation}
\mathbb{R}_{3,0}\simeq\mathbb{R}_{1,3}^{(0)}. \label{8}%
\end{equation}

\bigskip

The isomorphism is easily exhibited by putting $%
%TCIMACRO{\TeXButton{sigma}{\mbox{\boldmath{$\sigma$}}}}%
%BeginExpansion
\mbox{\boldmath{$\sigma$}}%
%EndExpansion
^{i}\mathbf{=m}^{i}\mathbf{m}^{0}$, $i=1,2,3$. Indeed, $\ $with $\delta
^{ij}=\mathrm{diag}(1,1,1)$, we have
\begin{equation}%
%TCIMACRO{\TeXButton{sigma}{\mbox{\boldmath{$\sigma$}}}}%
%BeginExpansion
\mbox{\boldmath{$\sigma$}}%
%EndExpansion
^{i}%
%TCIMACRO{\TeXButton{sigma}{\mbox{\boldmath{$\sigma$}}}}%
%BeginExpansion
\mbox{\boldmath{$\sigma$}}%
%EndExpansion
^{j}\mathbf{+}%
%TCIMACRO{\TeXButton{sigma}{\mbox{\boldmath{$\sigma$}}}}%
%BeginExpansion
\mbox{\boldmath{$\sigma$}}%
%EndExpansion
^{j}%
%TCIMACRO{\TeXButton{sigma}{\mbox{\boldmath{$\sigma$}}}}%
%BeginExpansion
\mbox{\boldmath{$\sigma$}}%
%EndExpansion
^{i}=2\delta^{ij}, \label{9}%
\end{equation}
which is the fundamental relation defining the algebra $\mathbb{R}_{3,0}$.
Elements of the Pauli algebra will be called Pauli numbers\footnote{Sometimes
they are also called `complex quaternions'. This last terminology will be
obvious in a while.}. As vector space we have that $%
%TCIMACRO{\dbigwedge }%
%BeginExpansion
{\displaystyle\bigwedge}
%EndExpansion
\mathbb{R}^{3,0}\hookrightarrow\mathbb{R}_{3,0}\subset\mathbb{R}_{1,3}$. So,
any Pauli number can be written as
\begin{equation}
\mathbf{P}=s+p^{i}%
%TCIMACRO{\TeXButton{sigma}{\mbox{\boldmath{$\sigma$}}}}%
%BeginExpansion
\mbox{\boldmath{$\sigma$}}%
%EndExpansion
^{i}+\frac{1}{2}p_{ij}^{i}%
%TCIMACRO{\TeXButton{sigma}{\mbox{\boldmath{$\sigma$}}}}%
%BeginExpansion
\mbox{\boldmath{$\sigma$}}%
%EndExpansion
^{i}%
%TCIMACRO{\TeXButton{sigma}{\mbox{\boldmath{$\sigma$}}}}%
%BeginExpansion
\mbox{\boldmath{$\sigma$}}%
%EndExpansion
^{j}+p\text{\textsc{i}}\mathbf{,} \label{10}%
\end{equation}
where $s,p_{i},p_{ij},p\in\mathbb{R}$ and $p_{ij}=-p_{ji}$ and also
\begin{equation}
\text{\textsc{i}}\mathbf{=}%
%TCIMACRO{\TeXButton{sigma}{\mbox{\boldmath{$\sigma$}}}}%
%BeginExpansion
\mbox{\boldmath{$\sigma$}}%
%EndExpansion
^{1}%
%TCIMACRO{\TeXButton{sigma}{\mbox{\boldmath{$\sigma$}}}}%
%BeginExpansion
\mbox{\boldmath{$\sigma$}}%
%EndExpansion
^{2}%
%TCIMACRO{\TeXButton{sigma}{\mbox{\boldmath{$\sigma$}}}}%
%BeginExpansion
\mbox{\boldmath{$\sigma$}}%
%EndExpansion
^{3}=\mathbf{m}^{5}. \label{11}%
\end{equation}

\bigskip Note that \textsc{i}$^{2}=-1$ and that \textsc{i} commutes with any
Pauli number. We can trivially verify that%
\begin{align}%
%TCIMACRO{\TeXButton{sigma}{\mbox{\boldmath{$\sigma$}}}}%
%BeginExpansion
\mbox{\boldmath{$\sigma$}}%
%EndExpansion
^{i}%
%TCIMACRO{\TeXButton{sigma}{\mbox{\boldmath{$\sigma$}}}}%
%BeginExpansion
\mbox{\boldmath{$\sigma$}}%
%EndExpansion
^{j}  &  =\text{\textsc{i}}\varepsilon_{k}^{i\text{ }j}%
%TCIMACRO{\TeXButton{sigma}{\mbox{\boldmath{$\sigma$}}}}%
%BeginExpansion
\mbox{\boldmath{$\sigma$}}%
%EndExpansion
^{k}+\delta^{ij},\label{12}\\
\mathbf{[}%
%TCIMACRO{\TeXButton{sigma}{\mbox{\boldmath{$\sigma$}}}}%
%BeginExpansion
\mbox{\boldmath{$\sigma$}}%
%EndExpansion
^{i}\mathbf{,}%
%TCIMACRO{\TeXButton{sigma}{\mbox{\boldmath{$\sigma$}}}}%
%BeginExpansion
\mbox{\boldmath{$\sigma$}}%
%EndExpansion
^{j}\mathbf{]}  &  \mathbf{\equiv}%
%TCIMACRO{\TeXButton{sigma}{\mbox{\boldmath{$\sigma$}}}}%
%BeginExpansion
\mbox{\boldmath{$\sigma$}}%
%EndExpansion
^{i}%
%TCIMACRO{\TeXButton{sigma}{\mbox{\boldmath{$\sigma$}}}}%
%BeginExpansion
\mbox{\boldmath{$\sigma$}}%
%EndExpansion
^{j}\mathbf{-}%
%TCIMACRO{\TeXButton{sigma}{\mbox{\boldmath{$\sigma$}}}}%
%BeginExpansion
\mbox{\boldmath{$\sigma$}}%
%EndExpansion
^{j}%
%TCIMACRO{\TeXButton{sigma}{\mbox{\boldmath{$\sigma$}}}}%
%BeginExpansion
\mbox{\boldmath{$\sigma$}}%
%EndExpansion
^{i}\mathbf{=}2%
%TCIMACRO{\TeXButton{sigma}{\mbox{\boldmath{$\sigma$}}}}%
%BeginExpansion
\mbox{\boldmath{$\sigma$}}%
%EndExpansion
^{i}\mathbf{\wedge}%
%TCIMACRO{\TeXButton{sigma}{\mbox{\boldmath{$\sigma$}}}}%
%BeginExpansion
\mbox{\boldmath{$\sigma$}}%
%EndExpansion
^{j}=2\text{\textsc{i}}\varepsilon_{k}^{i\text{ }j}%
%TCIMACRO{\TeXButton{sigma}{\mbox{\boldmath{$\sigma$}}}}%
%BeginExpansion
\mbox{\boldmath{$\sigma$}}%
%EndExpansion
^{k}.\nonumber
\end{align}

\bigskip

In that way, writing $\mathbb{R}_{3,0}=\mathbb{R}_{3,0}^{(0)}+\mathbb{R}%
_{3,0}^{(1)}$, any Pauli number can be written as%
\begin{equation}
\mathbf{P=Q}_{1}\mathbf{+}\text{\textsc{i}}\mathbf{Q}_{2},\hspace
{0.15in}\mathbf{Q}_{1}\in\mathbb{R}_{3,0}^{(0)},\hspace{0.15in}%
\text{\textsc{i}}\mathbf{Q}_{2}\in\mathbb{R}_{3,0}^{(1)}, \label{13}%
\end{equation}
with
\begin{align}
\mathbf{Q}_{1}  &  =a_{0}+a_{k}(\text{\textsc{i}}%
%TCIMACRO{\TeXButton{sigma}{\mbox{\boldmath{$\sigma$}}}}%
%BeginExpansion
\mbox{\boldmath{$\sigma$}}%
%EndExpansion
^{k}),\hspace{0.15in}a_{0}=s,\hspace{0.15in}a_{k}=\frac{1}{2}\varepsilon
_{k}^{i\text{ }j}p_{ij},\label{14}\\
\mathbf{Q}_{2}  &  =\text{\textsc{i}}\left(  b_{0}+b_{k}(\text{\textsc{i}}%
%TCIMACRO{\TeXButton{sigma}{\mbox{\boldmath{$\sigma$}}}}%
%BeginExpansion
\mbox{\boldmath{$\sigma$}}%
%EndExpansion
^{k}\right)  ),\hspace{0.15in}b_{0}=p,\hspace{0.15in}b_{k}=-p_{k}.\nonumber
\end{align}

\subsection{Quaternion Algebra}

Eqs.(\ref{14}) \ show that the quaternion algebra $\mathbb{R}_{0,2}%
=\mathbb{H}$ can be identified as the even subalgebra of $\mathbb{R}_{3,0}$, i.e.,%

\begin{equation}
\mathbb{R}_{0,2}=\mathbb{H\simeq R}_{3,0}^{(0)}. \label{15}%
\end{equation}

The statement is obvious once we identify the basis $\{1,\mathit{\hat{\imath}%
},\mathit{\hat{\jmath}}\mathbf{,\mathit{\hat{k}}\}}$ of $\mathbb{H}$ with
\begin{equation}
\{\mathbf{1,}\text{\textsc{i}}%
%TCIMACRO{\TeXButton{sigma}{\mbox{\boldmath{$\sigma$}}}}%
%BeginExpansion
\mbox{\boldmath{$\sigma$}}%
%EndExpansion
^{1}\mathbf{,}\text{\textsc{i}}%
%TCIMACRO{\TeXButton{sigma}{\mbox{\boldmath{$\sigma$}}}}%
%BeginExpansion
\mbox{\boldmath{$\sigma$}}%
%EndExpansion
^{2}\mathbf{,}\text{\textsc{i}}%
%TCIMACRO{\TeXButton{sigma}{\mbox{\boldmath{$\sigma$}}}}%
%BeginExpansion
\mbox{\boldmath{$\sigma$}}%
%EndExpansion
^{3}\}, \label{15'}%
\end{equation}
which are the generators of $\mathbb{R}_{3,0}^{(0)}$. We observe moreover that
the even subalgebra of the quaternions can be identified (in an obvious way)
with the complex field, i.e., $\mathbb{R}_{0,2}^{(0)}\simeq\mathbb{C}$.

\ \ Returning to Eq.(\ref{10}) we see that any $\mathbf{P}\in\mathbb{R}_{3,0}$
can also be written as
\begin{equation}
\mathbf{P=P}_{1}\mathbf{+\text{\textsc{i}}L}_{2}, \label{16}%
\end{equation}
where
\begin{align}
\mathbf{P}_{1}  &  =(s+p^{k}%
%TCIMACRO{\TeXButton{sigma}{\mbox{\boldmath{$\sigma$}}}}%
%BeginExpansion
\mbox{\boldmath{$\sigma$}}%
%EndExpansion
_{k})\in%
%TCIMACRO{\dbigwedge \nolimits^{0}}%
%BeginExpansion
{\displaystyle\bigwedge\nolimits^{0}}
%EndExpansion
\mathbb{R}^{3,0}\oplus%
%TCIMACRO{\dbigwedge \nolimits^{1}}%
%BeginExpansion
{\displaystyle\bigwedge\nolimits^{1}}
%EndExpansion
\mathbb{R}^{3,0}\equiv\mathbb{R\oplus}%
%TCIMACRO{\dbigwedge \nolimits^{1}}%
%BeginExpansion
{\displaystyle\bigwedge\nolimits^{1}}
%EndExpansion
\mathbb{R}^{3,0},\nonumber\\
\text{\textsc{i}}\mathbf{L}_{2}  &  =\text{\textsc{i}}\mathbf{(}%
p+\text{\textsc{i}}l^{k}%
%TCIMACRO{\TeXButton{sigma}{\mbox{\boldmath{$\sigma$}}}}%
%BeginExpansion
\mbox{\boldmath{$\sigma$}}%
%EndExpansion
_{k})\in%
%TCIMACRO{\dbigwedge \nolimits^{2}}%
%BeginExpansion
{\displaystyle\bigwedge\nolimits^{2}}
%EndExpansion
\mathbb{R}^{3,0}\oplus%
%TCIMACRO{\dbigwedge \nolimits^{3}}%
%BeginExpansion
{\displaystyle\bigwedge\nolimits^{3}}
%EndExpansion
\mathbb{R}^{3,0}, \label{17}%
\end{align}
with $l_{k}=-\varepsilon_{k}^{i\text{ }j}p_{ij}\in\mathbb{R}$. The important
fact that we want to recall here is that the subspaces $(\mathbb{R\oplus}%
%TCIMACRO{\dbigwedge \nolimits^{1}}%
%BeginExpansion
{\displaystyle\bigwedge\nolimits^{1}}
%EndExpansion
\mathbb{R}^{3,0})$ and $(%
%TCIMACRO{\dbigwedge \nolimits^{2}}%
%BeginExpansion
{\displaystyle\bigwedge\nolimits^{2}}
%EndExpansion
\mathbb{R}^{3,0}\oplus%
%TCIMACRO{\dbigwedge \nolimits^{3}}%
%BeginExpansion
{\displaystyle\bigwedge\nolimits^{3}}
%EndExpansion
\mathbb{R}^{3,0})$ do not close separately any algebra. In general, if
$\mathbf{A,C}\in(\mathbb{R\oplus}%
%TCIMACRO{\dbigwedge \nolimits^{1}}%
%BeginExpansion
{\displaystyle\bigwedge\nolimits^{1}}
%EndExpansion
\mathbb{R}^{3,0})$ then%
\begin{equation}
\mathbf{AC}\in\mathbb{R\oplus}%
%TCIMACRO{\dbigwedge \nolimits^{1}}%
%BeginExpansion
{\displaystyle\bigwedge\nolimits^{1}}
%EndExpansion
\mathbb{R}^{3,0}\oplus%
%TCIMACRO{\dbigwedge \nolimits^{2}}%
%BeginExpansion
{\displaystyle\bigwedge\nolimits^{2}}
%EndExpansion
\mathbb{R}^{3,0}. \label{18bis}%
\end{equation}

To continue, we introduce
\begin{equation}%
%TCIMACRO{\TeXButton{sigma}{\mbox{\boldmath{$\sigma$}}}}%
%BeginExpansion
\mbox{\boldmath{$\sigma$}}%
%EndExpansion
_{i}\mathbf{=m}_{i}\mathbf{m}_{0}=\mathbf{-}%
%TCIMACRO{\TeXButton{sigma}{\mbox{\boldmath{$\sigma$}}}}%
%BeginExpansion
\mbox{\boldmath{$\sigma$}}%
%EndExpansion
^{i},\hspace{0.15in}i=1,2,3. \label{19}%
\end{equation}

\bigskip Then, \textsc{i}$=-%
%TCIMACRO{\TeXButton{sigma}{\mbox{\boldmath{$\sigma$}}}}%
%BeginExpansion
\mbox{\boldmath{$\sigma$}}%
%EndExpansion
_{1}%
%TCIMACRO{\TeXButton{sigma}{\mbox{\boldmath{$\sigma$}}}}%
%BeginExpansion
\mbox{\boldmath{$\sigma$}}%
%EndExpansion
_{2}%
%TCIMACRO{\TeXButton{sigma}{\mbox{\boldmath{$\sigma$}}}}%
%BeginExpansion
\mbox{\boldmath{$\sigma$}}%
%EndExpansion
_{3}$ and the basis $\{1,\hat{\imath}$\textit{ }$,\hat{\jmath},\hat
{k}\mathbf{\}}$ of $\mathbb{H}$ can be identified with $\{1,\mathbf{-}%
$\textsc{i}$%
%TCIMACRO{\TeXButton{sigma}{\mbox{\boldmath{$\sigma$}}}}%
%BeginExpansion
\mbox{\boldmath{$\sigma$}}%
%EndExpansion
_{1}\mathbf{,-}$\textsc{i}$%
%TCIMACRO{\TeXButton{sigma}{\mbox{\boldmath{$\sigma$}}}}%
%BeginExpansion
\mbox{\boldmath{$\sigma$}}%
%EndExpansion
_{2}\mathbf{,-}$\textsc{i}$%
%TCIMACRO{\TeXButton{sigma}{\mbox{\boldmath{$\sigma$}}}}%
%BeginExpansion
\mbox{\boldmath{$\sigma$}}%
%EndExpansion
_{3}\}$.

Now, we already said that $\mathbb{R}_{3,0}\simeq\mathbb{C}\left(  2\right)
$. This permit us to represent the Pauli numbers by \ $2\times2$ complex
matrices, in the usual way ($\mathrm{i}=\sqrt{-1}$). We write $\mathbb{R}%
_{3,0}\ni\mathbf{P}\mapsto P\in\mathbb{C(}2)$, with
\begin{equation}%
\begin{array}
[c]{ccc}%
%TCIMACRO{\TeXButton{sigma}{\mbox{\boldmath{$\sigma$}}}}%
%BeginExpansion
\mbox{\boldmath{$\sigma$}}%
%EndExpansion
^{1} & \mapsto & \sigma^{1}=\left(
\begin{array}
[c]{cc}%
0 & 1\\
1 & 0
\end{array}
\right) \\%
%TCIMACRO{\TeXButton{sigma}{\mbox{\boldmath{$\sigma$}}}}%
%BeginExpansion
\mbox{\boldmath{$\sigma$}}%
%EndExpansion
^{2} & \mapsto & \sigma^{2}=\left(
\begin{array}
[c]{cc}%
0 & -\mathrm{i}\\
\mathrm{i} & 0
\end{array}
\right) \\%
%TCIMACRO{\TeXButton{sigma}{\mbox{\boldmath{$\sigma$}}}}%
%BeginExpansion
\mbox{\boldmath{$\sigma$}}%
%EndExpansion
^{3} & \mapsto & \sigma^{3}=\left(
\begin{array}
[c]{cc}%
1 & 0\\
0 & -1
\end{array}
\right)  .
\end{array}
\label{20}%
\end{equation}

\subsection{Minimal left and right ideals in the Pauli Algebra and Spinors}

It is not our intention to present the details of algebraic spinor theory here
(see, e.g., \cite{ficaro,roca,lounesto}). However, we will need to recall some
facts. The elements $\mathbf{e}_{\pm}=\frac{1}{2}(1+%
%TCIMACRO{\TeXButton{sigma}{\mbox{\boldmath{$\sigma$}}}}%
%BeginExpansion
\mbox{\boldmath{$\sigma$}}%
%EndExpansion
_{3})=\frac{1}{2}(1+\mathbf{m}_{3}\mathbf{m}_{0})\in\mathbb{R}_{1,3}%
^{(0)}\simeq\mathbb{R}_{3,0}$, $\mathbf{e}_{\pm}^{2}=\mathbf{e}_{\pm}$ are
minimal idempotents. They generate the minimal left and right ideals%
\begin{equation}
\mathbf{I}_{\pm}=\mathbb{R}_{1,3}^{(0)}\mathbf{e}_{\pm},\hspace{0.15in}%
\mathbf{R}_{\pm}\mathbf{=e}_{\pm}\mathbb{R}_{1,3}^{(0)}. \label{25}%
\end{equation}

From now on we write $\mathbf{e=e}_{+}$. It can be\ easily shown (see below)
that, e.g., $\mathbf{I=I}_{+}$ has the structure of a $2$-dimensional vector
space over the complex field \cite{ficaro,roca}, i.e., $\mathbf{I\simeq
}\mathbb{C}^{2}$. The elements of the vector space $\mathbf{I}$ are called
algebraic \textit{contravariant undotted spinors} and the elements of
$\mathbb{C}^{2}$ are the usual \ \textit{contravariant undotted spinors} used
in physics textbooks. They carry the $D^{(\frac{1}{2},0)\text{ }}$
representation of $Sl(2,\mathbb{C)}$ \cite{miller}. \ If \ $\mathbf{\varphi\in
I}$ we denote by $\varphi\in\mathbb{C}^{2}$ the usual matrix
representative\footnote{The matrix representation of elements of ideals are of
course, $2\times2$ complex matrices (see, \cite{ficaro}, for details). It
happens that both colums of that matrices have the \textit{same} information
and the representation by column matrices is enough here for our purposes.} of
$\mathbf{\varphi}$ is
\begin{equation}
\varphi=\left(
\begin{array}
[c]{c}%
\varphi^{1}\\
\varphi^{2}%
\end{array}
\right)  ,\hspace{0.15in}\varphi^{1},\varphi^{2}\in\mathbb{C}. \label{26}%
\end{equation}
\ We denote by $\mathbf{\dot{I}=e}\mathbb{R}_{1,3}^{(0)}$ the space of the
\ algebraic covariant dotted spinors. We have the isomorphism, $\mathbf{\dot
{I}\simeq(}\mathbb{C}^{2})^{\dagger}\simeq\mathbb{C}_{2}$, where $\dagger$
denotes Hermitian conjugation. The elements of $\mathbf{(}\mathbb{C}%
^{2})^{\dagger}$ are the usual contravariant spinor fields used in physics
textbooks. They carry the $D^{(0,\frac{1}{2})\text{ }}$ representation of
$Sl(2,\mathbb{C)}$ \cite{miller}. If \ $\overset{\cdot}{%
%TCIMACRO{\TeXButton{xi}{\mbox{\boldmath{$\xi$}}}}%
%BeginExpansion
\mbox{\boldmath{$\xi$}}%
%EndExpansion
}\in\mathbf{\dot{I}}$ its matrix representation \ in $\mathbf{(}\mathbb{C}%
^{2})^{\dagger}$ \ is a row matrix usually denoted by%
\begin{equation}
\dot{\xi}=\left(
\begin{array}
[c]{cc}%
\xi_{\dot{1}} & \xi_{\dot{2}}%
\end{array}
\right)  ,\hspace{0.15in}\xi_{\dot{1}},\xi_{\dot{2}}\in\mathbb{C}. \label{27}%
\end{equation}
The following representation of $\overset{\cdot}{%
%TCIMACRO{\TeXButton{xi}{\mbox{\boldmath{$\xi$}}}}%
%BeginExpansion
\mbox{\boldmath{$\xi$}}%
%EndExpansion
}\in\mathbf{\dot{I}}$ \ in $\mathbf{(}\mathbb{C}^{2})^{\dagger}$ is extremely
convenient. We say that to a covariant undotted spinor $\xi$ there corresponds
a covariant dotted spinor $\dot{\xi}$ given by
\begin{equation}
\mathbf{\dot{I}}\ni\overset{\cdot}{%
%TCIMACRO{\TeXButton{xi}{\mbox{\boldmath{$\xi$}}}}%
%BeginExpansion
\mbox{\boldmath{$\xi$}}%
%EndExpansion
}\mapsto\dot{\xi}=\bar{\xi}\varepsilon\in\mathbf{(}\mathbb{C}^{2})^{\dagger
},\hspace{0.15in}\bar{\xi}_{1},\bar{\xi}_{2}\in\mathbb{C}, \label{27'}%
\end{equation}
with%
\begin{equation}
\varepsilon=\left(
\begin{array}
[c]{cc}%
0 & 1\\
-1 & 0
\end{array}
\right)  . \label{27''}%
\end{equation}

\bigskip We can easily \ find a basis for $\mathbf{I}$ and $\mathbf{\dot{I}}$.
Indeed, since $\mathbf{I}=\mathbb{R}_{1,3}^{(0)}\mathbf{e}$ we have that any $%
%TCIMACRO{\TeXButton{bvarphi}{\mbox{\boldmath{$\varphi$}}}}%
%BeginExpansion
\mbox{\boldmath{$\varphi$}}%
%EndExpansion
\mathbf{\in I}$ can be written as%
\[%
%TCIMACRO{\TeXButton{bvarphi}{\mbox{\boldmath{$\varphi$}}}}%
%BeginExpansion
\mbox{\boldmath{$\varphi$}}%
%EndExpansion
\mathbf{=}%
%TCIMACRO{\TeXButton{bvarphi}{\mbox{\boldmath{$\varphi$}}}}%
%BeginExpansion
\mbox{\boldmath{$\varphi$}}%
%EndExpansion
^{1}%
%TCIMACRO{\TeXButton{vartheta}{\mbox{\boldmath{$\vartheta$}}}}%
%BeginExpansion
\mbox{\boldmath{$\vartheta$}}%
%EndExpansion
_{1}\mathbf{+}%
%TCIMACRO{\TeXButton{bvarphi}{\mbox{\boldmath{$\varphi$}}}}%
%BeginExpansion
\mbox{\boldmath{$\varphi$}}%
%EndExpansion
^{2}%
%TCIMACRO{\TeXButton{vartheta}{\mbox{\boldmath{$\vartheta$}}}}%
%BeginExpansion
\mbox{\boldmath{$\vartheta$}}%
%EndExpansion
_{2}%
\]
where
\begin{align}%
%TCIMACRO{\TeXButton{vartheta}{\mbox{\boldmath{$\vartheta$}}}}%
%BeginExpansion
\mbox{\boldmath{$\vartheta$}}%
%EndExpansion
_{1}  &  \mathbf{=}\mathbf{e,\hspace{0.15in}}%
%TCIMACRO{\TeXButton{vartheta}{\mbox{\boldmath{$\vartheta$}}}}%
%BeginExpansion
\mbox{\boldmath{$\vartheta$}}%
%EndExpansion
_{2}=%
%TCIMACRO{\TeXButton{sigma}{\mbox{\boldmath{$\sigma$}}}}%
%BeginExpansion
\mbox{\boldmath{$\sigma$}}%
%EndExpansion
_{1}\mathbf{e}\nonumber\\%
%TCIMACRO{\TeXButton{bvarphi}{\mbox{\boldmath{$\varphi$}}}}%
%BeginExpansion
\mbox{\boldmath{$\varphi$}}%
%EndExpansion
^{1}  &  =a+\mathbf{i}b,\hspace{0.15in}%
%TCIMACRO{\TeXButton{bvarphi}{\mbox{\boldmath{$\varphi$}}}}%
%BeginExpansion
\mbox{\boldmath{$\varphi$}}%
%EndExpansion
^{2}=c+\mathbf{i}d,\hspace{0.15in}a,b,c,d\in\mathbb{R.} \label{27A}%
\end{align}

Analogously we find that any $\overset{\cdot}{%
%TCIMACRO{\TeXButton{xi}{\mbox{\boldmath{$\xi$}}}}%
%BeginExpansion
\mbox{\boldmath{$\xi$}}%
%EndExpansion
}\in\mathbf{\dot{I}}$ can be written as
\begin{align}
\overset{\cdot}{%
%TCIMACRO{\TeXButton{xi}{\mbox{\boldmath{$\xi$}}}}%
%BeginExpansion
\mbox{\boldmath{$\xi$}}%
%EndExpansion
}  &  =%
%TCIMACRO{\TeXButton{xi}{\mbox{\boldmath{$\xi$}}}}%
%BeginExpansion
\mbox{\boldmath{$\xi$}}%
%EndExpansion
^{\dot{1}}\mathbf{s}^{\dot{1}}+%
%TCIMACRO{\TeXButton{xi}{\mbox{\boldmath{$\xi$}}}}%
%BeginExpansion
\mbox{\boldmath{$\xi$}}%
%EndExpansion
_{\dot{2}}\mathbf{s}^{\dot{2}}\nonumber\\
\mathbf{s}^{\dot{1}}  &  =\mathbf{e,}\hspace{0.15in}\mathbf{s}^{\dot{2}%
}=\mathbf{e}%
%TCIMACRO{\TeXButton{sigma}{\mbox{\boldmath{$\sigma$}}}}%
%BeginExpansion
\mbox{\boldmath{$\sigma$}}%
%EndExpansion
_{1}\mathbf{.} \label{27b}%
\end{align}

Defining the mapping
\begin{align}%
%TCIMACRO{\TeXButton{iota}{\mbox{\boldmath{$\iota$}}}}%
%BeginExpansion
\mbox{\boldmath{$\iota$}}%
%EndExpansion
&  :\mathbf{I}\otimes\mathbf{\dot{I}\rightarrow}\mathbb{R}_{1,3}^{(0)}%
\simeq\mathbb{R}_{3,0},\nonumber\\%
%TCIMACRO{\TeXButton{iota}{\mbox{\boldmath{$\iota$}}}}%
%BeginExpansion
\mbox{\boldmath{$\iota$}}%
%EndExpansion
(%
%TCIMACRO{\TeXButton{varphi}{\mbox{\boldmath{$\varphi$}}}}%
%BeginExpansion
\mbox{\boldmath{$\varphi$}}%
%EndExpansion
\mathbf{\otimes}\overset{\cdot}{%
%TCIMACRO{\TeXButton{xi}{\mbox{\boldmath{$\xi$}}}}%
%BeginExpansion
\mbox{\boldmath{$\xi$}}%
%EndExpansion
})  &  =%
%TCIMACRO{\TeXButton{varphi}{\mbox{\boldmath{$\varphi$}}}}%
%BeginExpansion
\mbox{\boldmath{$\varphi$}}%
%EndExpansion
\overset{\cdot}{%
%TCIMACRO{\TeXButton{xi}{\mbox{\boldmath{$\xi$}}}}%
%BeginExpansion
\mbox{\boldmath{$\xi$}}%
%EndExpansion
} \label{27c}%
\end{align}
we have%
\begin{align}
1  &  \equiv%
%TCIMACRO{\TeXButton{sigma}{\mbox{\boldmath{$\sigma$}}}}%
%BeginExpansion
\mbox{\boldmath{$\sigma$}}%
%EndExpansion
_{0}=%
%TCIMACRO{\TeXButton{iota}{\mbox{\boldmath{$\iota$}}}}%
%BeginExpansion
\mbox{\boldmath{$\iota$}}%
%EndExpansion
(\mathbf{s}_{1}\otimes\mathbf{s}^{\dot{1}}+\mathbf{s}_{2}\otimes
\mathbf{s}^{\dot{2}}),\nonumber\\%
%TCIMACRO{\TeXButton{sigma}{\mbox{\boldmath{$\sigma$}}}}%
%BeginExpansion
\mbox{\boldmath{$\sigma$}}%
%EndExpansion
_{1}  &  =-%
%TCIMACRO{\TeXButton{iota}{\mbox{\boldmath{$\iota$}}}}%
%BeginExpansion
\mbox{\boldmath{$\iota$}}%
%EndExpansion
(\mathbf{s}_{1}\otimes\mathbf{s}^{\dot{2}}+\mathbf{s}_{2}\otimes
\mathbf{s}^{\dot{1}}),\nonumber\\%
%TCIMACRO{\TeXButton{sigma}{\mbox{\boldmath{$\sigma$}}}}%
%BeginExpansion
\mbox{\boldmath{$\sigma$}}%
%EndExpansion
_{2}  &  =%
%TCIMACRO{\TeXButton{iota}{\mbox{\boldmath{$\iota$}}}}%
%BeginExpansion
\mbox{\boldmath{$\iota$}}%
%EndExpansion
[\mathbf{i}(\mathbf{s}_{1}\otimes\mathbf{s}^{\dot{2}}-\mathbf{s}_{2}%
\otimes\mathbf{s}^{\dot{1}})],\nonumber\\%
%TCIMACRO{\TeXButton{sigma}{\mbox{\boldmath{$\sigma$}}}}%
%BeginExpansion
\mbox{\boldmath{$\sigma$}}%
%EndExpansion
_{3}  &  =-%
%TCIMACRO{\TeXButton{iota}{\mbox{\boldmath{$\iota$}}}}%
%BeginExpansion
\mbox{\boldmath{$\iota$}}%
%EndExpansion
(\mathbf{s}_{1}\otimes\mathbf{s}^{\dot{1}}-\mathbf{s}_{2}\otimes
\mathbf{s}^{\dot{2}}). \label{27d}%
\end{align}

From this it follows that we have the identification%
\begin{equation}
\mathbb{R}_{3,0}\simeq\mathbb{R}_{1,3}^{(0)}\simeq\mathbb{C(}2\mathbb{)=}%
\mathbf{I}\otimes_{\mathbb{C}}\mathbf{\dot{I},} \label{27dd}%
\end{equation}
from where it follows that each Pauli number can be written as an appropriate
Clifford product of sums of algebraic contravariant undotted spinors and
algebraic\ covariant dotted spinors, and of course a representative of a Pauli
number in $\mathbb{C}^{2}$ can be written as an appropriate Kronecker product
of a complex column vector \ by a complex row vector.

Take an arbitrary $\mathbf{P\in}\mathbb{R}_{3,0}$ such that
\begin{equation}
\mathbf{P}=\frac{1}{j!}p_{\text{ }}^{_{\mathbf{k}_{1}\mathbf{k}_{2}%
...\mathbf{k}_{j}}}\text{ }%
%TCIMACRO{\TeXButton{sigma}{\mbox{\boldmath{$\sigma$}}}}%
%BeginExpansion
\mbox{\boldmath{$\sigma$}}%
%EndExpansion
_{\mathbf{k}_{1}\mathbf{k}_{2}\mathbf{...k}_{j}}, \label{27e}%
\end{equation}
where $p^{_{\mathbf{k}_{1}\mathbf{k}_{2}...\mathbf{k}_{j}}}\in\mathbb{R}$ and%
\begin{equation}%
%TCIMACRO{\TeXButton{sigma}{\mbox{\boldmath{$\sigma$}}}}%
%BeginExpansion
\mbox{\boldmath{$\sigma$}}%
%EndExpansion
_{_{\mathbf{k}_{1}\mathbf{k}_{2}...\mathbf{k}_{j}}}=%
%TCIMACRO{\TeXButton{sigma}{\mbox{\boldmath{$\sigma$}}}}%
%BeginExpansion
\mbox{\boldmath{$\sigma$}}%
%EndExpansion
_{\mathbf{k}_{1}}...%
%TCIMACRO{\TeXButton{sigma}{\mbox{\boldmath{$\sigma$}}}}%
%BeginExpansion
\mbox{\boldmath{$\sigma$}}%
%EndExpansion
_{\mathbf{k}_{j}},\hspace{0.15in}\text{and }%
%TCIMACRO{\TeXButton{sigma}{\mbox{\boldmath{$\sigma$}}}}%
%BeginExpansion
\mbox{\boldmath{$\sigma$}}%
%EndExpansion
_{0}\equiv1\in\mathbb{R}. \label{27f}%
\end{equation}
\ 

With the identification $\mathbb{R}_{3,0}\simeq\mathbb{R}_{1,3}^{(0)}%
\simeq\mathbf{I}\otimes_{\mathbb{C}}\mathbf{\dot{I}}$, we can write also
\begin{equation}
\mathbf{P=P}_{\text{ \ }\dot{B}}^{A}%
%TCIMACRO{\TeXButton{iota}{\mbox{\boldmath{$\iota$}}}}%
%BeginExpansion
\mbox{\boldmath{$\iota$}}%
%EndExpansion
(\mathbf{s}_{A}\otimes\mathbf{s}^{\dot{B}})=\mathbf{P}_{\text{ \ }\dot{B}}%
^{A}s_{B}\mathbf{s}^{\dot{B}}, \label{27g}%
\end{equation}
where the $\mathbf{P}_{\text{ \ }\dot{B}}^{A}=\mathbf{X}_{\text{ \ }\dot{B}%
}^{A}+\mathbf{iY}_{\text{ \ }\dot{B}}^{A}$, $\mathbf{X}_{\text{ \ }\dot{B}%
}^{A},\mathbf{Y}_{\text{ \ }\dot{B}}^{A}\in\mathbb{R}$.

Finally, the matrix representative of the Pauli number $\mathbf{P\in
}\mathbb{R}_{3,0}$\textbf{ }\ is $P\in\mathbb{C(}2)$ given by
\begin{equation}
P=P_{\text{ \ }\dot{B}}^{A}s_{A}s^{\dot{B}}, \label{27h}%
\end{equation}
with $P_{\text{ \ }\dot{B}}^{A}\in\mathbb{C}$ and
\begin{equation}%
\begin{array}
[c]{cc}%
s_{1}=\left(
\begin{array}
[c]{c}%
1\\
0
\end{array}
\right)  & s_{2}=\left(
\begin{array}
[c]{c}%
0\\
1
\end{array}
\right) \\
s^{\dot{1}}=\left(
\begin{array}
[c]{cc}%
1 & 0
\end{array}
\right)  & s^{\dot{2}}=\left(
\begin{array}
[c]{cc}%
0 & 1
\end{array}
\right)  .
\end{array}
\label{27i}%
\end{equation}

It is convenient for our purposes to introduce also covariant undotted spinors
and contravariant dotted spinors. Let $\varphi\in\mathbb{C}^{2}$ be given as
in Eq.(\ref{26}). We define the \textit{covariant} \ version of undotted
spinor $\varphi\in\mathbb{C}^{2}$ as $\varphi^{\ast}$ $\in(\mathbb{C}^{2}%
)^{t}\simeq\mathbb{C}_{2}$ such that%
\begin{align}
\varphi^{\ast}  &  =\left(  \varphi_{1},\varphi_{2}\right)  \equiv\varphi
_{A}s^{A},\nonumber\\
\varphi_{A}  &  =\varphi^{B}\varepsilon_{BA,\hspace{0.15in}}\varphi
^{B}=\varepsilon^{BA}\varphi_{A},\nonumber\\
s^{1}  &  =\left(
\begin{array}
[c]{cc}%
1 & 0
\end{array}
\right)  ,\hspace{0.15in}s^{2}=\left(
\begin{array}
[c]{cc}%
0 & 1
\end{array}
\right)  , \label{27j}%
\end{align}
where\footnote{The symbol \textrm{adiag} means the antidiagonal matrix.}
$\varepsilon_{AB}=\varepsilon^{AB}=\mathrm{adiag}(1,-1)$. We can write due to
the above identifications that there exists $\varepsilon\in\mathbb{C}(2)$
given by Eq.(\ref{27''}) which can be written also as
\begin{equation}
\varepsilon=\varepsilon^{AB}s_{A}\boxtimes s_{B}=\varepsilon_{AB}%
s^{A}\boxtimes s^{B}=\left(
\begin{array}
[c]{cc}%
0 & 1\\
-1 & 0
\end{array}
\right)  =\mathrm{i}\sigma_{2} \label{27K}%
\end{equation}
where $\boxtimes$ denote the \textit{Kronecker} product of matrices. We have,
e.g.,%
\begin{align}
s_{1}\boxtimes s_{2}  &  =\left(
\begin{array}
[c]{c}%
1\\
0
\end{array}
\right)  \boxtimes\left(
\begin{array}
[c]{c}%
0\\
1
\end{array}
\right)  =\left(
\begin{array}
[c]{c}%
1\\
0
\end{array}
\right)  \left(
\begin{array}
[c]{cc}%
0 & 1
\end{array}
\right)  =\left(
\begin{array}
[c]{cc}%
0 & 1\\
0 & 0
\end{array}
\right)  ,\nonumber\\
s^{1}\boxtimes s^{1}  &  =\left(
\begin{array}
[c]{cc}%
1 & 0
\end{array}
\right)  \boxtimes\left(
\begin{array}
[c]{cc}%
0 & 1
\end{array}
\right)  =\left(
\begin{array}
[c]{c}%
1\\
0
\end{array}
\right)  \left(
\begin{array}
[c]{cc}%
1 & 0
\end{array}
\right)  =\left(
\begin{array}
[c]{cc}%
1 & 0\\
0 & 0
\end{array}
\right)  . \label{27l}%
\end{align}

We now introduce the \textit{contravariant }version of the dotted spinor%
\[
\dot{\xi}=\left(
\begin{array}
[c]{cc}%
\xi_{\dot{1}} & \xi_{\dot{2}}%
\end{array}
\right)  \in\mathbb{C}_{2}%
\]
as being $\dot{\xi}^{\ast}\in\mathbb{C}^{2}$ such that%
\begin{align}
\dot{\xi}^{\ast}  &  =\left(
\begin{array}
[c]{c}%
\xi^{\dot{1}}\\
\xi^{\dot{2}}%
\end{array}
\right)  =\xi^{\dot{A}}s_{\dot{A}},\nonumber\\
\xi^{\dot{B}}  &  =\varepsilon^{\dot{B}\dot{A}}\xi_{\dot{A}},\hspace
{0.15in}\xi_{\dot{A}}=\varepsilon_{\dot{B}\dot{A}}\text{ }\xi^{\dot{B}%
},\nonumber\\
s_{\dot{1}}  &  =\left(
\begin{array}
[c]{c}%
1\\
0
\end{array}
\right)  ,s_{\dot{2}}=\left(
\begin{array}
[c]{c}%
0\\
1
\end{array}
\right)  , \label{27m}%
\end{align}
where \ $\varepsilon_{\dot{A}\dot{B}}=\varepsilon^{\dot{A}\dot{B}%
}=\mathrm{adiag}(1,-1)$. We can write due to the above identifications that
there exists $\dot{\varepsilon}\in\mathbb{C}(2)$ such that
\begin{equation}
\dot{\varepsilon}=\varepsilon^{\dot{A}\dot{B}}s_{\dot{A}}\boxtimes s_{\dot{B}%
}=\varepsilon_{\dot{A}\dot{B}}s^{\dot{A}}\boxtimes\dot{s}^{B}=\left(
\begin{array}
[c]{cc}%
0 & 1\\
-1 & 0
\end{array}
\right)  =\varepsilon. \label{27n}%
\end{equation}

Also, recall that even if \ $\{\mathbf{s}_{A}\}$,$\{\mathbf{s}_{\dot{A}}\}$
and $\{s^{\dot{A}}\}$,$\{s^{A}\}$ are bases of distinct spaces, we can
identify their matrix representations, as it is obvious from the above
formulas. So, we have $s_{A}\equiv s_{\dot{A}}$ and also $s^{\dot{A}}=s^{A}$.
This is the reason for the representation of a dotted covariant spinor as in
Eq.(\ref{27'}). Moreover, the above identifications permit us to write the
\textit{matrix} \textit{representation} of a Pauli number $\mathbf{P\in
}\mathbb{R}_{3,0}$ as, e.g.,
\begin{equation}
P=P_{AB}s^{A}\boxtimes s^{B} \label{27o}%
\end{equation}
besides the representation given by Eq.(\ref{27h}).

\section{Clifford and Spinor Bundles}

\subsection{Preliminaries}

To characterize in a rigorous mathematical way the \textit{basic} field
variables used in Sachs `unified' field theory \cite{s1,s2}, we shall\ need to
recall some results of the theory of spinor fields on Lorentzian spacetimes.
Here we follow the approach given in \cite{28,moro}.\footnote{Another
important reference on the subject of spinor fields is \cite{lami}, which
however only deals with the case of spinor fields on Riemannian manifolds.}

Recall that a Lorentzian manifold is a pair $(M,g)$, where $g\in\sec T^{2,0}M$
is a Lorentzian metric of signature $(1,3)$, i.e., for all $x\in M$,
$T_{x}M\simeq T_{x}^{\ast}M\simeq\mathbb{R}^{1,3}$, where $\mathbb{R}^{1,3}$
is the vector Minkowski space.

Recall that a Lorentzian spacetime is a pentuple $(M,g,D\mathbf{,\tau}%
_{g},\mathbf{\uparrow})$ where $(M,g,$\linebreak$\mathbf{\tau}_{g}%
,\mathbf{\uparrow})$ is an oriented Lorentzian manifold\footnote{Oriented by
the volume element $\mathbf{\tau}_{g}\in\sec%
%TCIMACRO{\dbigwedge \nolimits^{4}}%
%BeginExpansion
{\displaystyle\bigwedge\nolimits^{4}}
%EndExpansion
T^{\ast}M$.} \ which is also time oriented by an appropriated equivalence
relation\footnote{See \cite{sw} for details.} (denoted $\uparrow$) for the
timelike vectors at the tangent space $T_{x}M$, $\forall x\in M$.
$D$\textbf{\ }is a linear connection for $M$ such that $Dg=0$, $\mathbf{\Theta
}(D)=0$, $\mathcal{R}(D)\neq0$, where $\mathbf{\Theta}$ and $\mathcal{R}$ are
respectively the torsion and curvature tensors of $D$.

Now, Sachs theory uses spinor fields. These objects are sections of so-called
\textit{spinor bundles}, which only exist in \textit{spin manifolds}. The ones
of interest in Sachs theory are the matrix representation of the bundle of
dotted spinor fields, i.e., $S(M)=P_{\mathrm{Spin}_{1,3}^{e}}(M)\times
_{D^{(\frac{1}{2},0)}}\mathbb{C}^{2}$ and the matrix representation of the
bundle of undotted spinor fields (here denoted by) $\bar{S}%
(M)=P_{\mathrm{Spin}_{1,3}^{e}}(M)\times_{D^{(0,\frac{1}{2})}}\mathbb{C}_{2}$
. In the previous formula $D^{(\frac{1}{2},0)}$ and $D^{(0,\frac{1)}{2}}$
\ are the two fundamental non equivalent $2$-dimensional representations of
$Sl(2,\mathbb{C)\simeq}\mathrm{Spin}_{1,3}^{e}$, the universal covering group
of $\mathrm{SO}_{1,3}^{e}$, the restrict orthochronous Lorentz group.
$P_{\mathrm{Spin}_{1,3}^{e}}(M)$ is a principal bundle called the spin
structure bundle\footnote{It is a covering space of $P_{\mathrm{SO}_{1,3}^{e}%
}(M)$. See, e.g., \cite{moro} for details. Sections of \ $P_{\mathrm{Spin}%
_{1,3}^{e}}(M)$ are the so-called spin frames, i.e., a pair $(\Sigma,u)$ where
for any $x\in M$, $\Sigma(x)$ is an othonormal frame and $u(x)$ belongs to the
$\mathrm{Spin}_{1,3}^{e}$. For details see \cite{moro,28,rosw}.}. We recall
that it is a classical result \ (Geroch theorem \cite{g1}) that a
$4$-dimensional Lorentzian manifold is a spin manifold if and only if
$P_{\mathrm{SO}_{1,3}^{e}}(M)$ has a global section\footnote{In what follows
$P_{\mathrm{SO}_{1,3}^{e}}(M)$ denotes the principal bundle of oriented
\textit{Lorentz tetrads}. We presuppose that the reader is acquainted with the
structure of $P_{\mathrm{SO}_{1,3}^{e}}(M)$, whose sections are the time
oriented and oriented orthonormal frames, each one \ associated by a local
trivialization to a \textit{unique} element of $SO_{1,3}^{e}(M).$}, i.e., if
there exists a set\footnote{Called vierbein.} $\{\mathbf{e}_{\mathbf{0}%
},\mathbf{e}_{\mathbf{1}},\mathbf{e}_{\mathbf{2}},\mathbf{e}_{\mathbf{3}}\}$
of orthonormal fields defined for all $x\in M$. \ In other words, in order for
spinor fields to exist in a $4$-dimensional spacetime the orthonormal frame
bundle must be \textit{trivial.}

Now, the so-called tangent ($TM$) and cotangent ($T^{\ast}M$) bundles, the
tensor bundle ($\oplus_{r,s}\otimes_{s}^{r}TM)$ and the bundle of differential
forms for the spacetime are the bundles denoted by
\begin{align}
TM  &  =P_{\mathrm{SO}_{1,3}^{e}}(M)\times_{\rho_{_{1,3}}}\mathbb{R}%
^{1,3},\hspace{0.15in}T^{\ast}M=P_{\mathrm{SO}_{1,3}^{e}}(M)\times_{\rho
_{1,3}^{\ast}}\mathbb{R}^{1,3},\label{S1}\\
\oplus_{r,s}\otimes_{s}^{r}TM  &  =P_{\mathrm{SO}_{1,3}^{e}}(M)\times
_{\otimes_{s}^{r}\rho_{_{1,3}}}\mathbb{R}^{1,3},\hspace{0.15in}%
%TCIMACRO{\dbigwedge }%
%BeginExpansion
{\displaystyle\bigwedge}
%EndExpansion
T^{\ast}M=P_{\mathrm{SO}_{1,3}^{e}}(M)\times_{\Lambda_{\rho_{1,3}^{\ast}}^{k}}%
%TCIMACRO{\dbigwedge }%
%BeginExpansion
{\displaystyle\bigwedge}
%EndExpansion
\mathbb{R}^{1,3}.\nonumber
\end{align}

In Eqs.(\ref{S1})
\begin{equation}
\rho_{_{1,3}}:\mathrm{SO}_{1,3}^{e}\rightarrow\mathrm{SO}^{e}(\mathbb{R}%
^{1,3}) \label{S2}%
\end{equation}
\ is the standard vector representation of $\mathrm{SO}_{1,3}^{e}$ usually
denoted by \footnote{See, e.g., \cite{miller} if you need details.}
$D^{(\frac{1}{2},\frac{1}{2})}=$ $D^{(\frac{1}{2},0)}\otimes D^{\left(
0,\frac{1}{2}\right)  }$and $\rho_{1,3}^{\ast}$ is the dual (vector)
representation $\rho_{1,3}^{\ast}\left(  l)=\rho_{1,3}(l^{-1}\right)  ^{t}$.
\ Also $\otimes_{s}^{r}\rho_{_{1,3}}$ and $\Lambda_{\rho_{1,3}^{\ast}}^{k}$
are the induced tensor product and induced exterior power product
representations of $\mathrm{SO}_{1,3}^{e}$. We now briefly recall the
definition and some properties of Clifford bundle of multivector fields
\cite{28}. We have,%
\begin{align}
\mathcal{C\ell}(TM)  &  =P_{\mathrm{SO}_{1,3}^{e}}(M)\times_{c\ell
_{\rho_{_{1,3}}}}\mathbb{R}_{1,3}\nonumber\\
&  =P_{\mathrm{Spin}_{1,3}^{e}}(M)\times_{\mathrm{Ad}}\mathbb{R}_{1,3}.
\label{C3}%
\end{align}
Now, recall that \cite{lounesto} $\mathrm{Spin}_{1,3}^{e}\subset
\mathbb{R}_{1,3}^{(0)}$. Consider the $2$-$1$ homomorphism $\mathrm{h}%
:\mathrm{Spin}_{1,3}^{e}\rightarrow\mathrm{SO}_{1,3}^{e},\mathrm{h}(\pm u)=l$.
Then $c\ell_{_{\rho_{_{1,3}}}}$ is the following representation of
$\mathrm{SO}_{1,3}^{e}$,
\begin{align}
c\ell_{_{\rho_{_{1,3}}}}  &  :\mathrm{SO}_{1,3}^{e}\rightarrow\mathrm{Aut}%
(\mathbb{R}_{1,3}),\nonumber\\
c\ell_{_{\rho_{_{1,3}}}}(L)  &  =\mathrm{Ad}_{u}:\mathbb{R}_{1,3}%
\rightarrow\mathbb{R}_{1,3},\nonumber\\
\mathrm{Ad}_{u}(\mathbf{m)}  &  =u\mathbf{m}u^{-1} \label{C.4}%
\end{align}
i.e., it\ is the standard orthogonal transformation of $\mathbb{R}_{1,3}$
induced by an orthogonal transformation of $\mathbb{R}^{1,3}$. Note that
$\mathrm{Ad}_{u}$ act on vectors as the $D^{(\frac{1}{2},\frac{1}{2})}$
representation of $\mathrm{SO}_{1,3}^{e}$ and on multivectors as the induced
exterior power representation of that group. Indeed, observe, e.g., that for
$\mathbf{v\in}\mathbb{R}^{1,3}\subset\mathbb{R}_{1,3}$ we have in standard
notation
\[
L\mathbf{v=v}^{\nu}L_{\nu}^{\mu}\mathbf{m}_{\mathbf{\mu}}=\mathbf{v}^{\nu
}u\mathbf{m}_{\nu}u^{-1}=u\mathbf{v}u^{-1}.
\]

The proof of the second line of Eq.(\ref{C3}) is as follows. Consider the
representation%
\begin{align}
\mathrm{Ad}  &  :\mathrm{Spin}_{1,3}^{e}\rightarrow\mathrm{Aut}(\mathbb{R}%
_{1,3}),\nonumber\\
\mathrm{Ad}_{u}  &  :\mathbb{R}_{1,3}\rightarrow\mathbb{R}_{1,3}%
,\hspace{0.15in}\mathrm{Ad}_{u}\left(  m\right)  =umu^{-1}. \label{C5}%
\end{align}

Since $\mathrm{Ad}_{-1}=1$($=$ \textrm{identity) }the representation
$\mathrm{Ad}$ descends to a representation of $\mathrm{SO}_{1,3}^{e}$. This
representation is just $c\ell(\rho_{_{1,3}})$, from where the desired result follows.

Sections of $\mathcal{C\ell}(TM)$ can be called Clifford fields (of
multivectors). The sections of the even subbundle $\mathcal{C\ell}%
^{(0)}(TM)=P_{\mathrm{Spin}_{1,3}^{e}}(M)\times_{\mathrm{Ad}}\mathbb{R}%
_{1,3}^{(0)}$ may be called Pauli fields (of multivectors). Define the real
spinor bundles%

\begin{equation}
\mathcal{S}(M)=P_{\mathrm{Spin}_{1,3}^{e}}(M)\times_{l}\mathbf{I,\hspace
{0.15in}}\mathcal{\dot{S}}(M)=P_{\mathrm{Spin}_{1,3}^{e}}(M)\times
_{r}\mathbf{\dot{I}} \label{C.5a}%
\end{equation}
where $l$ stands for a left modular representation of $\mathrm{Spin}_{1,3}%
^{e}$ in $\mathbb{R}_{1,3}$ that mimics the $D^{(\frac{1}{2},0)}$
representation of $Sl(2,\mathbb{C)}$ and $r$ stands for a right modular
representation of $\mathrm{Spin}_{1,3}^{e}$ in $\mathbb{R}_{1,3}$ that mimics
the $D^{(0,\frac{1}{2})}$ representation of $Sl(2,\mathbb{C)}$.

Also recall that if $\bar{S}(M)$ is the bundle whose sections are the spinor
fields $\bar{\varphi}=(\bar{\varphi}_{1},\bar{\varphi}_{2})=\dot{\varphi
}\varepsilon=(\varphi^{\dot{1}},\varphi^{\dot{2}})$, then it is isomorphic to
the space of contravariant dotted spinors. We have,%
\begin{equation}
S(M)\mathbf{\simeq}P_{\mathrm{Spin}_{1,3}^{e}}(M)\times_{D^{(\frac{1}{2},0)}%
}\mathbb{C}^{2},\mathbf{\hspace{0.15in}}\dot{S}\left(  M\right)
\mathbf{\simeq}P_{\mathrm{Spin}_{1,3}^{e}}(M)\times_{D^{(0,\frac{1}{2})}%
}\mathbb{C}_{2}\simeq\bar{S}(M), \label{C6}%
\end{equation}
and from our playing with the Pauli algebra and dotted and undotted spinors in
section 2 we have that:
\begin{equation}
\mathcal{S}(M)\simeq S(M),\hspace{0.15in}\mathcal{\dot{S}}(M)\simeq\dot
{S}\left(  M\right)  \mathbf{\simeq}\bar{S}(M). \label{C.6bis}%
\end{equation}

Then, we have the obvious isomorphism%
\begin{align}
\mathcal{C\ell}^{(0)}(TM)  &  =P_{\mathrm{Spin}_{1,3}^{e}}(M)\times
_{\mathrm{Ad}}\mathbb{R}_{1,3}^{(0)}\nonumber\\
&  =P_{\mathrm{Spin}_{1,3}^{e}}(M)\times_{l\otimes r}\mathbf{I\otimes
}_{\mathbb{C}}\mathbf{\dot{I}}\nonumber\\
&  =\mathcal{S}(M)\otimes_{\mathbb{C}}\mathcal{\dot{S}}(M). \label{C7}%
\end{align}

Let us now introduce the following bundle,%
\begin{equation}
\mathbb{C}\ell^{(0)}(M)=P_{\mathrm{Spin}_{1,3}^{e}}(M)\times_{_{D^{(\frac
{1}{2}0)}\otimes D^{(0,\frac{1}{2})}}}\mathbb{C}(2). \label{C8'}%
\end{equation}
It is clear that%
\begin{equation}
\mathbb{C}\ell^{(0)}(M)=S(M)\otimes_{\mathbb{C}}\bar{S}(M)\simeq
\mathcal{C\ell}^{(0)}(M). \label{C8}%
\end{equation}

Finally, we consider the bundle
\begin{equation}
\mathcal{C\ell}^{(0)}(TM)\otimes%
%TCIMACRO{\dbigwedge }%
%BeginExpansion
{\displaystyle\bigwedge}
%EndExpansion
T^{\ast}M\simeq\mathbb{C}\ell^{(0)}(M)\otimes%
%TCIMACRO{\dbigwedge }%
%BeginExpansion
{\displaystyle\bigwedge}
%EndExpansion
T^{\ast}M. \label{C9}%
\end{equation}

Sections of $\mathcal{C\ell}^{(0)}(TM)\otimes%
%TCIMACRO{\dbigwedge }%
%BeginExpansion
{\displaystyle\bigwedge}
%EndExpansion
T^{\ast}M$ may be called \textit{Pauli valued differential forms} and sections
of $\mathbb{C}\ell^{(0)}(M)\otimes%
%TCIMACRO{\dbigwedge }%
%BeginExpansion
{\displaystyle\bigwedge}
%EndExpansion
T^{\ast}M$ may be called \textit{matrix Pauli valued differential forms}.
\textrm{ }

\bigskip\ Denote by $\mathcal{C\ell}_{(0,2)}^{\left(  0\right)  }(TM)$ $\ $the
seven dimensional subbundle $\left(  \mathbb{R\oplus}%
%TCIMACRO{\dbigwedge \nolimits^{2}}%
%BeginExpansion
{\displaystyle\bigwedge\nolimits^{2}}
%EndExpansion
TM\right)  \subset%
%TCIMACRO{\dbigwedge }%
%BeginExpansion
{\displaystyle\bigwedge}
%EndExpansion
TM\hookrightarrow\mathcal{C\ell}^{(0)}(TM)\subset\mathcal{C\ell}(TM)$. \ Now,
\ let $\langle x^{\mu}\rangle$ be the coordinate functions of a chart of the
maximal atlas of $M$. The fundamental field variable of Sachs theory can be
described as%
\[
\mathbf{Q=q}_{\mu}\otimes dx^{\mu}\equiv q_{\mu}dx^{\mu}\mathbf{\in}%
\sec\mathcal{C\ell}_{(0,2)}^{(0)}(TM)\otimes%
%TCIMACRO{\dbigwedge }%
%BeginExpansion
{\displaystyle\bigwedge}
%EndExpansion
T^{\ast}M\subset\sec\mathcal{C\ell}^{(0)}(TM)\otimes%
%TCIMACRO{\dbigwedge }%
%BeginExpansion
{\displaystyle\bigwedge}
%EndExpansion
T^{\ast}M
\]
i.e., a Pauli valued $1$-form obeying certain conditions to be presented
below. If we work (as Sachs did) with $\mathbb{C}\ell^{(0)}(M)\otimes%
%TCIMACRO{\dbigwedge }%
%BeginExpansion
{\displaystyle\bigwedge}
%EndExpansion
T^{\ast}M$, a representative of $\mathbf{Q}$ is $Q\in\sec\mathbb{C}\ell
^{(0)}(M)\otimes%
%TCIMACRO{\dbigwedge }%
%BeginExpansion
{\displaystyle\bigwedge}
%EndExpansion
T^{\ast}M$ such that\footnote{Note that a bold index (sub or superscript), say
$\mathbf{a}$ take the values $0,1,2,3$.}%
\begin{equation}
Q=q_{\mu}(x)dx^{\mu}=h_{\mu}^{\mathbf{a}}(x)dx^{\mu}\sigma_{\mathbf{a}},
\label{S4}%
\end{equation}
where $\sigma_{0}=\left(
\begin{array}
[c]{cc}%
1 & 0\\
0 & 1
\end{array}
\right)  $ and $\sigma_{j}$ ($j\mathbf{=}1,2,3$) are the Pauli matrices. \ We
observe that the notation anticipates the fact that in Sachs theory the
variables $h_{\mu}^{\mathbf{a}}(x)$ define the set $\{\theta^{\mathbf{a}%
}\}\equiv\{\theta^{\mathbf{0}},\theta^{\mathbf{1}},\theta^{\mathbf{2}}%
,\theta^{\mathbf{3}}\}$ with
\begin{equation}
\theta^{\mathbf{a}}=h_{\mu}^{\mathbf{a}}dx^{\mu}\in\sec%
%TCIMACRO{\dbigwedge }%
%BeginExpansion
{\displaystyle\bigwedge}
%EndExpansion
T^{\ast}M, \label{S5}%
\end{equation}
which is the dual basis of $\ \{\mathbf{e}_{\mathbf{a}}\}\equiv\{\mathbf{e}%
_{\mathbf{0}},\mathbf{e}_{\mathbf{1}},\mathbf{e}_{\mathbf{2}},\mathbf{e}%
_{\mathbf{3}}\}$, $\mathbf{e}_{\mathbf{a}}\in\sec TM$. \ We denote by
$\{e_{\mu}\}=\{e_{0},e_{1},e_{2},e_{3}\}$, a coordinate basis associated with
the local chart $\langle x^{\mu}\rangle$ covering $U\subset M$ . We have
\ $e_{\mu}=h_{\mu}^{\mathbf{a}}\mathbf{e}_{\mathbf{a}}\in\sec TM$, \ and the
set $\{e_{\mu}\}$ is the dual basis of $\{dx^{\mu}\}\equiv\{dx^{0}%
,dx^{1},dx^{2},dx^{3}\}$. We will also use the \textit{reciprocal basis} to a
given basis $\{\mathbf{e}_{\mathbf{a}}\}$, i.e., the set $\{\mathbf{e}%
^{\mathbf{a}}\}\equiv\{\mathbf{e}^{\mathbf{0}},\mathbf{e}^{\mathbf{1}%
},\mathbf{e}^{\mathbf{2}},\mathbf{e}^{\mathbf{3}}\},\mathbf{e}^{\mathbf{a}}%
\in\sec TM$, with $g(\mathbf{e}_{\mathbf{a}},e^{\mathbf{b}})=\delta
_{a}^{\mathbf{b}}$ \ and the \textit{reciprocal basis} to $\{\theta
^{\mathbf{a}}\}$, i.e., the set\ $\{\theta_{\mathbf{a}}\}=\{\theta
_{\mathbf{0}},\theta_{\mathbf{1}},\theta_{\mathbf{2}},\theta_{\mathbf{3}}\}$,
with $\theta_{\mathbf{a}}(e^{\mathbf{b}})=\delta_{a}^{\mathbf{b}}$. Recall
that since $\eta_{\mathbf{ab}}=g(\mathbf{e}_{\mathbf{a}},\mathbf{e}%
_{\mathbf{b}})$ ,\ we have%

\begin{equation}
g_{\mu\nu}=g\left(  e_{\mu},e_{\nu}\right)  =h_{\mu}^{\mathbf{a}}h_{\nu
}^{\mathbf{b}}\eta_{\mathbf{ab}}. \label{28new}%
\end{equation}

To continue, we define the%
\begin{equation}
\check{\sigma}_{0}=-\sigma_{0}\text{ and }\check{\sigma}_{\mathbf{j}}%
=\sigma_{\mathbf{j}},\mathbf{j}=1,2,3 \label{28newbis}%
\end{equation}
and
\begin{equation}
\check{Q}=\check{q}_{\mu}(x)dx^{\mu}=h_{\mu}^{\mathbf{a}}(x)dx^{\mu}%
\check{\sigma}_{\mathbf{a}}. \label{29}%
\end{equation}

Also, note that
\begin{equation}
\sigma_{\mathbf{a}}\check{\sigma}_{\mathbf{b}}+\sigma_{\mathbf{b}}%
\check{\sigma}_{\mathbf{a}}=-2\eta_{\mathbf{ab}}. \label{30}%
\end{equation}

Readers of Sachs books \cite{s1,s3} will recall that he said that $Q$ is a
representative of a \textit{quaternion}.\footnote{Note that Sachs represented
$Q$ by $d\mathsf{S}$, which is a very dangerous notation, which we avoid.}
From our previous discussion we see that this statement is \textit{wrong}%
.\footnote{Nevertheless the calculations done by Sachs in \cite{s1} are
correct because he worked always with the matrix representation of
$\mathbf{Q}$. However, his claim of having produce an unified field theory of
gravitation and electromagnetism is wrong as we shall prove in what follows.}
Sachs identification is a dangerous one, because the quaternions are a
division algebra, also-called a noncommutative field or skew-field and objects
like $\mathbf{Q=q}_{\mu}\otimes dx^{\mu}\mathbf{\in}\sec\mathcal{C\ell
}_{(0,2)}^{(0)}(TM)\otimes%
%TCIMACRO{\dbigwedge }%
%BeginExpansion
{\displaystyle\bigwedge}
%EndExpansion
T^{\ast}M\subset\sec\mathcal{C\ell}^{(0)}(TM)\otimes%
%TCIMACRO{\dbigwedge }%
%BeginExpansion
{\displaystyle\bigwedge}
%EndExpansion
T^{\ast}M$ are called \textit{paravector} fields. As it is clear from our
discussion they did not close a \textit{division} algebra.

Next we introduce a tensor product of sections $\mathbf{A,B}\in\sec
\mathcal{C\ell}^{(0)}(M)\otimes%
%TCIMACRO{\dbigwedge }%
%BeginExpansion
{\displaystyle\bigwedge}
%EndExpansion
T^{\ast}M$. Before we do that we recall that from now on%
\begin{equation}
\{1,%
%TCIMACRO{\TeXButton{sigma}{\mbox{\boldmath{$\sigma$}}}}%
%BeginExpansion
\mbox{\boldmath{$\sigma$}}%
%EndExpansion
_{\mathbf{k}},%
%TCIMACRO{\TeXButton{sigma}{\mbox{\boldmath{$\sigma$}}}}%
%BeginExpansion
\mbox{\boldmath{$\sigma$}}%
%EndExpansion
_{\mathbf{k}_{1}\mathbf{k}_{2}},%
%TCIMACRO{\TeXButton{sigma}{\mbox{\boldmath{$\sigma$}}}}%
%BeginExpansion
\mbox{\boldmath{$\sigma$}}%
%EndExpansion
_{\mathbf{123}}\}, \label{30'}%
\end{equation}
refers to a basis of $\mathcal{C\ell}^{(0)}(M)$, i.e., they are
fields.\footnote{We hope that in using (for symbol economy) the same notation
as in section 2 where the $\{1,%
%TCIMACRO{\TeXButton{sigma}{\mbox{\boldmath{$\sigma$}}}}%
%BeginExpansion
\mbox{\boldmath{$\sigma$}}%
%EndExpansion
_{\mathbf{k}},%
%TCIMACRO{\TeXButton{sigma}{\mbox{\boldmath{$\sigma$}}}}%
%BeginExpansion
\mbox{\boldmath{$\sigma$}}%
%EndExpansion
_{\mathbf{k}_{1}\mathbf{k}_{2}},%
%TCIMACRO{\TeXButton{sigma}{\mbox{\boldmath{$\sigma$}}}}%
%BeginExpansion
\mbox{\boldmath{$\sigma$}}%
%EndExpansion
_{\mathbf{123}}\}$ is a basis of $\mathbb{R}_{1,3}^{(0)}\simeq\mathbb{R}%
_{3.0}$ will produce no confusion.}

Recalling Eq.(\ref{27f}) we introduce the (obvious) notation
\begin{equation}
\mathbf{A}=\frac{1}{j!}a_{\mu}^{_{\mathbf{k}_{1}\mathbf{k}_{2}...\mathbf{k}%
_{j}}}\text{ }%
%TCIMACRO{\TeXButton{sigma}{\mbox{\boldmath{$\sigma$}}}}%
%BeginExpansion
\mbox{\boldmath{$\sigma$}}%
%EndExpansion
_{\mathbf{k}_{1}\mathbf{k}_{2}...\mathbf{k}_{j}}dx^{\mu},\hspace
{0.15in}\mathbf{B}=\frac{1}{l!}b_{\mu}^{_{\mathbf{k}_{1}\mathbf{k}%
_{2}...\mathbf{k}_{l}}}%
%TCIMACRO{\TeXButton{sigma}{\mbox{\boldmath{$\sigma$}}}}%
%BeginExpansion
\mbox{\boldmath{$\sigma$}}%
%EndExpansion
_{\mathbf{k}_{1}\mathbf{k}_{2}...\mathbf{k}_{l}}dx^{\mu}, \label{31}%
\end{equation}
where the $a_{\mu}^{_{\mathbf{k}_{1}\mathbf{k}_{2}...\mathbf{k}_{j}}},b_{\mu
}^{_{\mathbf{k}_{1}\mathbf{k}_{2}...\mathbf{k}_{j}}}$ are, in general,
\textit{real }scalar functions. Then, we define%
\begin{equation}
\mathbf{A}\otimes\mathbf{B}=\frac{1}{j!l!}a_{\mu}^{_{\mathbf{k}_{1}%
\mathbf{k}_{2}...\mathbf{k}_{j}}}b_{\nu}^{_{\mathbf{p}_{1}\mathbf{p}%
_{2}...\mathbf{p}_{l}}}%
%TCIMACRO{\TeXButton{sigma}{\mbox{\boldmath{$\sigma$}}}}%
%BeginExpansion
\mbox{\boldmath{$\sigma$}}%
%EndExpansion
_{\mathbf{k}_{1}\mathbf{k}_{2}...\mathbf{k}_{j}}%
%TCIMACRO{\TeXButton{sigma}{\mbox{\boldmath{$\sigma$}}}}%
%BeginExpansion
\mbox{\boldmath{$\sigma$}}%
%EndExpansion
_{\mathbf{p}_{1}\mathbf{p}_{2}...\mathbf{p}_{l}}dx^{\mu}\otimes dx^{\mu}.
\label{32}%
\end{equation}

Let us now compute the tensor product of $\mathbf{Q}\otimes\mathbf{\check{Q}}$
where $\mathbf{Q}\in\sec\mathcal{C\ell}_{(0,2)}^{(0)}(M)\otimes%
%TCIMACRO{\dbigwedge }%
%BeginExpansion
{\displaystyle\bigwedge}
%EndExpansion
T^{\ast}M$. \ We have,%

\begin{align}
\mathbf{Q}\otimes\mathbf{\check{Q}}  &  =\mathbf{q}_{\mu}(x)dx^{\mu}%
\otimes\mathbf{\check{q}}_{\nu}(x)dx^{v}=\mathbf{q}_{\mu}(x)\mathbf{\check{q}%
}_{\nu}(x)dx^{\mu}\otimes dx^{\nu}\nonumber\\
&  =\mathbf{q}_{\mu}(x)\mathbf{\check{q}}_{\nu}(x)\frac{1}{2}(dx^{\mu}\otimes
dx^{\nu}+dx^{\nu}\otimes dx^{\mu})\nonumber\\
&  +\frac{1}{2}\mathbf{q}_{\mu}(x)\mathbf{\check{q}}_{\nu}(x)(dx^{\mu}\otimes
dx^{\nu}-dx^{\nu}\otimes dx^{\mu})\nonumber\\
&  =\frac{1}{2}(\mathbf{q}_{\mu}(x)\mathbf{\check{q}}_{\nu}(x)+\mathbf{q}%
_{\nu}(x)\mathbf{\check{q}}_{\mu}(x))dx^{\mu}\otimes dx^{\nu}\nonumber\\
&  +\frac{1}{2}\mathbf{q}_{\mu}(x)\mathbf{\check{q}}_{\nu}(x)dx^{\mu}\wedge
dx^{\nu}\label{33}\\
&  =(-g_{\mu\nu}%
%TCIMACRO{\TeXButton{sigma}{\mbox{\boldmath{$\sigma$}}}}%
%BeginExpansion
\mbox{\boldmath{$\sigma$}}%
%EndExpansion
_{0})dx^{\mu}\otimes dx^{\nu}\nonumber\\
&  +\frac{1}{4}(\mathbf{q}_{\mu}(x)\mathbf{\check{q}}_{\nu}(x)-\mathbf{q}%
_{\nu}(x)\mathbf{\check{q}}_{\mu}(x))dx^{\mu}\wedge dx^{\nu}\nonumber\\
&  =-g_{\mu\nu}dx^{\mu}\otimes dx^{\nu}+\frac{1}{2}\mathbf{F}_{\mu\nu}%
^{\prime}dx^{\mu}\wedge dx^{\nu}.\nonumber
\end{align}

In writing Eq.(\ref{33}) we have used $dx^{\mu}\wedge dx^{\nu}\equiv dx^{\mu
}\otimes dx^{\nu}-dx^{\nu}\otimes dx^{\mu}$. Also, using
\begin{align}
g_{\mu\nu}  &  =\eta_{\mathbf{ab}}h_{\mu}^{\mathbf{a}}(x)h_{\nu}^{\mathbf{b}%
}(x),\hspace{0.15in}g=g_{\mu\nu}dx^{\mu}\otimes dx^{v}=\eta_{\mathbf{ab}%
}\theta^{\mathbf{a}}\otimes\theta^{\mathbf{b}}\nonumber\\
\mathbf{F}_{\mu\nu}^{\prime}  &  =\mathbf{F}_{\mu\nu}^{\prime k}\mathbf{i}%
%TCIMACRO{\TeXButton{sigma}{\mbox{\boldmath$\sigma$}}}%
%BeginExpansion
\mbox{\boldmath$\sigma$}%
%EndExpansion
_{k}\mathbf{=-}\frac{1}{2}(\varepsilon_{i\text{ }j}^{k}h_{\mu}^{i}(x)h_{\nu
}^{j}(x)\text{ })\mathbf{i}%
%TCIMACRO{\TeXButton{sigma}{\mbox{\boldmath$\sigma$}}}%
%BeginExpansion
\mbox{\boldmath$\sigma$}%
%EndExpansion
_{k};\hspace{0.15in}i,j,k=1,2,3,\nonumber\\
\mathbf{F}^{\prime}  &  =\frac{1}{2}\mathbf{F}_{\mu\nu}^{\prime}dx^{\mu}\wedge
dx^{v}=\frac{1}{2}(\mathbf{F}_{\text{ }\mu\nu}^{\prime ij}%
%TCIMACRO{\TeXButton{sigma}{\mbox{\boldmath$\sigma$}}}%
%BeginExpansion
\mbox{\boldmath$\sigma$}%
%EndExpansion
_{i}%
%TCIMACRO{\TeXButton{sigma}{\mbox{\boldmath$\sigma$}}}%
%BeginExpansion
\mbox{\boldmath$\sigma$}%
%EndExpansion
_{j})dx^{\mu}\wedge dx^{\nu}=(\frac{1}{2}\mathbf{F}_{\mu\nu}^{\prime
k}\mathbf{i}%
%TCIMACRO{\TeXButton{sigma}{\mbox{\boldmath$\sigma$}}}%
%BeginExpansion
\mbox{\boldmath$\sigma$}%
%EndExpansion
_{k})dx^{\mu}\wedge dx^{\nu}\nonumber\\
&  =-\varepsilon_{i\text{ }j}^{k}h_{\mu}^{i}(x)h_{\nu}^{j}(x)\text{ }dx^{\mu
}\wedge dx^{\nu}\mathbf{i\sigma}_{k}\in\sec%
%TCIMACRO{\dbigwedge \nolimits^{2}}%
%BeginExpansion
{\displaystyle\bigwedge\nolimits^{2}}
%EndExpansion
T^{\ast}M\otimes\mathcal{C\ell}_{(2)}^{(0)}\left(  M\right)  \label{33new}%
\end{align}
we can write Eq.(\ref{33}) as
\begin{align}
\mathbf{Q}\otimes\mathbf{\check{Q}}  &  \mathbf{=\mathbf{Q}}\overset
{s}{\mathbf{\otimes}}\mathbf{\check{Q}\mathbf{+}Q\wedge\check{Q}}\nonumber\\
&  =-g+\mathbf{F.} \label{33bis}%
\end{align}

We can also write%
\begin{equation}
\mathbf{Q}\otimes\mathbf{\check{Q}=-}\eta_{\mathbf{ab}}%
%TCIMACRO{\TeXButton{sigma}{\mbox{\boldmath{$\sigma$}}}}%
%BeginExpansion
\mbox{\boldmath{$\sigma$}}%
%EndExpansion
_{0}\theta^{\mathbf{a}}\otimes\theta^{\mathbf{b}}+\varepsilon_{i\text{ }j}%
^{k}\mathbf{i}%
%TCIMACRO{\TeXButton{sigma}{\mbox{\boldmath$\sigma$}}}%
%BeginExpansion
\mbox{\boldmath$\sigma$}%
%EndExpansion
_{k}\theta^{i}\wedge\theta^{j}. \label{33biss}%
\end{equation}

The above formulas show very clearly the mathematical nature of $\mathbf{F}$,
it is a $2$-form with values on the subspace of multivector Clifford fields,
i.e., $\mathbf{F:}%
%TCIMACRO{\dbigwedge \nolimits^{2}}%
%BeginExpansion
{\displaystyle\bigwedge\nolimits^{2}}
%EndExpansion
TM\hookrightarrow\mathcal{C\ell}_{(2)}^{(0)}(TM)\subset\mathcal{C\ell}%
^{(0)}(TM)$. Now, we write the formula for $Q\otimes\tilde{Q}$ where
$Q\in\mathbb{C}(2)\otimes%
%TCIMACRO{\dbigwedge ^{1}}%
%BeginExpansion
{\displaystyle\bigwedge^{1}}
%EndExpansion
T^{\ast}M$ given by Eq.(\ref{S4}) is the matrix representation of
$\mathbf{Q}\in\sec\mathcal{C\ell}_{(0,2)}^{(0)}(M)\otimes%
%TCIMACRO{\dbigwedge ^{1}}%
%BeginExpansion
{\displaystyle\bigwedge^{1}}
%EndExpansion
T^{\ast}M$.

We have,%
\begin{align}
Q\otimes\check{Q}  &  =Q\overset{s}{\mathbf{\otimes}}\tilde{Q}+Q\wedge
\check{Q}\nonumber\\
&  =(\mathbf{-}g_{\mu\nu}dx^{\mu}\otimes dx^{v})\sigma_{0}+(\varepsilon
_{i\text{ }j}^{k}v_{\mu}^{i}(x)v_{\nu}^{j}(x)\text{ }dx^{\mu}\wedge
dx^{v})(-\mathrm{i}\sigma_{k})\nonumber\\
&  =-g\sigma_{0}+\mathbf{F}^{\prime k}\mathrm{i}\sigma_{k}, \label{33again}%
\end{align}
with
\begin{equation}
\mathbf{F}^{\prime k}=\frac{1}{2}\mathbf{F}_{\mu\nu}^{\prime k}dx^{\mu}\wedge
dx^{v}=\varepsilon_{i\text{ }j}^{k}v_{\mu}^{i}(x)v_{\nu}^{j}(x)dx^{\mu}\wedge
dx^{\nu}. \label{36}%
\end{equation}

For future reference we also introduce
\begin{equation}
\mathbf{F}_{\mu\nu}^{\prime}=\mathbf{F}_{\mu\nu}^{\prime k}\mathrm{i}%
\sigma_{k}. \label{37}%
\end{equation}

\subsection{Covariant Derivatives of Spinor Fields}

We now briefly recall the concept of covariant spinor derivatives
\cite{choquet,lami,moro,28}. The idea is the following:

(i) Every connection on the principal bundle of orthonormal frames
$P_{\mathrm{SO}_{1,3}^{e}}(M)$ determines in a canonical way a unique
connection on the principal bundle $P_{\mathrm{Spin}_{1,3}^{e}}(M)$.

(ii) Let $D$ be a covariant derivative operator acting on sections of an
associate vector bundle to $P_{\mathrm{SO}_{1,3}^{e}}(M)$, say, the tensor
bundle \ $\tau M$ and let $D^{s}$ be the corresponding covariant spinor
derivative \ acting on sections of associate vector bundles to
$P_{\mathrm{Spin}_{1,3}^{e}}(M)$, say, e.g., the spinor bundles where
$\mathcal{P}(M)$ may be called \textit{Pauli spinor bundle}. Of course,
$\mathcal{P}(M)\simeq\mathcal{C\ell}^{(0)}(M)$. The matrix representations of
the above bundles are:%

\begin{align}
S\left(  M\right)   &  =P_{\mathrm{Spin}_{1,3}^{e}}(M)\times_{D^{(\frac{1}%
{2}0)}}\mathbb{C}^{2},\hspace{0.15in}\dot{S}(M)=P_{\mathrm{Spin}_{1,3}^{e}%
}(M)\times_{D^{(0,\frac{1}{2})}}\mathbb{C}_{2}\nonumber\\
P(M)  &  =S\left(  M\right)  \otimes\dot{S}(M)=P_{\mathrm{Spin}_{1,3}^{e}%
}(M)\times_{D^{(\frac{1}{2}0)}\otimes D^{(0,\frac{1}{2})}}\mathbb{C}%
^{2}\otimes\mathbb{C}_{2}, \label{D1}%
\end{align}
and $P(M)$ may be called \textit{matrix} \textit{Pauli spinor bundle}. Of
course, $P(M)\simeq\mathbb{C}\ell^{(0)}(M)$.

(iv) We have for $\mathbf{T}\in\sec%
%TCIMACRO{\dbigwedge }%
%BeginExpansion
{\displaystyle\bigwedge}
%EndExpansion
TM\hookrightarrow\mathcal{C\ell}^{(0)}(M)$ and $%
%TCIMACRO{\TeXButton{xi}{\mbox{\boldmath{$\xi$}}}}%
%BeginExpansion
\mbox{\boldmath{$\xi$}}%
%EndExpansion
\in\sec\mathcal{S}(M)$, $\overset{\cdot}{%
%TCIMACRO{\TeXButton{xi}{\mbox{\boldmath{$\xi$}}}}%
%BeginExpansion
\mbox{\boldmath{$\xi$}}%
%EndExpansion
}\in\sec\mathcal{\dot{S}}(M)$, $\mathbf{P}\in\sec$ $\mathcal{P}(M)$and
$\ \mathbf{v}\in\sec TM$ . Then,%

\begin{align}
D_{\mathbf{v}}^{s}(\mathbf{T}\otimes%
%TCIMACRO{\TeXButton{xi}{\mbox{\boldmath{$\xi$}}}}%
%BeginExpansion
\mbox{\boldmath{$\xi$}}%
%EndExpansion
)  &  =D_{\mathbf{v}}\mathbf{T}\otimes%
%TCIMACRO{\TeXButton{xi}{\mbox{\boldmath{$\xi$}}}}%
%BeginExpansion
\mbox{\boldmath{$\xi$}}%
%EndExpansion
+\mathbf{T\otimes}D_{\mathbf{v}}^{s}%
%TCIMACRO{\TeXButton{xi}{\mbox{\boldmath{$\xi$}}}}%
%BeginExpansion
\mbox{\boldmath{$\xi$}}%
%EndExpansion
,\nonumber\\
D_{\mathbf{v}}^{s}(\mathbf{T}\otimes\overset{\cdot}{%
%TCIMACRO{\TeXButton{xi}{\mbox{\boldmath{$\xi$}}}}%
%BeginExpansion
\mbox{\boldmath{$\xi$}}%
%EndExpansion
})  &  =D_{\mathbf{v}}\mathbf{T}\otimes\overset{\cdot}{%
%TCIMACRO{\TeXButton{xi}{\mbox{\boldmath{$\xi$}}}}%
%BeginExpansion
\mbox{\boldmath{$\xi$}}%
%EndExpansion
}+\mathbf{T\otimes}D_{\mathbf{v}}^{s}\overset{\cdot}{%
%TCIMACRO{\TeXButton{xi}{\mbox{\boldmath{$\xi$}}}}%
%BeginExpansion
\mbox{\boldmath{$\xi$}}%
%EndExpansion
}, \label{D.2}%
\end{align}
where%
\begin{align}
D_{\mathbf{v}}\mathbf{T}  &  =\partial_{\mathbf{v}}\mathbf{T}+\frac{1}{2}[%
%TCIMACRO{\TeXButton{omega}{\mbox{\boldmath{$\omega$}}}}%
%BeginExpansion
\mbox{\boldmath{$\omega$}}%
%EndExpansion
_{\mathbf{v}},\mathbf{T}],\nonumber\\
D_{\mathbf{v}}^{s}%
%TCIMACRO{\TeXButton{xi}{\mbox{\boldmath{$\xi$}}}}%
%BeginExpansion
\mbox{\boldmath{$\xi$}}%
%EndExpansion
&  =\partial_{\mathbf{v}}%
%TCIMACRO{\TeXButton{xi}{\mbox{\boldmath{$\xi$}}}}%
%BeginExpansion
\mbox{\boldmath{$\xi$}}%
%EndExpansion
+\frac{1}{2}%
%TCIMACRO{\TeXButton{omega}{\mbox{\boldmath{$\omega$}}}}%
%BeginExpansion
\mbox{\boldmath{$\omega$}}%
%EndExpansion
_{\mathbf{v}}%
%TCIMACRO{\TeXButton{xi}{\mbox{\boldmath{$\xi$}}}}%
%BeginExpansion
\mbox{\boldmath{$\xi$}}%
%EndExpansion
,\hspace{0.15in}\nonumber\\
D_{\mathbf{v}}^{s}\overset{\cdot}{%
%TCIMACRO{\TeXButton{xi}{\mbox{\boldmath{$\xi$}}}}%
%BeginExpansion
\mbox{\boldmath{$\xi$}}%
%EndExpansion
}  &  =\partial_{\mathbf{v}}\overset{\cdot}{%
%TCIMACRO{\TeXButton{xi}{\mbox{\boldmath{$\xi$}}}}%
%BeginExpansion
\mbox{\boldmath{$\xi$}}%
%EndExpansion
}-\frac{1}{2}\overset{\cdot}{%
%TCIMACRO{\TeXButton{xi}{\mbox{\boldmath{$\xi$}}}}%
%BeginExpansion
\mbox{\boldmath{$\xi$}}%
%EndExpansion
}%
%TCIMACRO{\TeXButton{omega}{\mbox{\boldmath{$\omega$}}}}%
%BeginExpansion
\mbox{\boldmath{$\omega$}}%
%EndExpansion
_{\mathbf{v}},\nonumber\\
D_{\mathbf{v}}P  &  =\partial_{\mathbf{v}}\mathbf{P}+\frac{1}{2}%
%TCIMACRO{\TeXButton{omega}{\mbox{\boldmath{$\omega$}}}}%
%BeginExpansion
\mbox{\boldmath{$\omega$}}%
%EndExpansion
_{\mathbf{v}}\mathbf{P}-\frac{1}{2}\mathbf{P}\text{ }%
%TCIMACRO{\TeXButton{omega}{\mbox{\boldmath{$\omega$}}}}%
%BeginExpansion
\mbox{\boldmath{$\omega$}}%
%EndExpansion
_{\mathbf{v}}=\partial_{\mathbf{v}}\mathbf{P}+\frac{1}{2}[%
%TCIMACRO{\TeXButton{omega}{\mbox{\boldmath{$\omega$}}}}%
%BeginExpansion
\mbox{\boldmath{$\omega$}}%
%EndExpansion
_{\mathbf{v}},\mathbf{P}]. \label{D.4}%
\end{align}

(v) For $\mathbf{T}\in\sec%
%TCIMACRO{\dbigwedge }%
%BeginExpansion
{\displaystyle\bigwedge}
%EndExpansion
TM\hookrightarrow\mathcal{C\ell}^{(0)}(TM)$ and $\xi\in\sec S(M)$, $\bar{\xi
}\in\sec\bar{S}(M)$, $P\in\sec$ $P(M)$and $\ \mathbf{v}\in\sec TM$ , we have
\begin{align}
D_{\mathbf{v}}^{s}(\mathbf{T}\otimes\xi)  &  =D_{\mathbf{v}}\mathbf{T}%
\otimes\xi+\mathbf{T}D_{\mathbf{v}}^{s}\xi,\label{D.2'}\\
D_{\mathbf{v}}^{s}(\mathbf{T}\otimes\bar{\xi})  &  =D_{\mathbf{v}}%
\mathbf{T}\otimes\bar{\xi}+\mathbf{T}D_{\mathbf{v}}^{s}\bar{\xi}\nonumber
\end{align}
\ \ 

and%
\begin{align}
D_{\mathbf{v}}\mathbf{T}  &  =\partial_{\mathbf{v}}\mathbf{T}+\frac{1}{2}[%
%TCIMACRO{\TeXButton{omega}{\mbox{\boldmath{$\omega$}}}}%
%BeginExpansion
\mbox{\boldmath{$\omega$}}%
%EndExpansion
_{\mathbf{v}},\mathbf{T}],\nonumber\\
D_{\mathbf{v}}^{s}\xi &  =\partial_{\mathbf{v}}\xi+\frac{1}{2}\Omega
_{\mathbf{v}}\xi,\hspace{0.15in}\nonumber\\
D_{\mathbf{v}}^{s}\dot{\xi}  &  =\partial_{\mathbf{v}}\dot{\xi}-\frac{1}%
{2}\dot{\xi}\Omega_{\mathbf{v}},\nonumber\\
D_{\mathbf{v}}P  &  =\partial_{\mathbf{v}}P+\frac{1}{2}\Omega_{\mathbf{v}%
}P-\frac{1}{2}P\text{ }\Omega_{\mathbf{v}}=\partial_{\mathbf{v}}P+\frac{1}%
{2}[\Omega_{\mathbf{v}},P]. \label{D.4'}%
\end{align}

In the above equations $%
%TCIMACRO{\TeXButton{omega}{\mbox{\boldmath{$\omega$}}}}%
%BeginExpansion
\mbox{\boldmath{$\omega$}}%
%EndExpansion
_{\mathbf{v}}\in\sec\mathcal{C\ell}^{(0)}(TM)$ and $\Omega_{\mathbf{v}}\in
\sec$ $P(M)$. Writing as usual, $\mathbf{v}=v^{\mathbf{a}}\mathbf{e}%
_{\mathbf{a}}$, \ $D_{\mathbf{e}_{\mathbf{a}}}e^{\mathbf{b}}=-\mathbf{\omega
}_{\mathbf{ac}}^{\mathbf{b}}e^{\mathbf{c}}$ , $\mathbf{\omega}_{\mathbf{abc}%
}=-\mathbf{\omega}_{\mathbf{cba}}=\eta_{\mathbf{ad}}\mathbf{\omega
}_{\mathbf{bc}}^{\mathbf{d}}$, $\mathbf{\omega}_{\text{.\textbf{b}}%
}^{\mathbf{a}\text{ }\mathbf{c}}=-\mathbf{\omega}_{\text{.\textbf{b}}%
}^{\mathbf{c}\text{ }\mathbf{a}}$, $\sigma_{\mathbf{b}}=\mathbf{e}%
_{\mathbf{b}}\mathbf{e}_{\mathbf{0}}$ and\footnote{Have in mind that \textsl{i
}is a \textit{Clifford field} here.} \textsl{i }$=-\sigma_{\mathbf{1}}%
\sigma_{\mathbf{2}}\sigma_{\mathbf{3}}$, we have
\begin{align}%
%TCIMACRO{\TeXButton{omega}{\mbox{\boldmath{$\omega$}}}}%
%BeginExpansion
\mbox{\boldmath{$\omega$}}%
%EndExpansion
_{\mathbf{e}_{\mathbf{a}}}  &  =\frac{1}{2}\mathbf{\omega}_{\mathbf{a}%
}^{\mathbf{bc}}\mathbf{e}_{\mathbf{b}}\mathbf{e}_{\mathbf{c}}=\frac{1}%
{2}\mathbf{\omega}_{\mathbf{a}}^{\mathbf{bc}}\mathbf{e}_{\mathbf{b}}%
\wedge\mathbf{e}_{\mathbf{c}}\nonumber\\
&  =\frac{1}{2}\mathbf{\omega}_{\mathbf{a}}^{\mathbf{bc}}%
%TCIMACRO{\TeXButton{sigma}{\mbox{\boldmath{$\sigma$}}}}%
%BeginExpansion
\mbox{\boldmath{$\sigma$}}%
%EndExpansion
_{\mathbf{b}}\overset{\vee}{%
%TCIMACRO{\TeXButton{sigma}{\mbox{\boldmath{$\sigma$}}}}%
%BeginExpansion
\mbox{\boldmath{$\sigma$}}%
%EndExpansion
}_{\mathbf{c}}\hspace{0.15in}\hspace{0.15in}\nonumber\\
&  =\frac{1}{2}(-2\mathbf{\omega}_{\mathbf{a}}^{\mathbf{0}i}\sigma
_{i}+\mathbf{\omega}_{\mathbf{a}}^{ji}\sigma_{i}\sigma_{j})\nonumber\\
&  =\frac{1}{2}(-2\mathbf{\omega}_{\mathbf{a}}^{\mathbf{0}i}\sigma
_{i}-i\ \varepsilon_{i\text{ }j}^{k}\mathbf{\omega}_{\mathbf{a}}^{ji}%
\sigma_{k})=%
%TCIMACRO{\TeXButton{Omega}{\mbox{\boldmath{$\Omega$}}}}%
%BeginExpansion
\mbox{\boldmath{$\Omega$}}%
%EndExpansion
_{\mathbf{a}}^{\mathbf{b}}\sigma_{\mathbf{b}}. \label{D.5}%
\end{align}

Note that the $%
%TCIMACRO{\TeXButton{Omega}{\mbox{\boldmath{$\Omega$}}}}%
%BeginExpansion
\mbox{\boldmath{$\Omega$}}%
%EndExpansion
_{\mathbf{a}}^{\mathbf{b}}$ are \ `formally' complex numbers. Also, observe
that we can write for the \ `formal' Hermitian conjugate $%
%TCIMACRO{\TeXButton{omega}{\mbox{\boldmath{$\omega$}}}}%
%BeginExpansion
\mbox{\boldmath{$\omega$}}%
%EndExpansion
_{\mathbf{e}_{\mathbf{a}}}^{\dagger}$ of $%
%TCIMACRO{\TeXButton{omega}{\mbox{\boldmath{$\omega$}}}}%
%BeginExpansion
\mbox{\boldmath{$\omega$}}%
%EndExpansion
_{\mathbf{e}_{\mathbf{a}}}$ of
\begin{equation}%
%TCIMACRO{\TeXButton{omega}{\mbox{\boldmath{$\omega$}}}}%
%BeginExpansion
\mbox{\boldmath{$\omega$}}%
%EndExpansion
_{\mathbf{e}_{\mathbf{a}}}^{\dagger}=-\mathbf{e}^{\mathbf{0}}%
%TCIMACRO{\TeXButton{omega}{\mbox{\boldmath{$\omega$}}}}%
%BeginExpansion
\mbox{\boldmath{$\omega$}}%
%EndExpansion
_{\mathbf{e}_{\mathbf{a}}}\mathbf{e}^{\mathbf{0}}. \label{D.5'}%
\end{equation}
Also, write $\Omega_{\mathbf{e}_{\mathbf{a}}}$ for the matrix representation
of $%
%TCIMACRO{\TeXButton{omega}{\mbox{\boldmath{$\omega$}}}}%
%BeginExpansion
\mbox{\boldmath{$\omega$}}%
%EndExpansion
_{\mathbf{e}_{\mathbf{a}}}$, i.e.,
\[
\Omega_{\mathbf{e}_{\mathbf{a}}}=\Omega_{\mathbf{a}}^{\mathbf{b}}%
\sigma_{\mathbf{b}},
\]
where $\Omega_{\mathbf{a}}^{\mathbf{b}}$ are complex numbers with the same
coefficients as the \ `formally' complex numbers $\mathbf{\Omega}_{\mathbf{a}%
}^{\mathbf{b}}$. We can easily verify that
\begin{equation}
\Omega_{\mathbf{e}_{\mathbf{a}}}=\varepsilon\Omega_{\mathbf{e}_{\mathbf{a}}%
}^{\dagger}\varepsilon. \label{D.7}%
\end{equation}

We can prove the third line of Eq.(\ref{D.4'}) as follows. First take the
Hermitian conjugation of \ the second line of Eq.(\ref{D.4'}), obtaining%
\[
D_{\mathbf{v}}\bar{\xi}=\partial_{\mathbf{v}}\bar{\xi}+\frac{1}{2}\bar{\xi
}\Omega_{\mathbf{v}}^{\dagger}.\hspace{0.15in}%
\]
Next multiply the above equation on the left by $\varepsilon$ and recall that
$\dot{\xi}=\bar{\xi}\varepsilon$ and Eq.(\ref{D.7}). We get
\begin{align*}
D_{\mathbf{v}}\dot{\xi}  &  =\partial_{\mathbf{v}}\dot{\xi}-\frac{1}{2}%
\dot{\xi}\varepsilon\Omega_{\mathbf{v}}^{\dagger}\varepsilon\\
&  =D_{\mathbf{v}}\dot{\xi}=\partial_{\mathbf{v}}\dot{\xi}-\frac{1}{2}\dot
{\xi}\Omega_{\mathbf{v}}.
\end{align*}
Note that this is compatible with the identification $\mathcal{C\ell}%
^{(0)}(TM)\simeq\mathcal{S}(M)\otimes_{\mathbb{C}}\mathcal{\dot{S}}(M)$ and
$\mathbb{C}\ell^{(0)}(M)\simeq S(M)\otimes_{\mathbb{C}}\dot{S}(M)$.

Note moreover that \ if $\mathbf{q}_{\mu}=e_{\mu}\mathbf{e}_{\mathbf{0}%
}=h_{\mu}^{\mathbf{a}}\mathbf{e}_{\mathbf{a}}\mathbf{e}_{\mathbf{0}}=h_{\mu
}^{\mathbf{a}}%
%TCIMACRO{\TeXButton{sigma}{\mbox{\boldmath{$\sigma$}}}}%
%BeginExpansion
\mbox{\boldmath{$\sigma$}}%
%EndExpansion
_{\mathbf{a}}$ $\in\mathcal{C\ell}^{(0)}(TM)\simeq\mathcal{S}(M)\otimes
_{\mathbb{C}}\mathcal{\dot{S}}(M)$ we have, \ %

\begin{equation}
D_{\mathbf{v}}\mathbf{q}_{\mu}=\partial_{\mathbf{v}}q_{\mu}+\frac{1}{2}%
%TCIMACRO{\TeXButton{omega}{\mbox{\boldmath{$\omega$}}}}%
%BeginExpansion
\mbox{\boldmath{$\omega$}}%
%EndExpansion
_{\mathbf{v}}\mathbf{q}_{\mu}+\frac{1}{2}\mathbf{q}_{\mu}%
%TCIMACRO{\TeXButton{omega}{\mbox{\boldmath{$\omega$}}}}%
%BeginExpansion
\mbox{\boldmath{$\omega$}}%
%EndExpansion
_{\mathbf{v}}^{\dagger}. \label{D.08}%
\end{equation}
For $q_{\mu}=h_{\mu}^{\mathbf{a}}\sigma_{\mathbf{a}}$ $\in\sec\mathbb{C}%
\ell^{(0)}(M)\simeq S(M)\otimes_{\mathbb{C}}\bar{S}(M),$ the matrix
representative of the $\mathbf{q}_{\mu}$ we have for any vector field
$\mathbf{v}\in\sec TM$ \
\begin{equation}
D_{\mathbf{v}}q_{\mu}=\partial_{\mathbf{v}}q_{\mu}+\frac{1}{2}\Omega
_{\mathbf{v}}q_{\mu}+\frac{1}{2}q_{\mu}\text{ }\Omega_{\mathbf{v}}^{\dagger}
\label{D.8}%
\end{equation}
which is the equation used by Sachs for the \textit{spinor} covariant
derivative of his \ `quaternion' fields. Note that M. Sachs in \cite{s1}
introduced also a kind of total covariant derivative for his \ `quaternion'
fields. That \ `derivative' denoted in this text by $D_{\mathbf{v}%
}^{\mathbf{S}}$ will be discussed below.

\subsection{Geometrical Meaning of $D_{e_{\nu}}q_{\mu}=\Gamma_{\nu\mu}%
^{\alpha}q_{\alpha}$}

We recall that Sachs wrote \footnote{See Eq.(3.69) in \cite{s1}.} that
\begin{equation}
D_{e_{\nu}}q_{\mu}=\Gamma_{\nu\mu}^{\alpha}q_{\alpha}, \label{D.9}%
\end{equation}
where $\Gamma_{\nu\mu}^{\alpha}$ are the connection coefficients of the
coordinate basis $\{e_{\mu}\}$, i.e.,
\begin{equation}
D_{e_{\nu}}e_{\mu}=\Gamma_{\nu\mu}^{\alpha}e_{\alpha} \label{D.9bis}%
\end{equation}

How, can Eq.(\ref{D.9}) be true? Well, let us calculate $D_{e_{\nu}}%
\mathbf{q}_{\mu}$ in $\mathcal{C\ell}(TM)$. We have,%

\begin{align}
D_{e_{\nu}}\mathbf{q}_{\mu}  &  =D_{e_{\nu}}(e_{\mu}\mathbf{e}_{\mathbf{0}%
})\nonumber\\
&  =(D_{e_{\nu}}e_{\mu})\mathbf{e}_{\mathbf{0}}+e_{\mu}(D_{e_{\nu}}%
\mathbf{e}_{\mathbf{0}})\nonumber\\
&  =\Gamma_{\nu\mu}^{\alpha}\mathbf{q}_{\alpha}+e_{\mu}(D_{e_{\nu}}%
\mathbf{e}_{\mathbf{0}}). \label{D.10}%
\end{align}

So, Eq.(\ref{D.9}) follows if, and only if
\begin{equation}
D_{e_{\nu}}\mathbf{e}_{\mathbf{0}}=0. \label{D.11}%
\end{equation}

To understand the physical meaning of Eq.(\ref{D.11}) let us recall the
following. In relativity theory reference frames are represented by time like
vector fields $Z\in\sec TM$ pointing to the future \cite{rosharif,sw}. If we
write the $\alpha_{\mathbf{Z}}=g(\mathbf{Z,)\in}%
%TCIMACRO{\dbigwedge \nolimits^{1}}%
%BeginExpansion
{\displaystyle\bigwedge\nolimits^{1}}
%EndExpansion
T^{\ast}M$ for the physically equivalent 1-form field we have the well known
\emph{decomposition}
\begin{equation}
D\alpha_{\mathbf{Z}}=\mathbf{a}_{\mathbf{Z}}\otimes\alpha_{\mathbf{Z}%
}+\mathbf{\varpi}_{\mathbf{Z}}+\mathbf{\sigma}_{\mathbf{Z}}+\frac{1}%
{3}E_{\mathbf{Z}}\mathbf{p}, \label{D.12}%
\end{equation}
where
\begin{equation}
\mathbf{p}=g-\alpha_{\mathbf{Z}}\otimes\alpha_{\mathbf{Z}} \label{D.13}%
\end{equation}
is called the projection tensor (and gives the metric of the rest space of an
instantaneous observer \cite{sw}), $\mathbf{a}_{\mathbf{Z}}=g(D_{\mathbf{Z}%
}\mathbf{Z,)}$ is the (form) acceleration of $\mathbf{Z}$, $\mathbf{\varpi
}_{\mathbf{Z}}$ is the rotation of $\mathbf{Z}$, $\mathbf{\sigma}_{\mathbf{Z}%
}$ is the shear of $\mathbf{Z}$ and $E_{\mathbf{Z}}$ is the expansion ratio of
$\mathbf{Z}$ . In a coordinate chart ($U,x^{\mu}$), writing $\mathbf{Z}%
=Z^{\mu}\partial/\partial x^{\mu}$ \ and $\mathbf{p}=(g_{\mu\nu}-Z_{\mu}%
Z_{\nu})dx^{\mu}\otimes dx^{\nu}$ we have
\begin{align}
\mathbf{\varpi}_{\mathbf{Z}\mu\nu}  &  =Z_{\left[  \alpha;\beta\right]
}p_{\mu}^{\alpha}p_{\nu}^{\beta},\nonumber\\
\mathbf{\sigma}_{\mathbf{Z}\alpha\beta}  &  =[Z_{\left(  \mu;\nu\right)
}-\frac{1}{3}\mathbf{E}_{\mathbf{Z}}h_{\mu\nu}]p_{\alpha}^{\mu}p_{\beta}^{\nu
},\nonumber\\
E_{\mathbf{Z}}  &  =Z^{\mu};_{\mu}. \label{D.14}%
\end{align}

Now, in Special Relativity where the space time manifold is the structure
$(M\mathcal{=R}^{4},g=\eta,D^{\eta},\tau_{\eta},\uparrow)$%
\footnote{$\mathbf{\eta}$ is a constant metric, i.e., there exists a chart
$\langle x^{\mu}\rangle$ of $\ M=\mathcal{R}^{4}$ such that $\mathbf{\eta
}(\partial/\partial x^{\mu},\partial/\partial x^{\nu})=\eta_{\mu\nu}$, the
numbers $\eta_{\mu\nu}$ forming a diagonal matrix with entries $(1,-1,-1,-1)$.
Also, $D^{\mathbf{\eta}}$ is the Levi-Civita connection of $\mathbf{\eta}$.}
an \emph{inertial reference frame }(\emph{IRF}) $\mathbf{I}\in\sec TM$ is
defined by $D^{\eta}\mathbf{I}=0$. \ We can show very easily (see, e.g.,
\cite{sw}) that in General Relativity Theory $\left(  \emph{GRT}\right)  $
\ where each gravitational field is modelled by a spacetime\footnote{More
precisely, by a diffeomorphism equivalence class of Lorentzian spacetimes.}
$(M,g,D,\tau_{g},\uparrow)$ there is \textit{in general} no shear free frame
$(\sigma_{\mathfrak{Q}}=0)$ on any open neighborhood $U$ \ of any given
spacetime point. The reason is clear in local coordinates $\langle x^{\mu
}\rangle$ covering $U$. Indeed, $\sigma_{\mathfrak{Q}}=0$ implies five
independent conditions on the components of the frame $\mathfrak{Q}$. Then, we
arrive at the conclusion that in a general spacetime model\footnote{We take
the opprotunity to correct an statement in \cite{rosharif}. There it is stated
that in General Relativity there are no inertial frames. Of, course, the
correct statement is that in a general spacetime model there are in general no
inertial frames. But, of course, there are spacetime models where there exist
frames $\mathfrak{Q}\in$ $\sec TU\subset\sec TM$ satisfying \ $D\mathfrak{Q}%
=0$. See below.} there is no frame $\mathfrak{Q}\in$ $\sec TU\subset\sec TM$
satisfying \ $D\mathfrak{Q}=0$, and in general there is no \emph{IRF} in any
model of \emph{GRT}.\medskip

The following question arises naturally: which characteristics a reference
frame on a \emph{GRT} spacetime model must have in order to reflect as much as
possible the properties of an \emph{IRF} of \emph{SRT}?

The answer to that question \cite{rosharif} is that there are two kind of
frames in \emph{GRT} such that each frame in one of these classes share some
important aspects of the \emph{IRFs} of \emph{SRT}. Both concepts are
important and it is important to distinguish between them in order to avoid
misunderstandings. These frames are the \emph{pseudo inertial} \emph{reference
frame }(\emph{PIRF}) and the and the local Lorentz reference frames
(\emph{LLRF}$\gamma$\emph{s}), but we don not need to enter the details here.

On the open set $U\subset M$ covered by a coordinate chart $\langle x^{\mu
}\rangle$ of the maximal atlas of $M$ multiplying Eq.(\ref{D.11}) by
$h_{\mathbf{a}}^{\nu}$ \ such that $\mathbf{e}_{\mathbf{a}}=h_{\mathbf{a}%
}^{\nu}e_{\nu}$, we get%
\begin{equation}
D_{\mathbf{e}_{\mathbf{a}}}\mathbf{e}_{\mathbf{0}}=0;\hspace{0.15cm}%
\mathbf{a}=0\mathbf{,}1,2,3.\text{ } \label{D.psi}%
\end{equation}

Then, it follows that
\begin{equation}
D_{X}\mathbf{e}_{\mathbf{0}}=0\text{, }\forall X\in\sec TM \label{D.new}%
\end{equation}
which characterizes $\mathbf{e}_{\mathbf{0}}$ as an inertial frame. This
imposes several restrictions on the spacetime described by the theory. Indeed,
if $\mathbf{Ric}$ is the Ricci tensor of the manifold modeling spacetime, we
have\footnote{See, exercise 3.2.12 of \cite{sw}.}%
\begin{equation}
\mathbf{Ric}(\mathbf{e}_{\mathbf{0}},X)=0\text{, }\forall X\in\sec TM.
\label{D.neww}%
\end{equation}

In particular, this condition cannot be realized in Einstein-de Sitter
spacetime. This fact is completely hidden in the matrix formalism used in
Sachs theory, where no restriction on the spacetime manifold (besides the one
of being a spin manifold) need to be imposed.

\subsection{Geometrical Meaning of $D_{e_{\mu}}\sigma_{\mathbf{i}}=0$ in
General Relativity}

We now discuss what happens in the usual theory of \ dotted and undotted two
component \textit{matrix} spinor fields in general relativity, as \ described,
e.g., in \cite{carmeli,penrose,penrindler}. In that formulation it is
postulated that the covariant spinor derivative of Pauli matrices must
satisfy
\begin{equation}
D_{e_{\mu}}\sigma_{i}=0,\hspace{0.15cm}i\mathbf{=}1,2,3 \label{D.15}%
\end{equation}

Eq.(\ref{D.15}) translate in our formalism as%
\begin{equation}
D_{e_{\mu}}%
%TCIMACRO{\TeXButton{sigma}{\mbox{\boldmath{$\sigma$}}}}%
%BeginExpansion
\mbox{\boldmath{$\sigma$}}%
%EndExpansion
_{i}=D_{e_{\mu}}\left(  \mathbf{e}_{i}\mathbf{e}_{\mathbf{0}}\right)  =0.
\label{D.16}%
\end{equation}
Differently from the case of Sachs theory, Eq.(\ref{D.16}) can be satisfied
if
\begin{equation}
D_{e_{\mu}}\mathbf{e}_{i}=\mathbf{e}_{i}(D_{e_{\mu}}\mathbf{e}_{0}%
)\mathbf{e}_{\mathbf{0}} \label{D.15bis}%
\end{equation}
or, writing $D_{e_{\mu}}\mathbf{e}_{\mathbf{a}}=\omega_{\mu\mathbf{a}%
}^{\mathbf{b}}\mathbf{e}_{\mathbf{b}}$,
\begin{equation}
\omega_{\mu i}^{\mathbf{b}}=\mathbf{e}^{\mathbf{b}}\lrcorner(\omega
_{\mu\mathbf{0}}^{\mathbf{a}}\mathbf{e}_{i}\mathbf{e}_{\mathbf{a}}%
\mathbf{e}_{\mathbf{0}}). \label{D.15biss}%
\end{equation}

This certainly implies some restrictions on possible spacetime models, but
that is the price in order to have spinor fields. At least we do not need to
necessarily have $D\mathbf{e}_{\mathbf{0}}=0$.

We analyze some possibilities of satisfying Eq.(\ref{D.15})

(\textbf{i}) Suppose that $\mathbf{e}_{\mathbf{0}}$ satisfy $D_{e_{\mu}%
}\mathbf{e}_{\mathbf{0}}=0$, i.e., $D\mathbf{e}_{\mathbf{0}}=0$.\textit{
}Then, a necessary and sufficient condition for the validity of Eq.(\ref{D.16}%
) is that%
\begin{equation}
D_{e_{\mu}}\mathbf{e}_{i}=0. \label{D.17}%
\end{equation}

\qquad Multiplying Eq.(\ref{D.17}) by $h_{\mathbf{a}}^{\mu}$ we get \
\begin{equation}
D_{\mathbf{e}_{\mathbf{a}}}e_{i}=0,\hspace{0.15cm}i\mathbf{=}1,2,3;\hspace
{0.15cm}\mathbf{a}=0,1,2,3 \label{D.newbis}%
\end{equation}

In particular,
\begin{equation}
D_{\mathbf{e}_{0}}\mathbf{e}_{i}=0,\hspace{0.15cm}i\mathbf{=}1,2,3
\label{D.18}%
\end{equation}

Eq.(\ref{D.18}) means that the fields $\mathbf{e}_{i}$ following each integral
line of $\mathbf{e}_{\mathbf{0}}$ are Fermi transported\footnote{An original
approach to the Fermi transport using Clifford bundle methods has been given
in \cite{rovapa}. There an equivalent spinor equation to the famous Darboux
equations of differential geometry is derived.} \cite{sw}. Physicists
interpret that equation saying that the $\mathbf{e}_{i}$ are physically
realizable by gyroscopic axes, which gives the local standard of no rotation.

The above conclusion sounds fine. However it follows from Eq.(\ref{D.new}) and
Eq.(\ref{D.newbis}) that
\begin{equation}
D_{\mathbf{e}_{\mathbf{a}}}\mathbf{e}_{\mathbf{b}}=0,\hspace{0.15cm}%
\mathbf{a=}0\mathbf{,}1,2,3;\hspace{0.15cm}\mathbf{b}=0,1,2,3. \label{D.19}%
\end{equation}

Recalling that existence of spinor fields implies that $\{\mathbf{e}%
_{\mathbf{a}}\}$ is a global tetrad \cite{g1}, Eq.(\ref{D.19}) implies that
the connection $D$ must be teleparallel. Then, under the above conditions the
curvature tensor of a spacetime admitting spinor fields must be\textit{ null}.
This, is in particular, the case of special relativity.

(\textbf{ii}) Suppose now that $\mathbf{e}_{\mathbf{0}}$ is a geodesic frame,
i.e., $D_{\mathbf{e}_{0}}\mathbf{e}_{\mathbf{0}}=0$. Then, $h_{\mathbf{0}%
}^{\nu}D_{e_{\nu}}\mathbf{e}_{0}=0$ and Eq. (\ref{D.15bis}) implies only that
\
\begin{equation}
D_{\mathbf{e}_{0}}\mathbf{e}_{i}=0;\hspace{0.15cm}i\mathbf{=}1,2,3
\label{D.18bis}%
\end{equation}
and we do not have \textit{any} inconsistency. If we take an integral line of
$\mathbf{e}_{\mathbf{0}}$, say $\gamma$, then the set $\{\left.
\mathbf{e}_{\mathbf{a}}\right\vert _{\gamma}\}$ may be called an
\textit{inertial moving frame} along $\gamma$. The set $\{\left.
\mathbf{e}_{\mathbf{a}}\right\vert _{\gamma}\}$ is also \emph{Fermi}
transported and since $\gamma$ is a geodesic worldline they define the
standard of \emph{no} rotation along $\gamma.$

In conclusion, a consistent definition of spinor fields in general relativity
using the Clifford and spin bundle formalism of this paper needs triviality of
the frame bundle, i.e., existence of a global tetrad, say $\{\mathbf{e}%
_{\mathbf{a}}\}$ and validity of Eq.(\ref{D.15bis}). A nice physical
interpretation follows moreover if the tetrad satisfies
\begin{equation}
D_{\mathbf{e}_{0}}\mathbf{e}_{\mathbf{a}}=0;\hspace{0.15cm}\mathbf{a=}%
0\mathbf{,}1,2,3. \label{D.18biss}%
\end{equation}

Of course, as it is the case in Sachs theory, the matrix formulation of spinor
fields do not impose any constrains in the possible spacetime models, besides
the one needed for the existence of a spinor structure. Saying that we have an
important comment.

\subsection{Covariant Derivative of the Dirac Gamma Matrices}

If we use a real spin bundle where we can formulate the Dirac equation, e.g.,
one where the typical fiber is the ideal of (algebraic) Dirac spinors, i.e.,
the ideal generated by a idempotent $\frac{1}{2}(1+E_{0})$, $E_{0}%
\in\mathbb{R}_{1,3}$, then no restriction is imposed on the global tetrad
field $\{\mathbf{e}_{\mathbf{a}}\}$ defining the spinor structure of spacetime
(see \cite{28,moro}). In particular, since%
\begin{equation}
D_{\mathbf{e}_{a}}\mathbf{e}_{\mathbf{b}}=\omega_{\mathbf{ab}}^{\mathbf{c}%
}\mathbf{e}_{\mathbf{c}}, \label{Dirac1}%
\end{equation}
we have,
\begin{equation}
D_{\mathbf{e}_{a}}\mathbf{e}_{\mathbf{b}}=\frac{1}{2}[%
%TCIMACRO{\TeXButton{omega}{\mbox{\boldmath{$\omega$}}}}%
%BeginExpansion
\mbox{\boldmath{$\omega$}}%
%EndExpansion
_{\mathbf{e}_{\mathbf{a}}},\mathbf{e}_{\mathbf{b}}] \label{Dirac2}%
\end{equation}

Then,
\begin{equation}
\omega_{\mathbf{ab}}^{\mathbf{c}}\mathbf{e}_{\mathbf{c}}-\frac{1}{2}%
%TCIMACRO{\TeXButton{omega}{\mbox{\boldmath{$\omega$}}}}%
%BeginExpansion
\mbox{\boldmath{$\omega$}}%
%EndExpansion
_{\mathbf{e}_{\mathbf{a}}}\mathbf{e}_{\mathbf{b}}+\frac{1}{2}\mathbf{e}%
_{\mathbf{b}}%
%TCIMACRO{\TeXButton{omega}{\mbox{\boldmath{$\omega$}}}}%
%BeginExpansion
\mbox{\boldmath{$\omega$}}%
%EndExpansion
_{\mathbf{e}_{\mathbf{a}}}=0. \label{Dirac3}%
\end{equation}

The matrix representation of the real spinor bundle, of course, sends
$\{\mathbf{e}_{\mathbf{a}}\}\mapsto\{\gamma_{\mathbf{a}}\}$, where the
$\gamma_{\mathbf{a}}$'s are the standard representation of the Dirac matrices.
Then, the matrix translation of Eq.(\ref{Dirac3}) is
\begin{equation}
\omega_{\mathbf{ab}}^{\mathbf{c}}\gamma_{\mathbf{c}}-\frac{1}{2}%
%TCIMACRO{\TeXButton{omega}{\mbox{\boldmath{$\omega$}}}}%
%BeginExpansion
\mbox{\boldmath{$\omega$}}%
%EndExpansion
_{\mathbf{e}_{\mathbf{a}}}\gamma_{\mathbf{b}}+\frac{1}{2}\gamma_{\mathbf{b}}%
%TCIMACRO{\TeXButton{omega}{\mbox{\boldmath{$\omega$}}}}%
%BeginExpansion
\mbox{\boldmath{$\omega$}}%
%EndExpansion
_{\mathbf{e}_{\mathbf{a}}}=0. \label{Dirac4}%
\end{equation}

For the matrix elements $\gamma_{\mathbf{b}B}^{A}$ we have
\begin{equation}
\omega_{\mathbf{ab}}^{\mathbf{c}}\gamma_{\mathbf{c}B}^{A}-\frac{1}{2}%
%TCIMACRO{\TeXButton{omega}{\mbox{\boldmath{$\omega$}}}}%
%BeginExpansion
\mbox{\boldmath{$\omega$}}%
%EndExpansion
_{\mathbf{e}_{\mathbf{a}}C}^{A}\gamma_{\mathbf{b}B}^{C}+\frac{1}{2}%
\gamma_{\mathbf{b}C}^{A}%
%TCIMACRO{\TeXButton{omega}{\mbox{\boldmath{$\omega$}}}}%
%BeginExpansion
\mbox{\boldmath{$\omega$}}%
%EndExpansion
_{\mathbf{e}_{\mathbf{a}}B}^{C}=0. \label{Dirac5}%
\end{equation}

In \cite{choquet} this last equation is confused with the covariant derivative
of $\gamma_{\mathbf{c}B}^{A}$. Indeed in an exercise in problem 4, Chapter
Vbis \cite{choquet} ask one to prove that

\begin{center}%
\begin{tabular}
[c]{|l|}\hline
$\nabla_{\mathbf{e}_{\mathbf{b}}}\gamma_{\mathbf{c}B}^{A}=\omega_{\mathbf{ab}%
}^{\mathbf{c}}\gamma_{\mathbf{c}B}^{A}-\frac{1}{2}%
%TCIMACRO{\TeXButton{omega}{\mbox{\boldmath{$\omega$}}}}%
%BeginExpansion
\mbox{\boldmath{$\omega$}}%
%EndExpansion
_{\mathbf{e}_{\mathbf{a}}C}^{A}\gamma_{\mathbf{b}B}^{C}+\frac{1}{2}%
\gamma_{\mathbf{b}C}^{A}%
%TCIMACRO{\TeXButton{omega}{\mbox{\boldmath{$\omega$}}}}%
%BeginExpansion
\mbox{\boldmath{$\omega$}}%
%EndExpansion
_{\mathbf{e}_{\mathbf{a}}B}^{C}=0.$\\\hline
\end{tabular}

\end{center}

Of course, the first member of the above equation does not define any
covariant derivative operator. Confusions as that one appears over and over
again in the literature, and of course, is also present in Sachs theory in a
small modified form, as shown in the next subsubsection.

\begin{center}

\end{center}

\subsection{$D_{e_{\nu}}^{\mathbf{S}}q_{\mu}=0$}

Now, taking into account Eq.(\ref{D.8}) and Eq.(\ref{D.9}) we can write:%
\begin{equation}
\partial_{\nu}\mathbf{q}_{\mu}+\frac{1}{2}%
%TCIMACRO{\TeXButton{omega}{\mbox{\boldmath{$\omega$}}}}%
%BeginExpansion
\mbox{\boldmath{$\omega$}}%
%EndExpansion
_{\mathbf{\nu}}\mathbf{q}_{\mu}+\frac{1}{2}\mathbf{q}_{\mu}%
%TCIMACRO{\TeXButton{omega}{\mbox{\boldmath{$\omega$}}}}%
%BeginExpansion
\mbox{\boldmath{$\omega$}}%
%EndExpansion
_{\mathbf{\nu}}-\Gamma_{\nu\mu}^{\alpha}\mathbf{q}_{\alpha}=0. \label{D.18'}%
\end{equation}

Sachs defined
\begin{equation}
D_{e_{\nu}}^{\mathbf{S}}\mathbf{q}_{\mu}=\partial_{\nu}\mathbf{q}_{\mu}%
+\frac{1}{2}%
%TCIMACRO{\TeXButton{omega}{\mbox{\boldmath{$\omega$}}}}%
%BeginExpansion
\mbox{\boldmath{$\omega$}}%
%EndExpansion
_{\mathbf{\nu}}\mathbf{q}_{\mu}+\frac{1}{2}\mathbf{q}_{\mu}%
%TCIMACRO{\TeXButton{omega}{\mbox{\boldmath{$\omega$}}}}%
%BeginExpansion
\mbox{\boldmath{$\omega$}}%
%EndExpansion
_{\mathbf{\nu}}-\Gamma_{\nu\mu}^{\alpha}\mathbf{q}_{\alpha} \label{D.18''}%
\end{equation}
from where
\begin{equation}
D_{e_{\nu}}^{\mathbf{S}}\mathbf{q}_{\mu}=0. \label{D.18'''}%
\end{equation}
Of course, the matrix representation of the last two equations are:%
\begin{align}
D_{e_{\nu}}^{\mathbf{S}}q_{\mu}  &  =\partial_{\nu}q_{\mu}+\frac{1}{2}%
\Omega_{\mathbf{\nu}}q_{\mu}+\frac{1}{2}q_{\mu}\text{ }\Omega_{\mathbf{\nu}%
}^{\dagger}-\Gamma_{\nu\mu}^{\alpha}q_{\alpha}.\nonumber\\
D_{e_{\nu}}^{\mathbf{S}}q_{\mu}  &  =0. \label{D.19'}%
\end{align}

Sachs call \footnote{See Eq.(3.69) in \cite{s1}.} $D_{e_{\nu}}^{\mathbf{S}%
}q_{\mu}$ the covariant derivative of a $q_{\mu}$ field. The nomination is an
\textit{unfortunate} one, since the equation $D_{e_{\nu}}^{\mathbf{S}}q_{\mu
}=0$ is a \textit{ trivial} identity and do not introduce any new connection
in the game.\footnote{The equation $D_{\nu}^{\mathbf{S}}\mathbf{q}_{\mu}=0$
(or its matrix representation) is a reminicescence of an analogous equation
for the components of tetrad fields often printed in physics textbooks and
confused with the metric compatibility condition of the connection. See,e.g.,
comments on page 76 of \cite{goeshu}.}

After this long exercise we can derive easily all formulas in chapters 3-6 of
\ \cite{s1} without using any matrix representation at all. In particular, for
future reference we collect some formulas,%

\begin{align}
\mathbf{q}^{\mu}\mathbf{\check{q}}_{\mu}  &  =-4,\hspace{0.15in}q^{\mu}%
\check{q}_{\mu}=-4\sigma_{0}\nonumber\\
\mathbf{q}_{\rho}^{\mu}%
%TCIMACRO{\TeXButton{omega}{\mbox{\boldmath{$\omega$}}}}%
%BeginExpansion
\mbox{\boldmath{$\omega$}}%
%EndExpansion
\mathbf{\check{q}}_{\mu}  &  =0,\hspace{0.15in}q^{\mu}\Omega_{\rho}\check
{q}_{\mu}=0,\nonumber\\%
%TCIMACRO{\TeXButton{omega}{\mbox{\boldmath{$\omega$}}}}%
%BeginExpansion
\mbox{\boldmath{$\omega$}}%
%EndExpansion
_{\rho}  &  =-\frac{1}{2}\mathbf{\check{q}}_{\mu}(\partial_{\rho}%
\mathbf{q}^{\mu}+\Gamma_{\rho\tau}^{\mu}\mathbf{q}^{\tau}),\hspace
{0.15in}\Omega_{\rho}=-\frac{1}{2}\check{q}_{\mu}(\partial_{\rho}q^{\mu
}+\Gamma_{\rho\tau}^{\mu}q^{\tau}) \label{D.20}%
\end{align}

Before we proceed, it is important to keep in mind that our \ `normalization'
of $%
%TCIMACRO{\TeXButton{omega}{\mbox{\boldmath{$\omega$}}}}%
%BeginExpansion
\mbox{\boldmath{$\omega$}}%
%EndExpansion
_{\rho}$ (and of $\Omega_{\rho}$) here \textit{differs} from Sachs one by a
factor of $1/2$. We prefer our normalization, since it is more natural and
avoid factors of $2$ when we perform contractions.

Before we discuss the equations of Sachs theory we think it is worth, using
Clifford algebra methods, to present a formulation of Einstein's gravitational
theory which resembles a gauge theory with group $Sl(2,\mathbb{C)}$ as the
gauge group. This formulation will then be compared with Sachs theory. Our
formulation permits to prove that contrary to his claims in \cite{s1,s2} he
did not produce any unified field theory of gravitation and electromagnetism.

\section{Recall of Some Facts of the Theory of Linear Connections}

\subsection{Preliminaries}

In the general theory of connections \cite{choquet, konu} a connection is a
1-form in the cotangent space of a principal bundle, with values in the Lie
algebra of a gauge group. In order to develop a theory of a linear
connection\footnote{In words, \ $\overset{\blacktriangle}{%
%TCIMACRO{\TeXButton{omega}{\mbox{\boldmath{$\omega$}}}}%
%BeginExpansion
\mbox{\boldmath{$\omega$}}%
%EndExpansion
}$ is a 1-form in the cotangent space of the bundle of ortonornal frames with
values in the Lie algebra $\mathrm{so}_{1,3}^{e}\simeq\mathrm{sl}%
(2,\mathbb{C)}$ of the group \textrm{SO}$_{_{1,3}}^{e}(M)$. \ }
\begin{equation}
\overset{\blacktriangle}{%
%TCIMACRO{\TeXButton{omega}{\mbox{\boldmath{$\omega$}}}}%
%BeginExpansion
\mbox{\boldmath{$\omega$}}%
%EndExpansion
}\in\sec T^{\ast}P_{\mathrm{SO}_{1,3}^{e}}(M)\otimes\mathrm{sl}(2,\mathbb{C)},
\label{9.00}%
\end{equation}
with an \textit{exterior} covariant derivative operator acting on sections of
associated vector bundles to the principal bundle $P_{\mathrm{SO}_{1,3}^{e}%
}(M)$ which reproduces moreover the well known results obtained with the usual
covariant derivative of tensor fields in the base manifold, we need to
introduce the concept of a \textit{soldering }form
\begin{equation}
\overset{\blacktriangle}{%
%TCIMACRO{\TeXButton{theta}{\mbox{\boldmath{$\theta$}}}}%
%BeginExpansion
\mbox{\boldmath{$\theta$}}%
%EndExpansion
}\in\sec T^{\ast}P_{\mathrm{SO}_{1,3}^{e}}(M)\otimes\mathbb{R}^{1,3}.
\label{9.01}%
\end{equation}
Let be $U\subset$ $M$ and $\pi_{1},\pi_{2}$ respectively the projections of
$T^{\ast}P_{\mathrm{SO}_{1,3}^{e}}(M)\otimes\mathbb{R}^{1,3}$ and
$P_{\mathrm{SO}_{1,3}^{e}}(M)$ to $M$, naturally associated to the projection
$\pi$ of $P_{\mathrm{SO}_{1,3}^{e}}(M)$. Let \
\begin{align}
\varsigma_{1}  &  :U\rightarrow\pi_{1}^{-1}(U)\subset T^{\ast}P_{\mathrm{SO}%
_{1,3}^{e}}(M)\otimes\mathbb{R}^{1,3},\nonumber\\
\varsigma_{2}  &  :U\rightarrow\pi_{2}^{-1}(U)\subset T^{\ast}P_{\mathrm{SO}%
_{1,3}^{e}}(M)\otimes\mathrm{sl}(2,\mathbb{C)}, \label{9.011}%
\end{align}
be two cross sections. We are interested in the study of the pullbacks $%
%TCIMACRO{\TeXButton{omega}{\mbox{\boldmath{$\omega$}}}}%
%BeginExpansion
\mbox{\boldmath{$\omega$}}%
%EndExpansion
=\varsigma_{2}^{\ast}$ $\overset{\blacktriangle}{%
%TCIMACRO{\TeXButton{omega}{\mbox{\boldmath{$\omega$}}}}%
%BeginExpansion
\mbox{\boldmath{$\omega$}}%
%EndExpansion
}$ and $%
%TCIMACRO{\TeXButton{theta}{\mbox{\boldmath{$\theta$}}}}%
%BeginExpansion
\mbox{\boldmath{$\theta$}}%
%EndExpansion
=\varsigma_{1}^{\ast}$ $\overset{\blacktriangle}{%
%TCIMACRO{\TeXButton{theta}{\mbox{\boldmath{$\theta$}}}}%
%BeginExpansion
\mbox{\boldmath{$\theta$}}%
%EndExpansion
}$ once we give a local trivialization of the respective bundles. As it is
well known \cite{konu}, we have in a local chart $\langle x^{\mu}\rangle$
covering $U,$
\begin{equation}%
%TCIMACRO{\TeXButton{theta}{\mbox{\boldmath{$\theta$}}}}%
%BeginExpansion
\mbox{\boldmath{$\theta$}}%
%EndExpansion
=e_{\mu}\otimes dx^{\mu}\equiv e_{\mu}dx^{\mu}\in\sec TM\otimes%
%TCIMACRO{\dbigwedge \nolimits^{1}}%
%BeginExpansion
{\displaystyle\bigwedge\nolimits^{1}}
%EndExpansion
T^{\ast}M\text{.} \label{9.1}%
\end{equation}

Now, we give the Clifford algebra structure to the tangent bundle, thus
generating the Clifford bundle $\mathcal{C\ell}(TM)=%
%TCIMACRO{\dbigcup \nolimits_{x}}%
%BeginExpansion
{\displaystyle\bigcup\nolimits_{x}}
%EndExpansion
\mathcal{C\ell}_{x}(M),$ with $\mathcal{C\ell}_{x}(M)\simeq\mathbb{R}_{1,3}$
introduced in Appendix A.

We recall moreover, a well known result \cite{lounesto}, namely, that for each
$x\in U\subset M$ the bivectors of $\mathcal{C\ell}(T_{x}M)$ generate under
the product defined by the commutator, the Lie algebra $\mathrm{sl}%
(2,\mathbb{C)}$. We thus are lead to define the representatives in
$\mathcal{C\ell}(TM)\otimes%
%TCIMACRO{\dbigwedge }%
%BeginExpansion
{\displaystyle\bigwedge}
%EndExpansion
T^{\ast}M$ for $%
%TCIMACRO{\TeXButton{theta}{\mbox{\boldmath{$\theta$}}}}%
%BeginExpansion
\mbox{\boldmath{$\theta$}}%
%EndExpansion
$ and for the the pullback $%
%TCIMACRO{\TeXButton{omega}{\mbox{\boldmath{$\omega$}}}}%
%BeginExpansion
\mbox{\boldmath{$\omega$}}%
%EndExpansion
$ of the connection \textit{in a given gauge} (that we represent with the same
symbols):
\begin{align}%
%TCIMACRO{\TeXButton{theta}{\mbox{\boldmath{$\theta$}}}}%
%BeginExpansion
\mbox{\boldmath{$\theta$}}%
%EndExpansion
&  =e_{\mu}dx^{\mu}=\mathbf{e}_{\mathbf{a}}\theta^{\mathbf{a}}\in\sec%
%TCIMACRO{\dbigwedge \nolimits^{1}}%
%BeginExpansion
{\displaystyle\bigwedge\nolimits^{1}}
%EndExpansion
TM\otimes%
%TCIMACRO{\dbigwedge \nolimits^{1}}%
%BeginExpansion
{\displaystyle\bigwedge\nolimits^{1}}
%EndExpansion
T^{\ast}M\hookrightarrow\mathcal{C\ell}(TM)\otimes%
%TCIMACRO{\dbigwedge \nolimits^{1}}%
%BeginExpansion
{\displaystyle\bigwedge\nolimits^{1}}
%EndExpansion
T^{\ast}M,\nonumber\\%
%TCIMACRO{\TeXButton{omega}{\mbox{\boldmath{$\omega$}}}}%
%BeginExpansion
\mbox{\boldmath{$\omega$}}%
%EndExpansion
&  =\frac{1}{2}\omega_{\mathbf{a}}^{\mathbf{bc}}\mathbf{e}_{\mathbf{b}%
}\mathbf{e}_{\mathbf{c}}\theta^{\mathbf{a}}\nonumber\\
&  =\frac{1}{2}\omega_{\mathbf{a}}^{\mathbf{bc}}(\mathbf{e}_{\mathbf{b}}%
\wedge\mathbf{e}_{\mathbf{c}})\otimes\theta^{\mathbf{a}}\in\sec%
%TCIMACRO{\dbigwedge \nolimits^{2}}%
%BeginExpansion
{\displaystyle\bigwedge\nolimits^{2}}
%EndExpansion
TM\otimes%
%TCIMACRO{\dbigwedge \nolimits^{1}}%
%BeginExpansion
{\displaystyle\bigwedge\nolimits^{1}}
%EndExpansion
T^{\ast}M\hookrightarrow\mathcal{C\ell}(TM)\otimes%
%TCIMACRO{\dbigwedge \nolimits^{1}}%
%BeginExpansion
{\displaystyle\bigwedge\nolimits^{1}}
%EndExpansion
T^{\ast}M. \label{9.2}%
\end{align}

Before we continue we must recall that whereas $%
%TCIMACRO{\TeXButton{theta}{\mbox{\boldmath{$\theta$}}}}%
%BeginExpansion
\mbox{\boldmath{$\theta$}}%
%EndExpansion
$ is a true tensor, $%
%TCIMACRO{\TeXButton{omega}{\mbox{\boldmath{$\omega$}}}}%
%BeginExpansion
\mbox{\boldmath{$\omega$}}%
%EndExpansion
$ is not a true tensor, since as it is well known, its \ `components' do not
have the tensor transformation properties. Note that the $\mathbf{\omega
}_{\mathbf{a}}^{\mathbf{bc}}$ are the \ `components' of the connection defined
by
\begin{equation}
D_{\mathbf{e}_{\mathbf{a}}}e^{\mathbf{b}}=-\mathbf{\omega}_{\mathbf{ac}%
}^{\mathbf{b}}e^{\mathbf{c}},\hspace{0.15in}\mathbf{\omega}_{\mathbf{ac}%
}^{\mathbf{b}}=-\mathbf{\omega}_{\mathbf{ca}}^{\mathbf{b}}, \label{9.2'}%
\end{equation}
where $D_{\mathbf{e}_{\mathbf{a}}}$ is a metric compatible covariant
derivative operator\footnote{After section 2.5, $D_{\mathbf{e}_{\mathbf{a}}}$
refers to the Levi-Civita covariant derivative operator.} defined on the
tensor bundle, that naturally acts on $\mathcal{C\ell}(TM)$ (see, e.g.,
\cite{cru}). Objects like $%
%TCIMACRO{\TeXButton{theta}{\mbox{\boldmath{$\theta$}}}}%
%BeginExpansion
\mbox{\boldmath{$\theta$}}%
%EndExpansion
$ and $%
%TCIMACRO{\TeXButton{omega}{\mbox{\boldmath{$\omega$}}}}%
%BeginExpansion
\mbox{\boldmath{$\omega$}}%
%EndExpansion
$ will be called Clifford valued differential forms (or Clifford valued forms,
for short)\footnote{Analogous, but non equivalent concepts have been
introduced in \cite{dimakis,vatorr,vatorr2,tucker}. In particular
\cite{dimakis} introduce clifforms, i.e., forms with values in a abstract
(internal) Clifford algebra $\mathbb{R}_{p,q}$ associated with a pair
$(\mathbb{R}^{n},g)$, where $n=p+q$ and $g$ is a bilinear form of signature
$(p,q)$ in $\mathbb{R}^{n}$. These objects \textit{differ }from the Clifford
valued differential forms used in this text., whith dispenses any abstract
(internal) space.}, and in the next sections we give a detailed account of the
algebra and calculus of that objects. But, before we start this project we
need to recall some concepts of the theory of linear connections.

\subsection{ Exterior Covariant Differential}

One of our objectives is to show how to describe, with our formalism
an\ exterior covariant differential (\textit{EXCD}) which acts naturally on
sections of Clifford valued differential forms (i.e., sections of
$\sec\mathcal{C\ell}(TM)\otimes%
%TCIMACRO{\dbigwedge }%
%BeginExpansion
{\displaystyle\bigwedge}
%EndExpansion
T^{\ast}M$ ) and which \textit{mimics} the action of the pullback of the
exterior covariant derivative operator acting on sections of a vector bundle
associated to the principal bundle $P_{\mathrm{SO}_{1,3}^{e}}(M)$, once a
linear metrical compatible connection is given. Our motivation for the
definition \ of the \textit{EXCD }is that with it, the calculations of
curvature bivectors, Bianchi identities, etc., use always the same formula. Of
course, we compare our definition, with other definitions of analogous, but
distinct concepts, already used in the literature, showing where they differ
from ours, and why we think that ours seems more appropriate. In particular,
with the \textit{EXCD }and its associated \textit{extended} covariant
derivative\textit{ (ECD)} we can write Einstein's equations in \ such a way
that the resulting equation looks like an equation for a gauge theory of the
group $Sl(2,\mathbb{C)}$. To achieve our goal, we recall below the well known
definition of the exterior covariant differential $\mathbf{d}^{E}$ acting on
arbitrary sections of \ a vector bundle $E(M)$ (associated to $P_{\mathrm{SO}%
_{1,3}^{e}}(M)$ and having as typical fiber a $l$-dimensional real vector
space) and on\ $\mathrm{end}E\left(  M\right)  =E\left(  M\right)  \otimes
E^{\ast}(M)$, the bundle of endomorphisms of $E\left(  M\right)  $. We recall
also the concept of absolute differential acting on sections of the tensor
bundle, for the particular case of $\bigwedge\nolimits^{l}TM$.

\begin{definition}
\label{extcovop}The exterior covariant differential operator $\mathbf{d}^{E}$
acting on sections of $E\left(  M\right)  $ and $\mathrm{end}E\left(
M\right)  $ is the mapping
\begin{equation}
\mathbf{d}^{E}\mathbf{:}\sec E\left(  M\right)  \rightarrow\sec E\left(
M\right)  \otimes%
%TCIMACRO{\dbigwedge \nolimits^{1}}%
%BeginExpansion
{\displaystyle\bigwedge\nolimits^{1}}
%EndExpansion
T^{\ast}M, \label{W1}%
\end{equation}
such that for any differentiable function $f:M\rightarrow\mathbb{R}$,
$A\in\sec E\left(  M\right)  $ and any $F\in\sec(\mathrm{end}E\left(
M\right)  \otimes%
%TCIMACRO{\dbigwedge \nolimits^{p}}%
%BeginExpansion
{\displaystyle\bigwedge\nolimits^{p}}
%EndExpansion
T^{\ast}M),$ $G\in\sec(\mathrm{end}E\left(  M\right)  \otimes%
%TCIMACRO{\dbigwedge \nolimits^{q}}%
%BeginExpansion
{\displaystyle\bigwedge\nolimits^{q}}
%EndExpansion
T^{\ast}M)$ we have:%
\begin{align}
\mathbf{d}^{E}\mathbf{(}fA)  &  =df\otimes A+f\mathbf{d}^{E}A,\nonumber\\
\mathbf{d}^{E}(F\otimes_{\wedge}A)  &  =\mathbf{d}^{E}F\otimes_{\wedge
}A+(-1)^{p}F\otimes_{\wedge}\mathbf{d}^{E}A,\nonumber\\
\mathbf{d}^{E}(F\otimes_{\wedge}G)  &  =\mathbf{d}^{E}F\otimes_{\wedge
}G+(-1)^{p}F\otimes_{\wedge}\mathbf{d}^{E}G. \label{W2}%
\end{align}

\end{definition}

In Eq.(\ref{W2}), writing $F=F^{a}\otimes f_{a}^{(p)}$, $G=G^{b}\otimes
g_{b}^{(q)}$where $F^{a}$, $G^{b}\in\sec(\mathrm{end}E\left(  M\right)  )$,
$f_{a}^{(p)}\in\sec%
%TCIMACRO{\dbigwedge \nolimits^{p}}%
%BeginExpansion
{\displaystyle\bigwedge\nolimits^{p}}
%EndExpansion
T^{\ast}M$ and $g_{b}^{(q)}\in\sec%
%TCIMACRO{\dbigwedge \nolimits^{q}}%
%BeginExpansion
{\displaystyle\bigwedge\nolimits^{q}}
%EndExpansion
T^{\ast}M$ we have
\begin{align}
F\otimes_{\wedge}A  &  =\left(  F^{a}\otimes f_{a}^{(p)}\right)
\otimes_{\wedge}A,\nonumber\\
F\otimes_{\wedge}G  &  =\left(  F^{a}\otimes f_{a}^{(p)}\right)
\otimes_{\wedge}G^{b}\otimes g_{b}^{(q)}. \label{W2'}%
\end{align}

In what follows, in order to simplify the notation we eventually use when
there is no possibility of confusion, the simplified (sloppy) notation%
\begin{align}
\left(  F^{a}A\right)  \otimes f_{a}^{(p)}  &  \equiv\left(  F^{a}A\right)
f_{a}^{(p)},\nonumber\\
\left(  F^{a}\otimes f_{a}^{(p)}\right)  \otimes_{\wedge}G^{b}\otimes
g_{b}^{(q)}  &  =\left(  F^{a}G^{b}\right)  f_{a}^{(p)}\wedge g_{b}^{\left(
q\right)  }, \label{W2''}%
\end{align}
\ where $F^{a}A\in\sec E\left(  M\right)  $ and $F^{a}G^{b}$means the
composition of the respective endomorphisms.

Let $U\subset M$ be an open subset of $M$, $\langle x^{\mu}\rangle$ a
coordinate functions of a maximal atlas of $M$, $\{e_{\mu}\}$ a coordinate
basis of $TU\subset TM$ and $\{s_{\mathbf{K}}\},$ $\mathbf{K=}1,2,...l$ a
basis for any $\sec E\left(  U\right)  \subset\sec E\left(  M\right)  $. Then,
a basis for any section of $E\left(  M\right)  \otimes%
%TCIMACRO{\dbigwedge \nolimits^{1}}%
%BeginExpansion
{\displaystyle\bigwedge\nolimits^{1}}
%EndExpansion
T^{\ast}M$ is given by $\{s_{\mathbf{K}}\otimes dx^{\mu}\}$.

\begin{definition}
The covariant derivative operator $D_{e_{\mu}}:\sec E\left(  M\right)
\rightarrow\sec E\left(  M\right)  $ is given by
\begin{equation}
\mathbf{d}^{E}A\doteq\left(  D_{e_{\mu}}A\right)  \otimes dx^{\mu}, \label{W3}%
\end{equation}
where, writing $A=A^{\mathbf{K}}\otimes s_{\mathbf{K}}$ we have%
\begin{equation}
D_{e_{\mu}}A=\partial_{\mu}A^{\mathbf{K}}\otimes s_{\mathbf{K}}+A^{\mathbf{K}%
}\otimes D_{e_{\mu}}s_{\mathbf{K}}. \label{W4}%
\end{equation}

\end{definition}

Now, let examine the case where $E\left(  M\right)  =TM\equiv%
%TCIMACRO{\dbigwedge \nolimits^{1}}%
%BeginExpansion
{\displaystyle\bigwedge\nolimits^{1}}
%EndExpansion
(TM)\hookrightarrow\mathcal{C\ell}(TM)$. Let $\{\mathbf{e}_{\mathbf{j}}\},$ be
an orthonormal basis of $TM$. Then, $\bigskip$using Eq.(\ref{W4})
\begin{align}
\mathbf{d}^{E}\mathbf{e}_{\mathbf{j}}  &  =(D_{e_{\mathbf{k}}}\mathbf{e}%
_{\mathbf{j}})\otimes\theta^{\mathbf{k}}\equiv\mathbf{e}_{\mathbf{k}}\otimes%
%TCIMACRO{\TeXButton{omega}{\mbox{\boldmath{$\omega$}}}}%
%BeginExpansion
\mbox{\boldmath{$\omega$}}%
%EndExpansion
_{\mathbf{j}}^{\mathbf{k}}\nonumber\\%
%TCIMACRO{\TeXButton{omega}{\mbox{\boldmath{$\omega$}}}}%
%BeginExpansion
\mbox{\boldmath{$\omega$}}%
%EndExpansion
_{\mathbf{j}}^{\mathbf{k}}  &  =%
%TCIMACRO{\TeXButton{omega}{\mbox{\boldmath{$\omega$}}}}%
%BeginExpansion
\mbox{\boldmath{$\omega$}}%
%EndExpansion
_{\mathbf{rj}}^{\mathbf{k}}\theta^{\mathbf{r}}, \label{W5}%
\end{align}
where the $%
%TCIMACRO{\TeXButton{omega}{\mbox{\boldmath{$\omega$}}}}%
%BeginExpansion
\mbox{\boldmath{$\omega$}}%
%EndExpansion
_{\mathbf{j}}^{\mathbf{k}}\in\sec%
%TCIMACRO{\dbigwedge \nolimits^{1}}%
%BeginExpansion
{\displaystyle\bigwedge\nolimits^{1}}
%EndExpansion
T^{\ast}M$ are the so-called \textit{connection 1-forms}.

Also, for $\mathbf{v=}v^{\mathbf{i}}\mathbf{e}_{\mathbf{i}}\in\sec TM$, we
have
\begin{align}
\mathbf{d}^{E}\mathbf{v}  &  =D_{\mathbf{e}_{\mathbf{i}}}\mathbf{v\otimes
}\theta^{\mathbf{i}}=\mathbf{e}_{\mathbf{i}}\otimes\mathbf{d}^{E}%
v^{\mathbf{i}},\nonumber\\
\mathbf{d}^{E}v^{\mathbf{i}}  &  =dv^{\mathbf{i}}+%
%TCIMACRO{\TeXButton{omega}{\mbox{\boldmath{$\omega$}}}}%
%BeginExpansion
\mbox{\boldmath{$\omega$}}%
%EndExpansion
_{\mathbf{k}}^{\mathbf{i}}v^{\mathbf{k}}. \label{9.5}%
\end{align}

\subsection{Absolute Differential}

Now, \ let \ $E\left(  M\right)  =TM\equiv%
%TCIMACRO{\dbigwedge \nolimits^{l}}%
%BeginExpansion
{\displaystyle\bigwedge\nolimits^{l}}
%EndExpansion
(TM)\hookrightarrow\mathcal{C\ell}(TM).$\ Recall that the usual
\textit{absolute differential} $D$ of $A\in\sec%
%TCIMACRO{\dbigwedge \nolimits^{l}}%
%BeginExpansion
{\displaystyle\bigwedge\nolimits^{l}}
%EndExpansion
TM\hookrightarrow\sec\mathcal{C\ell}\left(  TM\right)  $ is a mapping (see,
e.g., \cite{choquet})%
\begin{equation}
D\mathbf{:}\sec%
%TCIMACRO{\dbigwedge \nolimits^{l}}%
%BeginExpansion
{\displaystyle\bigwedge\nolimits^{l}}
%EndExpansion
TM\rightarrow\sec%
%TCIMACRO{\dbigwedge \nolimits^{l}}%
%BeginExpansion
{\displaystyle\bigwedge\nolimits^{l}}
%EndExpansion
TM\otimes%
%TCIMACRO{\dbigwedge \nolimits^{1}}%
%BeginExpansion
{\displaystyle\bigwedge\nolimits^{1}}
%EndExpansion
T^{\ast}M, \label{W.12}%
\end{equation}
such that for any differentiable $A\in\sec%
%TCIMACRO{\dbigwedge \nolimits^{l}}%
%BeginExpansion
{\displaystyle\bigwedge\nolimits^{l}}
%EndExpansion
TM$ we have%
\begin{equation}
DA=\left(  D_{\mathbf{e}_{\mathbf{i}}}A\right)  \otimes\theta^{i},
\label{W.14}%
\end{equation}
where $D_{e_{\mathbf{i}}}A$ is the standard covariant derivative of $A\in\sec%
%TCIMACRO{\dbigwedge \nolimits^{l}}%
%BeginExpansion
{\displaystyle\bigwedge\nolimits^{l}}
%EndExpansion
TM\hookrightarrow\sec\mathcal{C\ell}\left(  TM\right)  $. Also, for any
differentiable function $f:M\rightarrow\mathbb{R}$, and differentiable
$A\in\sec%
%TCIMACRO{\dbigwedge \nolimits^{l}}%
%BeginExpansion
{\displaystyle\bigwedge\nolimits^{l}}
%EndExpansion
TM$ we have%

\begin{equation}
D\mathbf{(}fA)=df\otimes A+fDA. \label{9.13}%
\end{equation}

Now, if we suppose that the orthonormal basis $\{\mathbf{e}_{\mathbf{j}}\}$ of
$TM$ is such that each $\mathbf{e}_{\mathbf{j}}\in\sec%
%TCIMACRO{\dbigwedge \nolimits^{1}}%
%BeginExpansion
{\displaystyle\bigwedge\nolimits^{1}}
%EndExpansion
TM$ $\hookrightarrow\sec\mathcal{C\ell}\left(  TM\right)  ,$ we can find
easily using the Clifford algebra structure of the space of multivectors that
Eq.(\ref{W5}) can be written as:
\begin{align}
D\mathbf{e}_{\mathbf{j}}  &  =(D_{\mathbf{e}_{\mathbf{k}}}\mathbf{e}%
_{\mathbf{j}})\theta^{\mathbf{k}}=\frac{1}{2}[%
%TCIMACRO{\TeXButton{omega}{\mbox{\boldmath{$\omega$}}}}%
%BeginExpansion
\mbox{\boldmath{$\omega$}}%
%EndExpansion
,\mathbf{e}_{\mathbf{j}}]=-\mathbf{e}_{\mathbf{j}}\lrcorner%
%TCIMACRO{\TeXButton{omega}{\mbox{\boldmath{$\omega$}}}}%
%BeginExpansion
\mbox{\boldmath{$\omega$}}%
%EndExpansion
\nonumber\\%
%TCIMACRO{\TeXButton{omega}{\mbox{\boldmath{$\omega$}}}}%
%BeginExpansion
\mbox{\boldmath{$\omega$}}%
%EndExpansion
&  =\frac{1}{2}\omega_{\mathbf{k}}^{\mathbf{ab}}\mathbf{e}_{\mathbf{a}}%
\wedge\mathbf{e}_{\mathbf{b}}\otimes\theta^{\mathbf{k}}\nonumber\\
&  \equiv\frac{1}{2}\omega_{\mathbf{k}}^{\mathbf{ab}}\mathbf{e}_{\mathbf{a}%
}\mathbf{e}_{\mathbf{b}}\otimes\theta^{\mathbf{k}}\in\sec%
%TCIMACRO{\dbigwedge \nolimits^{2}}%
%BeginExpansion
{\displaystyle\bigwedge\nolimits^{2}}
%EndExpansion
TM\otimes%
%TCIMACRO{\dbigwedge \nolimits^{1}}%
%BeginExpansion
{\displaystyle\bigwedge\nolimits^{1}}
%EndExpansion
T^{\ast}M\hookrightarrow\sec\mathcal{C\ell}(TM)\otimes%
%TCIMACRO{\dbigwedge \nolimits^{1}}%
%BeginExpansion
{\displaystyle\bigwedge\nolimits^{1}}
%EndExpansion
T^{\ast}M, \label{9.9}%
\end{align}
where $%
%TCIMACRO{\TeXButton{omega}{\mbox{\boldmath{$\omega$}}}}%
%BeginExpansion
\mbox{\boldmath{$\omega$}}%
%EndExpansion
$ is the \textit{representative} of the connection in a given gauge.

The general case is given by the following proposition.

\begin{proposition}
For $A\in\sec%
%TCIMACRO{\dbigwedge \nolimits^{l}}%
%BeginExpansion
{\displaystyle\bigwedge\nolimits^{l}}
%EndExpansion
TM\hookrightarrow\sec\mathcal{C\ell}\left(  TM\right)  $%
\begin{equation}
DA=dA+\frac{1}{2}[%
%TCIMACRO{\TeXButton{omega}{\mbox{\boldmath{$\omega$}}}}%
%BeginExpansion
\mbox{\boldmath{$\omega$}}%
%EndExpansion
,A]. \label{W.15}%
\end{equation}

\end{proposition}

\begin{proof}
The proof is a simple calculation, left to the reader.
\end{proof}

Eq.(\ref{W.15}) can now be extended by linearity for an arbitrary
nonhomogeneous multivector $A\in\sec\mathcal{C\ell}\left(  TM\right)
$.\medskip

\begin{remark}
We see that when $E(M)=%
%TCIMACRO{\dbigwedge \nolimits^{l}}%
%BeginExpansion
{\displaystyle\bigwedge\nolimits^{l}}
%EndExpansion
TM\hookrightarrow\sec\mathcal{C\ell}\left(  TM\right)  $ the absolute
differential $D$ can be identified with the exterior covariant derivative
$\mathbf{d}^{E}$.
\end{remark}

We proceed now to find an appropriate\textit{ exterior }covariant differential
which acts naturally on Clifford valued differential forms, i.e., objects that
are sections of \ $\mathcal{C\ell}(TM)\otimes%
%TCIMACRO{\dbigwedge }%
%BeginExpansion
{\displaystyle\bigwedge}
%EndExpansion
T^{\ast}M$ $(\equiv%
%TCIMACRO{\dbigwedge }%
%BeginExpansion
{\displaystyle\bigwedge}
%EndExpansion
T^{\ast}M\ \otimes\mathcal{C\ell}(TM))$ (see next section).\ Note that we
cannot simply use the above definition by using $E\left(  M\right)
=\mathcal{C\ell}(TM)$ and \textrm{end}$E\left(  M\right)  =$ \textrm{end}%
$\mathcal{C\ell}(TM)$, because \textrm{end}$\mathcal{C\ell}(TM)\neq
\mathcal{C\ell}(TM)\otimes%
%TCIMACRO{\dbigwedge }%
%BeginExpansion
{\displaystyle\bigwedge}
%EndExpansion
T^{\ast}M$. Instead, we must use the above theory and possible applications as
a guide in order to find an appropriate definition. Let us see how this can be done.

\section{Clifford Valued Differential Forms}

\begin{definition}
A \textit{homogeneous} multivector valued differential form of type $(l,p)$ is
a section of $%
%TCIMACRO{\dbigwedge \nolimits^{l}}%
%BeginExpansion
{\displaystyle\bigwedge\nolimits^{l}}
%EndExpansion
TM\otimes%
%TCIMACRO{\dbigwedge \nolimits^{p}}%
%BeginExpansion
{\displaystyle\bigwedge\nolimits^{p}}
%EndExpansion
T^{\ast}M\hookrightarrow\mathcal{C\ell}(TM)\otimes%
%TCIMACRO{\dbigwedge }%
%BeginExpansion
{\displaystyle\bigwedge}
%EndExpansion
T^{\ast}M$, for $0\leq l\leq4$, $0\leq p\leq4$. A section of $\mathcal{C\ell
}(TM)\otimes%
%TCIMACRO{\dbigwedge }%
%BeginExpansion
{\displaystyle\bigwedge}
%EndExpansion
T^{\ast}M$ such that the multivector part is non homogeneous is called a
Clifford valued differential form.
\end{definition}

We recall, that any $A\in\sec%
%TCIMACRO{\dbigwedge \nolimits^{l}}%
%BeginExpansion
{\displaystyle\bigwedge\nolimits^{l}}
%EndExpansion
TM\otimes%
%TCIMACRO{\dbigwedge \nolimits^{p}}%
%BeginExpansion
{\displaystyle\bigwedge\nolimits^{p}}
%EndExpansion
T^{\ast}M$ $\hookrightarrow\sec\mathcal{C\ell}(TM)\otimes%
%TCIMACRO{\dbigwedge \nolimits^{p}}%
%BeginExpansion
{\displaystyle\bigwedge\nolimits^{p}}
%EndExpansion
T^{\ast}M$ can always be written as%
\begin{align}
A  &  =m_{(l)}\otimes\psi^{(p)}\equiv\frac{1}{l!}m_{(l)}^{\mathbf{i}%
_{1}...\mathbf{i}_{l}}\mathbf{e}_{\mathbf{i}_{1}}...\mathbf{e}_{\mathbf{i}%
_{l}}\otimes\psi^{(p)}\nonumber\\
&  =\frac{1}{p!}m_{(l)}\otimes\psi_{\mathbf{j}_{1}...\mathbf{j}_{p}}%
^{(p)}\theta^{\mathbf{j}_{1}}\wedge...\wedge\theta^{\mathbf{j}_{p}}\nonumber\\
&  =\frac{1}{l!p!}m_{(l)}^{\mathbf{i}_{1}...\mathbf{i}_{l}}\mathbf{e}%
_{\mathbf{i}_{1}}...\mathbf{e}_{\mathbf{i}_{l}}\otimes\psi_{\mathbf{j}%
_{1}....\mathbf{j}_{p}}^{(p)}\theta^{\mathbf{j}_{1}}\wedge...\wedge
\theta^{\mathbf{j}_{p}}\label{9.6}\\
&  =\frac{1}{l!p!}A_{\mathbf{j}_{1}...\mathbf{j}_{p}}^{\mathbf{i}%
_{1}...\mathbf{i}_{l}}\mathbf{e}_{\mathbf{i}_{1}}...\mathbf{e}_{\mathbf{i}%
_{l}}\otimes\theta^{\mathbf{i}_{1}}\wedge...\wedge\theta^{\mathbf{i}_{p}%
}.\nonumber
\end{align}
\medskip

\begin{definition}
The $\otimes_{\wedge}$ product of $A=\overset{m}{A}\otimes\psi^{(p)}\in
\sec\mathcal{C\ell}(TM)\otimes%
%TCIMACRO{\dbigwedge \nolimits^{p}}%
%BeginExpansion
{\displaystyle\bigwedge\nolimits^{p}}
%EndExpansion
T^{\ast}M$ and $B=\overset{m}{B}\otimes\chi^{(p)}\in\sec\mathcal{C\ell
}(TM)\otimes%
%TCIMACRO{\dbigwedge \nolimits^{q}}%
%BeginExpansion
{\displaystyle\bigwedge\nolimits^{q}}
%EndExpansion
T^{\ast}M$ is the mapping: \
\begin{align}
\hspace{-1cm}\otimes_{\wedge}  &  :\sec\mathcal{C\ell}(TM)\otimes%
%TCIMACRO{\dbigwedge \nolimits^{l}}%
%BeginExpansion
{\displaystyle\bigwedge\nolimits^{l}}
%EndExpansion
T^{\ast}M\times\sec\mathcal{C\ell}(TM)\otimes%
%TCIMACRO{\dbigwedge \nolimits^{p}}%
%BeginExpansion
{\displaystyle\bigwedge\nolimits^{p}}
%EndExpansion
T^{\ast}M\nonumber\\
&  \rightarrow\sec\mathcal{C\ell}(TM)\otimes%
%TCIMACRO{\dbigwedge \nolimits^{l+p}}%
%BeginExpansion
{\displaystyle\bigwedge\nolimits^{l+p}}
%EndExpansion
T^{\ast}M,\nonumber\\
A\otimes_{\wedge}B  &  =\overset{m}{A}\overset{m}{B}\otimes\psi^{\left(
p\right)  }\wedge\chi^{(q)}. \label{9.6PROD}%
\end{align}
\smallskip
\end{definition}

\begin{definition}
The \textit{commutator} $[A,B]$ of \ $A\in\sec%
%TCIMACRO{\dbigwedge \nolimits^{l}}%
%BeginExpansion
{\displaystyle\bigwedge\nolimits^{l}}
%EndExpansion
TM\otimes%
%TCIMACRO{\dbigwedge \nolimits^{p}}%
%BeginExpansion
{\displaystyle\bigwedge\nolimits^{p}}
%EndExpansion
T^{\ast}M\hookrightarrow\sec\mathcal{C\ell}(TM)\otimes%
%TCIMACRO{\dbigwedge \nolimits^{p}}%
%BeginExpansion
{\displaystyle\bigwedge\nolimits^{p}}
%EndExpansion
T^{\ast}M$ and $B\in%
%TCIMACRO{\dbigwedge \nolimits^{m}}%
%BeginExpansion
{\displaystyle\bigwedge\nolimits^{m}}
%EndExpansion
TM\otimes%
%TCIMACRO{\dbigwedge \nolimits^{q}}%
%BeginExpansion
{\displaystyle\bigwedge\nolimits^{q}}
%EndExpansion
T^{\ast}M\hookrightarrow\sec\mathcal{C\ell}(TM)\otimes%
%TCIMACRO{\dbigwedge \nolimits^{q}}%
%BeginExpansion
{\displaystyle\bigwedge\nolimits^{q}}
%EndExpansion
T^{\ast}M$\ \ is the mapping:%
\begin{align}
\lbrack\hspace{0.15in},\hspace{0.15in}]  &  :\sec%
%TCIMACRO{\dbigwedge \nolimits^{l}}%
%BeginExpansion
{\displaystyle\bigwedge\nolimits^{l}}
%EndExpansion
TM\otimes%
%TCIMACRO{\dbigwedge \nolimits^{p}}%
%BeginExpansion
{\displaystyle\bigwedge\nolimits^{p}}
%EndExpansion
T^{\ast}M\times\sec%
%TCIMACRO{\dbigwedge \nolimits^{m}}%
%BeginExpansion
{\displaystyle\bigwedge\nolimits^{m}}
%EndExpansion
TM\otimes%
%TCIMACRO{\dbigwedge \nolimits^{q}}%
%BeginExpansion
{\displaystyle\bigwedge\nolimits^{q}}
%EndExpansion
T^{\ast}M\nonumber\\
&  \rightarrow\sec((%
%TCIMACRO{\dsum \limits_{k=|l-m|}^{|l+m|}}%
%BeginExpansion
{\displaystyle\sum\limits_{k=|l-m|}^{|l+m|}}
%EndExpansion%
%TCIMACRO{\dbigwedge \nolimits^{k}}%
%BeginExpansion
{\displaystyle\bigwedge\nolimits^{k}}
%EndExpansion
T^{\ast}M)\otimes%
%TCIMACRO{\dbigwedge \nolimits^{p+q}}%
%BeginExpansion
{\displaystyle\bigwedge\nolimits^{p+q}}
%EndExpansion
T^{\ast}M)\nonumber\\
\lbrack A,B]  &  =A\otimes_{\wedge}B-\left(  -1\right)  ^{pq}B\otimes_{\wedge
}A. \label{9.7}%
\end{align}
Writing $A=\frac{1}{l!}A^{\mathbf{j}_{1}..\mathbf{.j}_{l}}\mathbf{e}%
_{\mathbf{j}_{1}}\mathbf{...e}_{\mathbf{j}_{l}}\psi^{\left(  p\right)  }$,
$B=\frac{1}{m!}B^{\mathbf{i}_{1}...\mathbf{i}_{m}}\mathbf{e}_{\mathbf{i}_{1}%
}\mathbf{...e}_{\mathbf{i}_{m}}\chi^{(q)}$, with $\psi^{(p)}\in\sec%
%TCIMACRO{\dbigwedge \nolimits^{p}}%
%BeginExpansion
{\displaystyle\bigwedge\nolimits^{p}}
%EndExpansion
T^{\ast}M$ and $\chi^{(q)}\in\sec%
%TCIMACRO{\dbigwedge \nolimits^{q}}%
%BeginExpansion
{\displaystyle\bigwedge\nolimits^{q}}
%EndExpansion
T^{\ast}M$, we have
\begin{equation}
\lbrack A,B]=\frac{1}{l!m!}A^{\mathbf{j}_{1}..\mathbf{.j}_{l}}B^{\mathbf{i}%
_{1}...\mathbf{i}_{m}}\left[  \mathbf{e}_{\mathbf{j}_{1}}\mathbf{...e}%
_{\mathbf{j}_{l}}\mathbf{,e}_{\mathbf{i}_{1}}\mathbf{...e}_{\mathbf{i}_{m}%
}\right]  \psi^{\left(  p\right)  }\wedge\chi^{(q)}, \label{9.7BIS}%
\end{equation}
\ The definition of\ the commutator is extended by linearity to arbitrary
sections of $\mathcal{C\ell}(TM)\otimes%
%TCIMACRO{\dbigwedge }%
%BeginExpansion
{\displaystyle\bigwedge}
%EndExpansion
T^{\ast}M$.
\end{definition}

Now, we have the proposition.

\begin{proposition}
Let \ $A\in\sec\mathcal{C\ell}(TM)\otimes%
%TCIMACRO{\dbigwedge \nolimits^{p}}%
%BeginExpansion
{\displaystyle\bigwedge\nolimits^{p}}
%EndExpansion
T^{\ast}M$, $B\in\sec\mathcal{C\ell}(TM)\otimes%
%TCIMACRO{\dbigwedge \nolimits^{q}}%
%BeginExpansion
{\displaystyle\bigwedge\nolimits^{q}}
%EndExpansion
T^{\ast}M$, $C\in A\in\sec\mathcal{C\ell}(TM)\otimes%
%TCIMACRO{\dbigwedge \nolimits^{r}}%
%BeginExpansion
{\displaystyle\bigwedge\nolimits^{r}}
%EndExpansion
T^{\ast}M$. Then,%
\begin{equation}
\lbrack A,B]=(-1)^{1+pq}[B,A], \label{p1}%
\end{equation}
and%
\begin{equation}
(-1)^{pr}\left[  \left[  A,B\right]  ,C\right]  +(-1)^{qp}\left[  \left[
B,C\right]  ,A\right]  +(-)^{rq}\left[  \left[  C,A\right]  ,B\right]  =0.
\label{p2}%
\end{equation}

\end{proposition}

\begin{proof}
It follows directly from a simple calculation, left to the reader.
\end{proof}

Eq.(\ref{p2}) may be called the \textit{graded Jacobi identity }%
\cite{bleecker}.

\begin{corollary}
Let $A^{(2)}\in\sec%
%TCIMACRO{\dbigwedge \nolimits^{2}}%
%BeginExpansion
{\displaystyle\bigwedge\nolimits^{2}}
%EndExpansion
(TM)\otimes%
%TCIMACRO{\dbigwedge \nolimits^{p}}%
%BeginExpansion
{\displaystyle\bigwedge\nolimits^{p}}
%EndExpansion
T^{\ast}M$ and \ $B\in\sec%
%TCIMACRO{\dbigwedge \nolimits^{r}}%
%BeginExpansion
{\displaystyle\bigwedge\nolimits^{r}}
%EndExpansion
(TM)\otimes%
%TCIMACRO{\dbigwedge \nolimits^{q}}%
%BeginExpansion
{\displaystyle\bigwedge\nolimits^{q}}
%EndExpansion
T^{\ast}M$. Then,%
\begin{equation}
\lbrack A^{(2)},B]=C, \label{p2bis}%
\end{equation}
where $C\in\sec%
%TCIMACRO{\dbigwedge \nolimits^{r}}%
%BeginExpansion
{\displaystyle\bigwedge\nolimits^{r}}
%EndExpansion
(TM)\otimes%
%TCIMACRO{\dbigwedge \nolimits^{p+q}}%
%BeginExpansion
{\displaystyle\bigwedge\nolimits^{p+q}}
%EndExpansion
T^{\ast}M$.
\end{corollary}

\begin{proof}
It follows from \ a direct \ calculation, left to the reader.
\end{proof}

\begin{proposition}
Let \ $\omega\in\sec%
%TCIMACRO{\dbigwedge \nolimits^{2}}%
%BeginExpansion
{\displaystyle\bigwedge\nolimits^{2}}
%EndExpansion
(TM)\otimes%
%TCIMACRO{\dbigwedge \nolimits^{1}}%
%BeginExpansion
{\displaystyle\bigwedge\nolimits^{1}}
%EndExpansion
T^{\ast}M$ , $A\in\sec%
%TCIMACRO{\dbigwedge \nolimits^{l}}%
%BeginExpansion
{\displaystyle\bigwedge\nolimits^{l}}
%EndExpansion
(TM)\otimes%
%TCIMACRO{\dbigwedge \nolimits^{p}}%
%BeginExpansion
{\displaystyle\bigwedge\nolimits^{p}}
%EndExpansion
T^{\ast}M$.$B\in\sec%
%TCIMACRO{\dbigwedge \nolimits^{m}}%
%BeginExpansion
{\displaystyle\bigwedge\nolimits^{m}}
%EndExpansion
(TM)\otimes%
%TCIMACRO{\dbigwedge \nolimits^{q}}%
%BeginExpansion
{\displaystyle\bigwedge\nolimits^{q}}
%EndExpansion
T^{\ast}M$. Then, we have%
\begin{equation}
(p+q)[\omega,A\otimes_{\wedge}B]=p[\omega,A]\otimes_{\wedge}B+(-1)^{p}%
qA\otimes_{\wedge}[\omega,B]. \label{p.2eureka}%
\end{equation}

\end{proposition}

\begin{proof}
\ Write, $\omega=\frac{1}{2}\omega_{i}^{ab}\mathbf{e}_{\mathbf{a}}%
\mathbf{e}_{\mathbf{b}}\theta^{i},A=\frac{1}{m!}A^{j_{1}...j_{l}}%
\mathbf{e}_{\mathbf{j}_{1}}...\mathbf{e}_{\mathbf{j}_{l}}A^{(p)},B=\frac
{1}{m!}B^{i_{1}...i_{m}}\mathbf{e}_{\mathbf{i}_{1}}...\mathbf{e}%
_{\mathbf{i}_{m}}B^{(q)}$.

\hspace{0.2cm}Then,
\begin{align*}
(p+q)[\omega,A\otimes_{\wedge}B]  &  =(p+q)\frac{1}{2l!m!}\omega_{i}%
^{ab}A^{j_{1}...j_{l}}B^{i_{1}...i_{m}}[\mathbf{e}_{\mathbf{a}}\mathbf{e}%
_{\mathbf{b}},\mathbf{e}_{\mathbf{j}_{1}}...\mathbf{e}_{\mathbf{j}_{l}%
}\mathbf{e}_{\mathbf{i}_{1}}...\mathbf{e}_{\mathbf{i}_{m}}]\otimes\theta
^{i}\wedge A^{(p)}\wedge B^{(q)}\\
&  =p\frac{1}{2l!m!}\omega_{i}^{ab}A^{j_{1}...j_{l}}B^{i_{1}...i_{m}%
}[\mathbf{e}_{\mathbf{a}}\mathbf{e}_{\mathbf{b}},\mathbf{e}_{\mathbf{j}_{1}%
}...\mathbf{e}_{\mathbf{j}_{l}}\mathbf{e}_{\mathbf{i}_{1}}...\mathbf{e}%
_{\mathbf{i}_{m}}]\otimes\theta^{i}\wedge A^{(p)}\wedge B^{(q)}\\
&  +q\frac{1}{2l!m!}\omega_{i}^{ab}A^{j_{1}...j_{l}}B^{i_{1}...i_{m}%
}[\mathbf{e}_{\mathbf{a}}\mathbf{e}_{\mathbf{b}},\mathbf{e}_{\mathbf{j}_{1}%
}...\mathbf{e}_{\mathbf{j}_{l}}\mathbf{e}_{\mathbf{i}_{1}}...\mathbf{e}%
_{\mathbf{i}_{m}}]\otimes\theta^{i}\wedge A^{(p)}\wedge B^{(q)}\\
&  =pA[\omega,A]\otimes_{\wedge}B+(-1)^{p}qA\otimes_{\wedge}[\omega,B].
\end{align*}

\end{proof}

\begin{definition}
The action of the differential operator $d$ acting on%
\[
A\in\sec%
%TCIMACRO{\dbigwedge \nolimits^{l}}%
%BeginExpansion
{\displaystyle\bigwedge\nolimits^{l}}
%EndExpansion
TM\otimes%
%TCIMACRO{\dbigwedge \nolimits^{p}}%
%BeginExpansion
{\displaystyle\bigwedge\nolimits^{p}}
%EndExpansion
T^{\ast}M\hookrightarrow\sec\mathcal{C\ell}(TM)\otimes%
%TCIMACRO{\dbigwedge \nolimits^{p}}%
%BeginExpansion
{\displaystyle\bigwedge\nolimits^{p}}
%EndExpansion
T^{\ast}M,
\]
is given by:%
\begin{align}
dA  &  \circeq\mathbf{e}_{\mathbf{j}_{1}}...\mathbf{e}_{\mathbf{j}_{l}}\otimes
dA^{\mathbf{j}_{1}...\mathbf{j}_{l}}\label{9.8}\\
&  =\mathbf{e}_{\mathbf{j}_{1}}...\mathbf{e}_{\mathbf{j}_{l}}\otimes d\frac
{1}{p!}A_{\mathbf{i}_{1}...\mathbf{i}_{p}}^{\mathbf{j}_{1}...\mathbf{j}_{l}%
}\theta^{\mathbf{i}_{1}}\wedge...\wedge\theta^{\mathbf{i}_{p}}.\nonumber
\end{align}
\ 
\end{definition}

We have the important\medskip\ proposition.

\begin{proposition}
Let $A\in\sec\mathcal{C\ell}(TM)\otimes%
%TCIMACRO{\dbigwedge \nolimits^{p}}%
%BeginExpansion
{\displaystyle\bigwedge\nolimits^{p}}
%EndExpansion
T^{\ast}M$ and $B\in\sec\mathcal{C\ell}(TM)\otimes%
%TCIMACRO{\dbigwedge \nolimits^{q}}%
%BeginExpansion
{\displaystyle\bigwedge\nolimits^{q}}
%EndExpansion
T^{\ast}M$. Then,
\begin{equation}
d[A,B]=[dA,B]+(-1)^{p}[A,dB]. \label{p3}%
\end{equation}

\end{proposition}

\begin{proof}
The proof of that proposition is a simple calculation, left to the reader.
\end{proof}

We now define the exterior covariant differential operator (\textit{EXCD)}
$\mathbf{D}$ and the \textit{extended} covariant derivative (\textit{ECD})
$\mathbf{D}_{\mathbf{e}_{\mathbf{r}}}$ acting on a Clifford valued form
$\ \mathcal{A}\in\sec%
%TCIMACRO{\dbigwedge \nolimits^{l}}%
%BeginExpansion
{\displaystyle\bigwedge\nolimits^{l}}
%EndExpansion
TM\otimes%
%TCIMACRO{\dbigwedge \nolimits^{p}}%
%BeginExpansion
{\displaystyle\bigwedge\nolimits^{p}}
%EndExpansion
T^{\ast}M\hookrightarrow\sec\mathcal{C\ell}\left(  TM\right)  $ $\otimes%
%TCIMACRO{\dbigwedge \nolimits^{p}}%
%BeginExpansion
{\displaystyle\bigwedge\nolimits^{p}}
%EndExpansion
T^{\ast}M$, as follows.

\subsection{Exterior Covariant Differential of Clifford Valued Forms}

\begin{definition}
The exterior covariant differential of $\mathcal{A}$ is the mapping :%
\begin{align}
\mathbf{D}  &  \mathbf{:}\sec%
%TCIMACRO{\dbigwedge \nolimits^{l}}%
%BeginExpansion
{\displaystyle\bigwedge\nolimits^{l}}
%EndExpansion
TM\otimes%
%TCIMACRO{\dbigwedge \nolimits^{p}}%
%BeginExpansion
{\displaystyle\bigwedge\nolimits^{p}}
%EndExpansion
T^{\ast}M\rightarrow\sec[(%
%TCIMACRO{\dbigwedge \nolimits^{l}}%
%BeginExpansion
{\displaystyle\bigwedge\nolimits^{l}}
%EndExpansion
TM\otimes%
%TCIMACRO{\dbigwedge \nolimits^{p}}%
%BeginExpansion
{\displaystyle\bigwedge\nolimits^{p}}
%EndExpansion
T^{\ast}M)\otimes_{\wedge}%
%TCIMACRO{\dbigwedge \nolimits^{1}}%
%BeginExpansion
{\displaystyle\bigwedge\nolimits^{1}}
%EndExpansion
T^{\ast}M]\nonumber\\
&  \subset\sec%
%TCIMACRO{\dbigwedge \nolimits^{l}}%
%BeginExpansion
{\displaystyle\bigwedge\nolimits^{l}}
%EndExpansion
TM\otimes%
%TCIMACRO{\dbigwedge \nolimits^{p+1}}%
%BeginExpansion
{\displaystyle\bigwedge\nolimits^{p+1}}
%EndExpansion
T^{\ast}M,\text{ }\nonumber\\
\mathbf{D}\mathcal{A}  &  =d\mathcal{A}+\frac{p}{2}[%
%TCIMACRO{\TeXButton{omega}{\mbox{\boldmath{$\omega$}}}}%
%BeginExpansion
\mbox{\boldmath{$\omega$}}%
%EndExpansion
,\mathcal{A}],\text{ if \ }\mathcal{A\in}\sec%
%TCIMACRO{\dbigwedge \nolimits^{l}}%
%BeginExpansion
{\displaystyle\bigwedge\nolimits^{l}}
%EndExpansion
TM\otimes%
%TCIMACRO{\dbigwedge \nolimits^{p}}%
%BeginExpansion
{\displaystyle\bigwedge\nolimits^{p}}
%EndExpansion
T^{\ast}M,\text{ }l,p\geq1. \label{W.21BIS}%
\end{align}

\end{definition}

\begin{proposition}
Let be $\mathcal{A}\in\sec%
%TCIMACRO{\dbigwedge \nolimits^{l}}%
%BeginExpansion
{\displaystyle\bigwedge\nolimits^{l}}
%EndExpansion
TM\otimes%
%TCIMACRO{\dbigwedge \nolimits^{p}}%
%BeginExpansion
{\displaystyle\bigwedge\nolimits^{p}}
%EndExpansion
T^{\ast}M\hookrightarrow\sec\mathcal{C\ell}\left(  TM\right)  $ $\otimes%
%TCIMACRO{\dbigwedge \nolimits^{p}}%
%BeginExpansion
{\displaystyle\bigwedge\nolimits^{p}}
%EndExpansion
T^{\ast}M$, $\mathcal{B}\in\sec%
%TCIMACRO{\dbigwedge \nolimits^{m}}%
%BeginExpansion
{\displaystyle\bigwedge\nolimits^{m}}
%EndExpansion
TM\otimes%
%TCIMACRO{\dbigwedge \nolimits^{q}}%
%BeginExpansion
{\displaystyle\bigwedge\nolimits^{q}}
%EndExpansion
T^{\ast}M\hookrightarrow\sec\mathcal{C\ell}\left(  TM\right)  $ $\otimes%
%TCIMACRO{\dbigwedge \nolimits^{q}}%
%BeginExpansion
{\displaystyle\bigwedge\nolimits^{q}}
%EndExpansion
T^{\ast}M$. Then, the exterior differential satisfies
\begin{equation}
\mathbf{D}(\mathcal{A}\otimes_{\wedge}\mathcal{B)=}\mathbf{D}\mathcal{A}%
\otimes_{\wedge}\mathcal{B+}(-1)^{p}\mathcal{A}\otimes_{\wedge}\mathbf{D}%
\mathcal{B} \label{W.16bisss}%
\end{equation}

\end{proposition}

\begin{proof}
It follows directly from the definition if we take into account the properties
of the product \ $\otimes_{\wedge}$ and Eq.(\ref{p.2eureka}).
\end{proof}

\subsection{Extended Covariant Derivative of Clifford Valued Forms}

\begin{definition}
\ The \textit{ extended covariant derivative operator is the mapping}%
\[
\mathbf{D}_{e_{\mathbf{r}}}:\sec%
%TCIMACRO{\dbigwedge \nolimits^{l}}%
%BeginExpansion
{\displaystyle\bigwedge\nolimits^{l}}
%EndExpansion
TM\otimes%
%TCIMACRO{\dbigwedge \nolimits^{p}}%
%BeginExpansion
{\displaystyle\bigwedge\nolimits^{p}}
%EndExpansion
T^{\ast}M\rightarrow\sec%
%TCIMACRO{\dbigwedge \nolimits^{l}}%
%BeginExpansion
{\displaystyle\bigwedge\nolimits^{l}}
%EndExpansion
TM\otimes%
%TCIMACRO{\dbigwedge \nolimits^{p}}%
%BeginExpansion
{\displaystyle\bigwedge\nolimits^{p}}
%EndExpansion
T^{\ast}M,
\]
\textbf{ }such that for any $\mathcal{A}\in\sec%
%TCIMACRO{\dbigwedge \nolimits^{l}}%
%BeginExpansion
{\displaystyle\bigwedge\nolimits^{l}}
%EndExpansion
TM\otimes%
%TCIMACRO{\dbigwedge \nolimits^{p}}%
%BeginExpansion
{\displaystyle\bigwedge\nolimits^{p}}
%EndExpansion
T^{\ast}M\hookrightarrow\sec\mathcal{C\ell}\left(  TM\right)  $ $\otimes%
%TCIMACRO{\dbigwedge \nolimits^{p}}%
%BeginExpansion
{\displaystyle\bigwedge\nolimits^{p}}
%EndExpansion
T^{\ast}M$, $l,$ $p\geq1$, we have
\end{definition}

\begin{equation}
\mathbf{D}\mathcal{A=(}\mathbf{D}_{\mathbf{e}_{\mathbf{r}}}\mathcal{A)\otimes
}_{\wedge}\theta^{r}. \label{EXTCODER}%
\end{equation}

We can immediately verify that
\begin{equation}
\mathbf{D}_{\mathbf{e}_{\mathbf{r}}}\mathcal{A=}\mathbf{\partial
}_{e_{\mathbf{r}}}\mathcal{A}+\frac{p}{2}[%
%TCIMACRO{\TeXButton{omega}{\mbox{\boldmath{$\omega$}}}}%
%BeginExpansion
\mbox{\boldmath{$\omega$}}%
%EndExpansion
_{\mathbf{r}},\mathcal{A}],
\end{equation}
and, of course, in general\footnote{For a Clifford algebra formula for the
calculation of $D_{e_{\mathbf{r}}}\mathcal{A}$, $\mathcal{A\in}\sec%
%TCIMACRO{\dbigwedge \nolimits^{p}}%
%BeginExpansion
{\displaystyle\bigwedge\nolimits^{p}}
%EndExpansion
T^{\ast}M$ see Eq.(\ref{der1}).
\par
{}} \
\begin{equation}
\mathbf{D}_{\mathbf{e}_{\mathbf{r}}}\mathcal{A\neq}D_{e_{\mathbf{r}}%
}\mathcal{A} \label{OK}%
\end{equation}

Let us write explicitly some important cases which will appear latter.

\subsubsection{Case\textbf{\ }$p=1$}

Let $\mathcal{A}\in\sec%
%TCIMACRO{\dbigwedge \nolimits^{l}}%
%BeginExpansion
{\displaystyle\bigwedge\nolimits^{l}}
%EndExpansion
TM\otimes%
%TCIMACRO{\dbigwedge \nolimits^{1}}%
%BeginExpansion
{\displaystyle\bigwedge\nolimits^{1}}
%EndExpansion
T^{\ast}M\hookrightarrow\sec\mathcal{C\ell}\left(  TM\right)  $ $\otimes%
%TCIMACRO{\dbigwedge \nolimits^{1}}%
%BeginExpansion
{\displaystyle\bigwedge\nolimits^{1}}
%EndExpansion
T^{\ast}M$. Then,
\begin{equation}
\mathbf{D}\mathcal{A}=d\mathcal{A}+\frac{1}{2}[%
%TCIMACRO{\TeXButton{omega}{\mbox{\boldmath{$\omega$}}}}%
%BeginExpansion
\mbox{\boldmath{$\omega$}}%
%EndExpansion
,\mathcal{A}], \label{18}%
\end{equation}
and%
\begin{equation}
\mathbf{D}_{e_{\mathbf{k}}}\mathcal{A}=\partial_{e_{\mathbf{r}}}%
\mathcal{A}+\frac{1}{2}[%
%TCIMACRO{\TeXButton{omega}{\mbox{\boldmath{$\omega$}}}}%
%BeginExpansion
\mbox{\boldmath{$\omega$}}%
%EndExpansion
_{\mathbf{k}},\mathcal{A}]. \label{W.18a}%
\end{equation}

\subsubsection{Case $p=2$}

\ Let $\mathcal{F}\in\sec%
%TCIMACRO{\dbigwedge \nolimits^{l}}%
%BeginExpansion
{\displaystyle\bigwedge\nolimits^{l}}
%EndExpansion
TM\otimes%
%TCIMACRO{\dbigwedge \nolimits^{2}}%
%BeginExpansion
{\displaystyle\bigwedge\nolimits^{2}}
%EndExpansion
T^{\ast}M\hookrightarrow\sec\mathcal{C\ell}\left(  TM\right)  $ $\otimes%
%TCIMACRO{\dbigwedge ^{2}}%
%BeginExpansion
{\displaystyle\bigwedge^{2}}
%EndExpansion
T^{\ast}M$. Then, \
\begin{equation}
\mathbf{D}\mathcal{F}=d\mathcal{F}+[%
%TCIMACRO{\TeXButton{omega}{\mbox{\boldmath{$\omega$}}}}%
%BeginExpansion
\mbox{\boldmath{$\omega$}}%
%EndExpansion
,\mathcal{F}], \label{W.19}%
\end{equation}
and%

\begin{equation}
\mathbf{D}_{e_{\mathbf{r}}}\mathcal{F}=\partial_{e_{\mathbf{r}}}\mathcal{F}+[%
%TCIMACRO{\TeXButton{omega}{\mbox{\boldmath{$\omega$}}}}%
%BeginExpansion
\mbox{\boldmath{$\omega$}}%
%EndExpansion
_{\mathbf{r}},\mathcal{F}]. \label{W.19bis}%
\end{equation}

\subsection{Cartan Exterior Differential}

Recall that \cite{frankel} \textit{Cartan} defined the exterior covariant
differential $\ $of $\mathfrak{C}=e_{\mathbf{i}}\otimes\mathfrak{C}%
^{\mathbf{i}}\in$\ sec$%
%TCIMACRO{\dbigwedge \nolimits^{1}}%
%BeginExpansion
{\displaystyle\bigwedge\nolimits^{1}}
%EndExpansion
TM\otimes%
%TCIMACRO{\dbigwedge \nolimits^{p}}%
%BeginExpansion
{\displaystyle\bigwedge\nolimits^{p}}
%EndExpansion
T^{\ast}M$ as a mapping%
\begin{align}
\mathbf{D}^{c}  &  \mathbf{:}\
%TCIMACRO{\dbigwedge \nolimits^{1}}%
%BeginExpansion
{\displaystyle\bigwedge\nolimits^{1}}
%EndExpansion
TM\otimes%
%TCIMACRO{\dbigwedge \nolimits^{p}}%
%BeginExpansion
{\displaystyle\bigwedge\nolimits^{p}}
%EndExpansion
T^{\ast}M\longrightarrow\
%TCIMACRO{\dbigwedge \nolimits^{1}}%
%BeginExpansion
{\displaystyle\bigwedge\nolimits^{1}}
%EndExpansion
TM\otimes%
%TCIMACRO{\dbigwedge \nolimits^{p+1}}%
%BeginExpansion
{\displaystyle\bigwedge\nolimits^{p+1}}
%EndExpansion
T^{\ast}M,\nonumber\\
\mathbf{D}^{c}\mathfrak{C}  &  =\mathbf{D}^{c}\mathbf{(}e_{\mathbf{i}}%
\otimes\mathfrak{C}^{\mathbf{i}})=e_{\mathbf{i}}\otimes d\mathfrak{C}%
^{\mathbf{i}}+\mathbf{D}^{c}e_{\mathbf{i}}\wedge\mathfrak{C}^{\mathbf{i}%
},\label{cartan1}\\
\mathbf{D}^{c}e_{\mathbf{j}}  &  =(D_{e_{\mathbf{k}}}e_{\mathbf{j}}%
)\theta^{\mathbf{k}}\nonumber
\end{align}
which in view of Eq.(\ref{9.8}) and Eq.(\ref{9.9}) can be written as
\begin{equation}
\mathbf{D}^{c}\mathfrak{C}=\mathbf{D}^{c}\mathbf{(}e_{\mathbf{i}}%
\otimes\mathfrak{C}^{\mathbf{i}})=d\mathfrak{C}+\frac{1}{2}[%
%TCIMACRO{\TeXButton{omega}{\mbox{\boldmath{$\omega$}}}}%
%BeginExpansion
\mbox{\boldmath{$\omega$}}%
%EndExpansion
,\mathfrak{C}]. \label{cartan2}%
\end{equation}

So, we have, for $p>1$, the following relation between the exterior covariant
differential $\mathbf{D}$ and Cartan's exterior differential ($p>1$)
\begin{equation}
\mathbf{D}\mathfrak{C=}\mathbf{D}^{c}\mathfrak{C}+\frac{p-1}{2}[%
%TCIMACRO{\TeXButton{omega}{\mbox{\boldmath{$\omega$}}}}%
%BeginExpansion
\mbox{\boldmath{$\omega$}}%
%EndExpansion
,\mathfrak{C}]. \label{cartan3}%
\end{equation}

Note moreover that when $\mathfrak{C}^{(1)}=e_{\mathbf{i}}\otimes
\mathfrak{C}^{\mathbf{i}}\in$\ sec$%
%TCIMACRO{\dbigwedge \nolimits^{1}}%
%BeginExpansion
{\displaystyle\bigwedge\nolimits^{1}}
%EndExpansion
TM\otimes%
%TCIMACRO{\dbigwedge \nolimits^{1}}%
%BeginExpansion
{\displaystyle\bigwedge\nolimits^{1}}
%EndExpansion
T^{\ast}M$, we have
\begin{equation}
\mathbf{D}\mathfrak{C}^{(1)}\mathfrak{=}\mathbf{D}^{c}\mathfrak{C}^{(1)}.
\label{cartan4}%
\end{equation}

We end this section with two observations:

(i) There are other approaches to the concept of exterior covariant
differential acting on sections of a vector bundle $E\otimes%
%TCIMACRO{\dbigwedge \nolimits^{p}}%
%BeginExpansion
{\displaystyle\bigwedge\nolimits^{p}}
%EndExpansion
T^{\ast}M$ and also in sections of $\mathrm{end}(E)$ $\otimes%
%TCIMACRO{\dbigwedge \nolimits^{p}}%
%BeginExpansion
{\displaystyle\bigwedge\nolimits^{p}}
%EndExpansion
T^{\ast}M$, as e.g., in
\cite{baezm,beentucker,frankel,goeshu,nasen,palais,sternberg}. Not all are
completely equivalent among themselves and to the one presented above. Our
definitions, we think, have the merit of mimicking coherently the pullback
under a local section of the covariant differential acting on sections of
vector bundles associated to a given principal bundle as used in gauge
theories. Indeed, this consistence will be checked in several situations below.

(ii) Some authors, e.g., \cite{beentucker,thirring2} find convenient to
introduce the concept of \textit{exterior covariant derivative} of \ indexed
$p$-forms, which are objects like the curvature $2$-forms (see below) or the
connection 1-forms introduced above. We do not use such concept in this paper.

\subsection{Torsion and Curvature}

Let $%
%TCIMACRO{\TeXButton{theta}{\mbox{\boldmath{$\theta$}}}}%
%BeginExpansion
\mbox{\boldmath{$\theta$}}%
%EndExpansion
=e_{\mu}dx^{\mu}=\mathbf{e}_{\mathbf{a}}\theta^{\mathbf{a}}\in\sec%
%TCIMACRO{\dbigwedge \nolimits^{1}}%
%BeginExpansion
{\displaystyle\bigwedge\nolimits^{1}}
%EndExpansion
TM\otimes%
%TCIMACRO{\dbigwedge \nolimits^{1}}%
%BeginExpansion
{\displaystyle\bigwedge\nolimits^{1}}
%EndExpansion
T^{\ast}M\hookrightarrow\mathcal{C\ell}(TM)\otimes%
%TCIMACRO{\dbigwedge \nolimits^{1}}%
%BeginExpansion
{\displaystyle\bigwedge\nolimits^{1}}
%EndExpansion
T^{\ast}M$ \ and \ $%
%TCIMACRO{\TeXButton{omega}{\mbox{\boldmath{$\omega$}}}}%
%BeginExpansion
\mbox{\boldmath{$\omega$}}%
%EndExpansion
=\frac{1}{2}\left(  \omega_{\mathbf{a}}^{\mathbf{bc}}\mathbf{e}_{\mathbf{b}%
}\wedge\mathbf{e}_{\mathbf{c}}\right)  \otimes\theta^{\mathbf{a}}\equiv
\frac{1}{2}\omega_{\mathbf{a}}^{\mathbf{bc}}\mathbf{e}_{\mathbf{b}}%
\mathbf{e}_{\mathbf{c}}\theta^{\mathbf{a}}\in\sec%
%TCIMACRO{\dbigwedge \nolimits^{2}}%
%BeginExpansion
{\displaystyle\bigwedge\nolimits^{2}}
%EndExpansion
M\otimes%
%TCIMACRO{\dbigwedge \nolimits^{1}}%
%BeginExpansion
{\displaystyle\bigwedge\nolimits^{1}}
%EndExpansion
T^{\ast}M\hookrightarrow\mathcal{C\ell}(TM)\otimes%
%TCIMACRO{\dbigwedge \nolimits^{1}}%
%BeginExpansion
{\displaystyle\bigwedge\nolimits^{1}}
%EndExpansion
T^{\ast}M$ \ be respectively the \textit{representatives} of a soldering form
and a connection on the \textit{basis manifold. }Then, following the standard
procedure \cite{konu}, the \textit{torsion} of the connection and the
\textit{curvature} of the connection on the basis manifold are defined by
\begin{equation}%
%TCIMACRO{\TeXButton{Theta}{\mbox{\boldmath{$\Theta$}}}}%
%BeginExpansion
\mbox{\boldmath{$\Theta$}}%
%EndExpansion
=\mathbf{D}%
%TCIMACRO{\TeXButton{theta}{\mbox{\boldmath{$\theta$}}}}%
%BeginExpansion
\mbox{\boldmath{$\theta$}}%
%EndExpansion
\in\sec%
%TCIMACRO{\dbigwedge \nolimits^{1}}%
%BeginExpansion
{\displaystyle\bigwedge\nolimits^{1}}
%EndExpansion
TM\otimes%
%TCIMACRO{\dbigwedge \nolimits^{2}}%
%BeginExpansion
{\displaystyle\bigwedge\nolimits^{2}}
%EndExpansion
T^{\ast}M\hookrightarrow\mathcal{C\ell}(TM)\otimes%
%TCIMACRO{\dbigwedge \nolimits^{2}}%
%BeginExpansion
{\displaystyle\bigwedge\nolimits^{2}}
%EndExpansion
T^{\ast}M, \label{9.15}%
\end{equation}
and
\begin{equation}
\mathcal{R=}\mathbf{D}%
%TCIMACRO{\TeXButton{omega}{\mbox{\boldmath{$\omega$}}}}%
%BeginExpansion
\mbox{\boldmath{$\omega$}}%
%EndExpansion
\in\sec%
%TCIMACRO{\dbigwedge \nolimits^{2}}%
%BeginExpansion
{\displaystyle\bigwedge\nolimits^{2}}
%EndExpansion
M\otimes%
%TCIMACRO{\dbigwedge \nolimits^{2}}%
%BeginExpansion
{\displaystyle\bigwedge\nolimits^{2}}
%EndExpansion
T^{\ast}M\hookrightarrow\mathcal{C\ell}(TM)\otimes%
%TCIMACRO{\dbigwedge \nolimits^{2}}%
%BeginExpansion
{\displaystyle\bigwedge\nolimits^{2}}
%EndExpansion
T^{\ast}M. \label{9.16}%
\end{equation}
We now calculate $%
%TCIMACRO{\TeXButton{Theta}{\mbox{\boldmath{$\Theta$}}}}%
%BeginExpansion
\mbox{\boldmath{$\Theta$}}%
%EndExpansion
$ and $\mathbf{D}\mathcal{R}$. We have,%

\begin{equation}
\mathbf{D}%
%TCIMACRO{\TeXButton{theta}{\mbox{\boldmath{$\theta$}}}}%
%BeginExpansion
\mbox{\boldmath{$\theta$}}%
%EndExpansion
=\mathbf{D(}\mathbf{e}_{\mathbf{a}}\theta^{\mathbf{a}})=\mathbf{e}%
_{\mathbf{a}}d\theta^{\mathbf{a}}+\frac{1}{2}[%
%TCIMACRO{\TeXButton{omega}{\mbox{\boldmath{$\omega$}}}}%
%BeginExpansion
\mbox{\boldmath{$\omega$}}%
%EndExpansion
_{\mathbf{a}},\mathbf{e}_{\mathbf{d}}]\theta^{\mathbf{a}}\wedge\theta
^{\mathbf{d}} \label{9.17}%
\end{equation}
and since $\frac{1}{2}[%
%TCIMACRO{\TeXButton{omega}{\mbox{\boldmath{$\omega$}}}}%
%BeginExpansion
\mbox{\boldmath{$\omega$}}%
%EndExpansion
_{\mathbf{a}},\mathbf{e}_{\mathbf{d}}]=-\mathbf{e}_{\mathbf{d}}\lrcorner%
%TCIMACRO{\TeXButton{omega}{\mbox{\boldmath{$\omega$}}}}%
%BeginExpansion
\mbox{\boldmath{$\omega$}}%
%EndExpansion
_{\mathbf{a}}=\omega_{\mathbf{ad}}^{\mathbf{c}}\mathbf{e}_{\mathbf{c}}$ we have%

\begin{equation}
\mathbf{D(}\mathbf{e}_{\mathbf{a}}\theta^{\mathbf{a}})=\mathbf{e}_{\mathbf{a}%
}[d\theta^{\mathbf{a}}+\omega_{\mathbf{bd}}^{\mathbf{a}}\theta^{\mathbf{b}%
}\wedge\theta^{\mathbf{d}}]=\mathbf{e}_{\mathbf{a}}%
%TCIMACRO{\TeXButton{Theta}{\mbox{\boldmath{$\Theta$}}}}%
%BeginExpansion
\mbox{\boldmath{$\Theta$}}%
%EndExpansion
^{\mathbf{a}}, \label{9.18}%
\end{equation}
and we recognize%
\begin{equation}%
%TCIMACRO{\TeXButton{Theta}{\mbox{\boldmath{$\Theta$}}}}%
%BeginExpansion
\mbox{\boldmath{$\Theta$}}%
%EndExpansion
^{\mathbf{a}}=d\theta^{\mathbf{a}}+\omega_{\mathbf{bd}}^{\mathbf{a}}%
\theta^{\mathbf{b}}\wedge\theta^{\mathbf{d}}, \label{9.19}%
\end{equation}
as \textit{Cartan's first structure equation}.

For a torsion free connection, the torsion 2-forms $%
%TCIMACRO{\TeXButton{Theta}{\mbox{\boldmath{$\Theta$}}}}%
%BeginExpansion
\mbox{\boldmath{$\Theta$}}%
%EndExpansion
^{\mathbf{a}}=0$, and it follows that $%
%TCIMACRO{\TeXButton{Theta}{\mbox{\boldmath{$\Theta$}}}}%
%BeginExpansion
\mbox{\boldmath{$\Theta$}}%
%EndExpansion
=0$. A metrical compatible connection ($\mathbf{D}g=0$) satisfying $%
%TCIMACRO{\TeXButton{Theta}{\mbox{\boldmath{$\Theta$}}}}%
%BeginExpansion
\mbox{\boldmath{$\Theta$}}%
%EndExpansion
^{\mathbf{a}}=0$ is called a Levi-Civita connection. In the remaining of this
paper we \textit{restrict} ourself to that case.

Now, according to Eq.(\ref{W.21BIS}) we have,
\begin{equation}
\mathbf{D}\mathcal{R}=d\mathcal{R}+[%
%TCIMACRO{\TeXButton{omega}{\mbox{\boldmath{$\omega$}}}}%
%BeginExpansion
\mbox{\boldmath{$\omega$}}%
%EndExpansion
,\mathcal{R}]. \label{9.20}%
\end{equation}

Now, taking into account that
\begin{equation}
\mathcal{R}=d%
%TCIMACRO{\TeXButton{omega}{\mbox{\boldmath{$\omega$}}}}%
%BeginExpansion
\mbox{\boldmath{$\omega$}}%
%EndExpansion
+\frac{1}{2}[%
%TCIMACRO{\TeXButton{omega}{\mbox{\boldmath{$\omega$}}}}%
%BeginExpansion
\mbox{\boldmath{$\omega$}}%
%EndExpansion
,%
%TCIMACRO{\TeXButton{omega}{\mbox{\boldmath{$\omega$}}}}%
%BeginExpansion
\mbox{\boldmath{$\omega$}}%
%EndExpansion
], \label{9.21}%
\end{equation}
and that from Eqs.(\ref{p1}).(\ref{p2}) and (\ref{p3}) it follows that%

\begin{align}
d[%
%TCIMACRO{\TeXButton{omega}{\mbox{\boldmath{$\omega$}}}}%
%BeginExpansion
\mbox{\boldmath{$\omega$}}%
%EndExpansion
,%
%TCIMACRO{\TeXButton{omega}{\mbox{\boldmath{$\omega$}}}}%
%BeginExpansion
\mbox{\boldmath{$\omega$}}%
%EndExpansion
]  &  =[d%
%TCIMACRO{\TeXButton{omega}{\mbox{\boldmath{$\omega$}}}}%
%BeginExpansion
\mbox{\boldmath{$\omega$}}%
%EndExpansion
,%
%TCIMACRO{\TeXButton{omega}{\mbox{\boldmath{$\omega$}}}}%
%BeginExpansion
\mbox{\boldmath{$\omega$}}%
%EndExpansion
]-[%
%TCIMACRO{\TeXButton{omega}{\mbox{\boldmath{$\omega$}}}}%
%BeginExpansion
\mbox{\boldmath{$\omega$}}%
%EndExpansion
,d%
%TCIMACRO{\TeXButton{omega}{\mbox{\boldmath{$\omega$}}}}%
%BeginExpansion
\mbox{\boldmath{$\omega$}}%
%EndExpansion
],\nonumber\\
\lbrack d%
%TCIMACRO{\TeXButton{omega}{\mbox{\boldmath{$\omega$}}}}%
%BeginExpansion
\mbox{\boldmath{$\omega$}}%
%EndExpansion
,%
%TCIMACRO{\TeXButton{omega}{\mbox{\boldmath{$\omega$}}}}%
%BeginExpansion
\mbox{\boldmath{$\omega$}}%
%EndExpansion
]  &  =-[%
%TCIMACRO{\TeXButton{omega}{\mbox{\boldmath{$\omega$}}}}%
%BeginExpansion
\mbox{\boldmath{$\omega$}}%
%EndExpansion
,d%
%TCIMACRO{\TeXButton{omega}{\mbox{\boldmath{$\omega$}}}}%
%BeginExpansion
\mbox{\boldmath{$\omega$}}%
%EndExpansion
],\nonumber\\
\lbrack\lbrack%
%TCIMACRO{\TeXButton{omega}{\mbox{\boldmath{$\omega$}}}}%
%BeginExpansion
\mbox{\boldmath{$\omega$}}%
%EndExpansion
,%
%TCIMACRO{\TeXButton{omega}{\mbox{\boldmath{$\omega$}}}}%
%BeginExpansion
\mbox{\boldmath{$\omega$}}%
%EndExpansion
],%
%TCIMACRO{\TeXButton{omega}{\mbox{\boldmath{$\omega$}}}}%
%BeginExpansion
\mbox{\boldmath{$\omega$}}%
%EndExpansion
]  &  =0, \label{9.22}%
\end{align}
we have immediately%
\begin{equation}
\mathbf{D}\mathcal{R}=d\mathcal{R}+[%
%TCIMACRO{\TeXButton{omega}{\mbox{\boldmath{$\omega$}}}}%
%BeginExpansion
\mbox{\boldmath{$\omega$}}%
%EndExpansion
,\mathcal{R}]=0. \label{9.23}%
\end{equation}

Eq.(\ref{9.23}) is known as the \textit{Bianchi identity}.

Note that
\begin{align}
\mathcal{R}  &  =\frac{1}{4}R_{\mu\nu}^{\mathbf{ab}}\mathbf{e}_{\mathbf{a}%
}\wedge\mathbf{e}_{\mathbf{b}}\otimes(dx^{\mu}\wedge dx^{\nu})\nonumber\\
&  \equiv\frac{1}{4}\mathcal{R}_{\mathbf{cd}}^{\mathbf{ab}}\mathbf{e}%
_{\mathbf{a}}\mathbf{e}_{\mathbf{b}}\otimes\theta^{\mathbf{c}}\wedge
\theta^{\mathbf{d}}=\frac{1}{4}R_{\rho\sigma}^{\alpha\beta}e_{\alpha}e_{\beta
}\otimes dx^{\rho}\wedge dx^{\sigma}\nonumber\\
&  =\frac{1}{4}R_{\mathcal{\mu\nu\rho\sigma}}e^{\mu}e^{\nu}\otimes dx^{\rho
}\wedge dx^{\sigma}, \label{9.24}%
\end{align}
where $R_{\mathcal{\mu\nu\rho\sigma}}$ are the components of the curvature
tensor, also known in differential geometry as the Riemann tensor. We recall
the well known symmetries%
\begin{align}
R_{\mathcal{\mu\nu\rho\sigma}}  &  =-R_{\mathcal{\nu\mu\rho\sigma}%
},\nonumber\\
R_{\mathcal{\mu\nu\rho\sigma}}  &  =-R_{\mathcal{\mu\nu\sigma\rho}%
},\nonumber\\
R_{\mathcal{\mu\nu\rho\sigma}}  &  =R_{\mathcal{\rho\sigma\mu\nu}}.
\label{9.42}%
\end{align}

\bigskip We also write Eq.(\ref{9.24}) as
\begin{align}
\mathcal{R}  &  =\frac{1}{4}R_{\mathbf{cd}}^{\mathbf{ab}}\mathbf{e}%
_{\mathbf{a}}\mathbf{e}_{\mathbf{b}}\otimes(\theta^{\mathbf{c}}\wedge
\theta^{\mathbf{d}})=\frac{1}{2}\mathbf{R}_{\mu\nu}dx^{\mu}\wedge
dx^{\mathbf{\nu}}\nonumber\\
&  =\frac{1}{2}\mathcal{R}_{\mathbf{b}\ }^{\mathbf{a}}\mathbf{e}_{\mathbf{a}%
}\mathbf{e}^{\mathbf{b}}, \label{9.43}%
\end{align}
with
\begin{align}
\mathbf{R}_{\mu\nu}  &  =\frac{1}{2}R_{\mu\nu}^{\mathbf{ab}}\mathbf{e}%
_{\mathbf{a}}\mathbf{e}_{\mathbf{b}}=\frac{1}{2}R_{\mu\nu}^{\mathbf{ab}%
}\mathbf{e}_{\mathbf{a}}\wedge\mathbf{e}_{\mathbf{b}}\in\sec%
%TCIMACRO{\dbigwedge \nolimits^{2}}%
%BeginExpansion
{\displaystyle\bigwedge\nolimits^{2}}
%EndExpansion
TM\hookrightarrow\mathcal{C\ell}(TM),\nonumber\\
\mathcal{R}^{\mathbf{ab}}  &  =\frac{1}{2}R_{\mu\nu}^{\mathbf{ab}}dx^{\mu
}\wedge dx^{\mathbf{\nu}}\in\sec%
%TCIMACRO{\dbigwedge \nolimits^{2}}%
%BeginExpansion
{\displaystyle\bigwedge\nolimits^{2}}
%EndExpansion
T^{\ast}M, \label{9.44}%
\end{align}
where $\mathbf{R}_{\mu\nu}$ will be called curvature bivectors and the
$\mathcal{R}_{\mathbf{b}}^{\mathbf{a}}$ are called after Cartan the curvature
$2$-forms. The $\mathcal{R}_{\mathbf{b}}^{\mathbf{a}}$ satisfy
\textit{Cartan's second structure equation}%
\begin{equation}
\mathcal{R}_{\mathbf{b}}^{\mathbf{a}}=d%
%TCIMACRO{\TeXButton{omega}{\mbox{\boldmath{$\omega$}}}}%
%BeginExpansion
\mbox{\boldmath{$\omega$}}%
%EndExpansion
_{\mathbf{b}}^{\mathbf{a}}+%
%TCIMACRO{\TeXButton{omega}{\mbox{\boldmath{$\omega$}}}}%
%BeginExpansion
\mbox{\boldmath{$\omega$}}%
%EndExpansion
_{\mathbf{c}}^{\mathbf{a}}\wedge%
%TCIMACRO{\TeXButton{omega}{\mbox{\boldmath{$\omega$}}}}%
%BeginExpansion
\mbox{\boldmath{$\omega$}}%
%EndExpansion
_{\mathbf{d}}^{\mathbf{c}}, \label{9.45}%
\end{equation}
which follows calculating $d\mathcal{R}$ from Eq.(\ref{9.21}). \ Now, we can
also write,
\begin{align}
\mathbf{D}\mathcal{R}  &  =d\mathcal{R}+[%
%TCIMACRO{\TeXButton{omega}{\mbox{\boldmath{$\omega$}}}}%
%BeginExpansion
\mbox{\boldmath{$\omega$}}%
%EndExpansion
,\mathcal{R}]\nonumber\\
&  =\frac{1}{2}\{d(\frac{1}{2}R_{\mu\nu}^{\mathbf{ab}}\mathbf{e}_{\mathbf{a}%
}\mathbf{e}_{\mathbf{b}}dx^{\mu}\wedge dx^{\nu})+\frac{1}{2}[%
%TCIMACRO{\TeXButton{omega}{\mbox{\boldmath{$\omega$}}}}%
%BeginExpansion
\mbox{\boldmath{$\omega$}}%
%EndExpansion
_{\rho},\mathbf{R}_{\mu\nu}]\}dx^{\rho}\wedge dx^{\mu}\wedge dx^{\nu
}\nonumber\\
&  =\frac{1}{2}\{\partial_{\rho}\mathbf{R}_{\mu\nu}+[%
%TCIMACRO{\TeXButton{omega}{\mbox{\boldmath{$\omega$}}}}%
%BeginExpansion
\mbox{\boldmath{$\omega$}}%
%EndExpansion
_{\rho},\mathbf{R}_{\mu\nu}]\}dx^{\rho}\wedge dx^{\mu}\wedge dx^{\nu
}\label{9.25}\\
&  =\frac{1}{2}\mathbf{D}_{e_{\rho}}\mathbf{R}_{\mu\nu}dx^{\rho}\wedge
dx^{\mu}\wedge dx^{\nu}\nonumber\\
&  =\frac{1}{3!}\left(  \mathbf{D}_{e_{\rho}}\mathbf{R}_{\mu\nu}%
+\mathbf{D}_{e_{\mu}}\mathbf{R}_{\nu\rho}+\mathbf{D}_{e_{\nu}}\mathbf{R}%
_{\rho\mu}\right)  dx^{\rho}\wedge dx^{\mu}\wedge dx^{\nu}=0,\nonumber
\end{align}
\textit{ }\ from where it follows that
\begin{equation}
\mathbf{D}_{e_{\rho}}\mathbf{R}_{\mu\nu}+\mathbf{D}_{e_{\mu}}\mathbf{R}%
_{\nu\rho}+\mathbf{D}_{e_{\nu}}\mathbf{R}_{\rho\mu}=0. \label{9.28}%
\end{equation}

\begin{remark}
Eq.(\ref{9.28}) \ is called in Physics textbooks on gauge theories (see, e.g.,
\cite{nasen,ryder}) Bianchi identity. Note that physicists call the extended
covariant derivative operator
\begin{equation}
\mathbf{D}_{e_{\rho}}\equiv\mathbf{D}_{\rho}=\partial_{\rho}+[%
%TCIMACRO{\TeXButton{omega}{\mbox{\boldmath{$\omega$}}}}%
%BeginExpansion
\mbox{\boldmath{$\omega$}}%
%EndExpansion
_{\rho},], \label{9.26}%
\end{equation}
acting on the curvature bivectors as the \ `\textit{covariant derivative'.
Note however that, as detailed above, this operator is not the usual covariant
derivative operator }$D_{\mathbf{e}_{a}}$ acting on sections of the tensor bundle.
\end{remark}

We now find the explicit expression for the curvature bivectors $\mathbf{R}%
_{\mu\nu}$ in terms \ of the connections bivectors $%
%TCIMACRO{\TeXButton{omega}{\mbox{\boldmath{$\omega$}}}}%
%BeginExpansion
\mbox{\boldmath{$\omega$}}%
%EndExpansion
_{\mu}=%
%TCIMACRO{\TeXButton{omega}{\mbox{\boldmath{$\omega$}}}}%
%BeginExpansion
\mbox{\boldmath{$\omega$}}%
%EndExpansion
(e_{\mu})$,\ which will be used latter. First recall that by definition%
\begin{equation}
\mathbf{R}_{\mu\nu}=\mathcal{R}(e_{\mu},e_{\nu})=-\mathcal{R}(e_{\nu},e_{\mu
})=-\mathbf{R}_{\mu\nu}. \label{9.29}%
\end{equation}

Now, observe that using Eqs.(\ref{p1}), (\ref{p2}) and (\ref{p3}) we can
easily show that%
\begin{align}
\lbrack%
%TCIMACRO{\TeXButton{omega}{\mbox{\boldmath{$\omega$}}}}%
%BeginExpansion
\mbox{\boldmath{$\omega$}}%
%EndExpansion
,%
%TCIMACRO{\TeXButton{omega}{\mbox{\boldmath{$\omega$}}}}%
%BeginExpansion
\mbox{\boldmath{$\omega$}}%
%EndExpansion
](e_{\mu},e_{\nu})  &  =2[%
%TCIMACRO{\TeXButton{omega}{\mbox{\boldmath{$\omega$}}}}%
%BeginExpansion
\mbox{\boldmath{$\omega$}}%
%EndExpansion
(e_{\mu}),%
%TCIMACRO{\TeXButton{omega}{\mbox{\boldmath{$\omega$}}}}%
%BeginExpansion
\mbox{\boldmath{$\omega$}}%
%EndExpansion
(e_{\nu})]\nonumber\\
&  =2[%
%TCIMACRO{\TeXButton{omega}{\mbox{\boldmath{$\omega$}}}}%
%BeginExpansion
\mbox{\boldmath{$\omega$}}%
%EndExpansion
_{\mu},%
%TCIMACRO{\TeXButton{omega}{\mbox{\boldmath{$\omega$}}}}%
%BeginExpansion
\mbox{\boldmath{$\omega$}}%
%EndExpansion
_{\nu}]. \label{9.30}%
\end{align}

Using Eqs. (\ref{9.21}), (\ref{9.29}) and (\ref{9.30}) we get
\begin{equation}
\mathbf{R}_{\mu\nu}=\partial_{\mu}%
%TCIMACRO{\TeXButton{omega}{\mbox{\boldmath{$\omega$}}}}%
%BeginExpansion
\mbox{\boldmath{$\omega$}}%
%EndExpansion
_{v}-\partial_{v}%
%TCIMACRO{\TeXButton{omega}{\mbox{\boldmath{$\omega$}}}}%
%BeginExpansion
\mbox{\boldmath{$\omega$}}%
%EndExpansion
_{\mu}+[%
%TCIMACRO{\TeXButton{omega}{\mbox{\boldmath{$\omega$}}}}%
%BeginExpansion
\mbox{\boldmath{$\omega$}}%
%EndExpansion
_{\mu},%
%TCIMACRO{\TeXButton{omega}{\mbox{\boldmath{$\omega$}}}}%
%BeginExpansion
\mbox{\boldmath{$\omega$}}%
%EndExpansion
_{\nu}]. \label{9.40}%
\end{equation}

\subsection{Some Useful Formulas}

\begin{proposition}
Let $A\in\sec%
%TCIMACRO{\dbigwedge \nolimits^{p}}%
%BeginExpansion
{\displaystyle\bigwedge\nolimits^{p}}
%EndExpansion
TM\hookrightarrow\sec\mathcal{C\ell}(TM)$ and $\mathcal{R}$ the curvature of
the connection \ as defined in Eq.(\ref{9.16}). Then,
\begin{equation}
\mathbf{D}^{2}A=\frac{1}{2}[\mathcal{R},A]. \label{9.T1}%
\end{equation}

\end{proposition}

\begin{proof}
The first member is%

\begin{align}
\mathbf{D}^{2}A  &  =\mathbf{DD}A=\mathbf{D(}dA+\frac{1}{2}[%
%TCIMACRO{\TeXButton{omega}{\mbox{\boldmath{$\omega$}}}}%
%BeginExpansion
\mbox{\boldmath{$\omega$}}%
%EndExpansion
,A])\nonumber\\
&  =d^{2}A+\frac{1}{2}[%
%TCIMACRO{\TeXButton{omega}{\mbox{\boldmath{$\omega$}}}}%
%BeginExpansion
\mbox{\boldmath{$\omega$}}%
%EndExpansion
,dA]+\frac{1}{2}d[%
%TCIMACRO{\TeXButton{omega}{\mbox{\boldmath{$\omega$}}}}%
%BeginExpansion
\mbox{\boldmath{$\omega$}}%
%EndExpansion
,A]+\frac{1}{4}[%
%TCIMACRO{\TeXButton{omega}{\mbox{\boldmath{$\omega$}}}}%
%BeginExpansion
\mbox{\boldmath{$\omega$}}%
%EndExpansion
,[%
%TCIMACRO{\TeXButton{omega}{\mbox{\boldmath{$\omega$}}}}%
%BeginExpansion
\mbox{\boldmath{$\omega$}}%
%EndExpansion
,A]]. \label{9.T2}%
\end{align}

Now, as can be easily verified,%
\begin{equation}
d[%
%TCIMACRO{\TeXButton{omega}{\mbox{\boldmath{$\omega$}}}}%
%BeginExpansion
\mbox{\boldmath{$\omega$}}%
%EndExpansion
,A]=[d%
%TCIMACRO{\TeXButton{omega}{\mbox{\boldmath{$\omega$}}}}%
%BeginExpansion
\mbox{\boldmath{$\omega$}}%
%EndExpansion
,A]-[%
%TCIMACRO{\TeXButton{omega}{\mbox{\boldmath{$\omega$}}}}%
%BeginExpansion
\mbox{\boldmath{$\omega$}}%
%EndExpansion
,dA], \label{9.T13}%
\end{equation}%
\begin{equation}
\lbrack%
%TCIMACRO{\TeXButton{omega}{\mbox{\boldmath{$\omega$}}}}%
%BeginExpansion
\mbox{\boldmath{$\omega$}}%
%EndExpansion
,[%
%TCIMACRO{\TeXButton{omega}{\mbox{\boldmath{$\omega$}}}}%
%BeginExpansion
\mbox{\boldmath{$\omega$}}%
%EndExpansion
,A]]=[[%
%TCIMACRO{\TeXButton{omega}{\mbox{\boldmath{$\omega$}}}}%
%BeginExpansion
\mbox{\boldmath{$\omega$}}%
%EndExpansion
,%
%TCIMACRO{\TeXButton{omega}{\mbox{\boldmath{$\omega$}}}}%
%BeginExpansion
\mbox{\boldmath{$\omega$}}%
%EndExpansion
],A], \label{9.T4}%
\end{equation}
\
\begin{equation}
\frac{1}{4}[%
%TCIMACRO{\TeXButton{omega}{\mbox{\boldmath{$\omega$}}}}%
%BeginExpansion
\mbox{\boldmath{$\omega$}}%
%EndExpansion
,[%
%TCIMACRO{\TeXButton{omega}{\mbox{\boldmath{$\omega$}}}}%
%BeginExpansion
\mbox{\boldmath{$\omega$}}%
%EndExpansion
,A]]=\frac{1}{2}[%
%TCIMACRO{\TeXButton{omega}{\mbox{\boldmath{$\omega$}}}}%
%BeginExpansion
\mbox{\boldmath{$\omega$}}%
%EndExpansion
\otimes_{\wedge}%
%TCIMACRO{\TeXButton{omega}{\mbox{\boldmath{$\omega$}}}}%
%BeginExpansion
\mbox{\boldmath{$\omega$}}%
%EndExpansion
,A]]. \label{9.T5}%
\end{equation}

Using \ these equations in Eq.(\ref{9.T2}) we have,%

\[
\mathbf{D}^{2}A=\frac{1}{2}[d%
%TCIMACRO{\TeXButton{omega}{\mbox{\boldmath{$\omega$}}}}%
%BeginExpansion
\mbox{\boldmath{$\omega$}}%
%EndExpansion
+%
%TCIMACRO{\TeXButton{omega}{\mbox{\boldmath{$\omega$}}}}%
%BeginExpansion
\mbox{\boldmath{$\omega$}}%
%EndExpansion
\otimes_{\wedge}%
%TCIMACRO{\TeXButton{omega}{\mbox{\boldmath{$\omega$}}}}%
%BeginExpansion
\mbox{\boldmath{$\omega$}}%
%EndExpansion
,A]=\frac{1}{2}[\mathcal{R},A].
\]

\end{proof}

In particular, when $a\in\sec%
%TCIMACRO{\dbigwedge \nolimits^{1}}%
%BeginExpansion
{\displaystyle\bigwedge\nolimits^{1}}
%EndExpansion
TM\hookrightarrow\sec\mathcal{C\ell}(TM)$ we have
\begin{equation}
\mathbf{D}^{2}a=\mathcal{R}\llcorner a \label{9.TC}%
\end{equation}

Also, we can show using the previous result that if \ $\mathcal{A}\in
\sec\mathcal{C\ell}(TM)\otimes%
%TCIMACRO{\dbigwedge \nolimits^{1}}%
%BeginExpansion
{\displaystyle\bigwedge\nolimits^{1}}
%EndExpansion
T^{\ast}M$ it holds%
\begin{equation}
\mathbf{D}^{2}\mathcal{A}=\frac{1}{2}[\mathcal{R},\mathcal{A}]. \label{9.TC1}%
\end{equation}

It is a useful test of the consistence of our formalism to derive once again
that $\mathbf{D}\mathcal{R}=0$, by calculating $\mathbf{D}^{3}A$ for $A\in\sec%
%TCIMACRO{\dbigwedge \nolimits^{r}}%
%BeginExpansion
{\displaystyle\bigwedge\nolimits^{r}}
%EndExpansion
TM\hookrightarrow\sec\mathcal{C\ell}(TM)$. We have:%
\begin{equation}
\mathbf{D}^{3}A=\mathbf{D}\left(  \mathbf{D}^{2}A\right)  =\mathbf{D}%
^{2}(\mathbf{D}A). \label{9.TC2}%
\end{equation}

Now, using the above formulas and recalling Eq.(\ref{9.T1}), we can write:
\begin{align}
\mathbf{D}^{3}A  &  =\mathbf{D}(\mathbf{D}^{2}A)=\frac{1}{2}\mathbf{D}%
[\mathcal{R},A]\nonumber\\
&  =\frac{1}{2}\mathbf{D}(\mathcal{R}\otimes_{\wedge}A-A\otimes_{\wedge
}\mathcal{R})\nonumber\\
&  =\frac{1}{2}(\mathbf{D}\mathcal{R}\otimes_{\wedge}A+\mathcal{R}%
\otimes_{\wedge}\mathbf{D}A-\mathbf{D}A\otimes_{\wedge}\mathcal{R}%
+(-1)^{1+r}A\otimes_{\wedge}\mathbf{D}\mathcal{R} \label{9.TC3}%
\end{align}
and%
\begin{align}
\mathbf{D}^{3}A  &  =\mathbf{D}^{2}(\mathbf{D}A)=\frac{1}{2}[\mathcal{R}%
,\mathbf{D}A]\nonumber\\
&  =\frac{1}{2}\left(  \mathcal{R}\otimes_{\wedge}\mathbf{D}A-\mathbf{D}%
A\otimes_{\wedge}\mathcal{R}\right)  . \label{9.TC4}%
\end{align}

Comparing Eqs.(\ref{9.TC3}) and (\ref{9.TC4}) we get that%
\begin{equation}
\mathbf{D}\mathcal{R}\otimes_{\wedge}A+(-1)^{1+r}A\otimes_{\wedge}%
\mathbf{D}\mathcal{R}=[\mathbf{D}\mathcal{R},A]=0, \label{9.TC5}%
\end{equation}
from where it follows that $\mathbf{D}\mathcal{R}=0$, as it may be.

\section{ General Relativity as a $Sl\left(  2,\mathbb{C}\right)  $ Gauge
Theory}

\subsection{The Nonhomogeneous Field Equations}

The analogy of the fields $\mathbf{R}_{\mu\nu}=\frac{1}{2}R_{\mu\nu
}^{\mathbf{ab}}\mathbf{e}_{\mathbf{a}}\mathbf{e}_{\mathbf{b}}=\frac{1}%
{2}R_{\mu\nu}^{\mathbf{ab}}\mathbf{e}_{\mathbf{a}}\wedge\mathbf{e}%
_{\mathbf{b}}\in\sec%
%TCIMACRO{\dbigwedge \nolimits^{2}}%
%BeginExpansion
{\displaystyle\bigwedge\nolimits^{2}}
%EndExpansion
TM\hookrightarrow\mathcal{C\ell}(TM)$ with the gauge fields of particle fields
is so appealing that it is irresistible to propose some kind of a
$Sl(2,\mathbb{C)}$ formulation for the gravitational field. And indeed this
has already been done, and the interested reader may consult, e.g.,
\cite{carmeli,mielke}. \ Here, we observe that despite the similarities, the
gauge theories of particle physics are in general formulated in flat Minkowski
spacetime and the theory here must be for a field on a general Lorentzian
spacetime. This introduces additional complications, but it is not our purpose
to discuss that issue with all attention it deserves here. Indeed, for our
purposes in this paper we will need only to recall some facts.

To start, recall that in gauge theories besides the homogenous field equations
given by Bianchi's identities, we also have the nonhomogeneous field equation.
This equation, in analogy to the nonhomogeneous equation for the
electromagnetic field (see Eq.(\ref{1.9}) in Appendix A) is written here as%

\begin{equation}
\mathbf{D\star}\mathcal{R}=d\mathbf{\star}\mathcal{R+}\frac{1}{2}[%
%TCIMACRO{\TeXButton{omega}{\mbox{\boldmath{$\omega$}}}}%
%BeginExpansion
\mbox{\boldmath{$\omega$}}%
%EndExpansion
,\mathbf{\star}\mathcal{R}]=-\star\mathcal{J}\mathbf{,} \label{10.0}%
\end{equation}
where the $\mathcal{J}\in\sec%
%TCIMACRO{\dbigwedge \nolimits^{2}}%
%BeginExpansion
{\displaystyle\bigwedge\nolimits^{2}}
%EndExpansion
TM\otimes%
%TCIMACRO{\dbigwedge \nolimits^{1}}%
%BeginExpansion
{\displaystyle\bigwedge\nolimits^{1}}
%EndExpansion
T^{\ast}M\hookrightarrow\mathcal{C\ell}(TM)\otimes%
%TCIMACRO{\dbigwedge \nolimits^{1}}%
%BeginExpansion
{\displaystyle\bigwedge\nolimits^{1}}
%EndExpansion
T^{\ast}M$ is a \ `current', which, if the theory is to be one equivalent to
General Relativity, must be in some way related with the energy momentum
tensor in Einstein theory. In order to write from this equation an equation
for the curvature bivectors, it is very useful to imagine that \ $%
%TCIMACRO{\dbigwedge }%
%BeginExpansion
{\displaystyle\bigwedge}
%EndExpansion
T^{\ast}M\hookrightarrow\mathcal{C\ell}(T^{\ast}M)$, the Clifford bundle of
differential forms, for in that case the powerful calculus described in the
Appendix A can be used. So, we write:
\begin{align}%
%TCIMACRO{\TeXButton{omega}{\mbox{\boldmath{$\omega$}}}}%
%BeginExpansion
\mbox{\boldmath{$\omega$}}%
%EndExpansion
&  \in\sec%
%TCIMACRO{\dbigwedge \nolimits^{2}}%
%BeginExpansion
{\displaystyle\bigwedge\nolimits^{2}}
%EndExpansion
TM\otimes%
%TCIMACRO{\dbigwedge \nolimits^{1}}%
%BeginExpansion
{\displaystyle\bigwedge\nolimits^{1}}
%EndExpansion
T^{\ast}M\hookrightarrow\mathcal{C\ell}(TM)\otimes%
%TCIMACRO{\dbigwedge \nolimits^{1}}%
%BeginExpansion
{\displaystyle\bigwedge\nolimits^{1}}
%EndExpansion
T^{\ast}M\hookrightarrow\mathcal{C\ell}(TM)\otimes\mathcal{C\ell}(T^{\ast
}M),\nonumber\\
\mathcal{R}  &  \mathcal{=}\mathbf{D}%
%TCIMACRO{\TeXButton{omega}{\mbox{\boldmath{$\omega$}}}}%
%BeginExpansion
\mbox{\boldmath{$\omega$}}%
%EndExpansion
\in\sec%
%TCIMACRO{\dbigwedge \nolimits^{2}}%
%BeginExpansion
{\displaystyle\bigwedge\nolimits^{2}}
%EndExpansion
TM\otimes%
%TCIMACRO{\dbigwedge \nolimits^{2}}%
%BeginExpansion
{\displaystyle\bigwedge\nolimits^{2}}
%EndExpansion
T^{\ast}M\hookrightarrow\mathcal{C\ell}(TM)\otimes%
%TCIMACRO{\dbigwedge \nolimits^{2}}%
%BeginExpansion
{\displaystyle\bigwedge\nolimits^{2}}
%EndExpansion
T^{\ast}M\hookrightarrow\mathcal{C\ell}(TM)\otimes\mathcal{C\ell}(T^{\ast
}M)\nonumber\\
\mathcal{J}  &  \mathbf{=J_{\nu}\otimes}\theta^{\nu}\mathbf{\equiv J}_{\nu
}\theta^{\nu}\in\sec%
%TCIMACRO{\dbigwedge \nolimits^{2}}%
%BeginExpansion
{\displaystyle\bigwedge\nolimits^{2}}
%EndExpansion
TM\otimes%
%TCIMACRO{\dbigwedge \nolimits^{1}}%
%BeginExpansion
{\displaystyle\bigwedge\nolimits^{1}}
%EndExpansion
T^{\ast}M\hookrightarrow\mathcal{C\ell}(TM)\otimes\mathcal{C\ell}(T^{\ast}M).
\label{10.01}%
\end{align}

Now, using Eq.(\ref{a.hodge}) for the Hodge star operator given in the
Appendix A.3 and the relation between the operators $d=%
%TCIMACRO{\TeXButton{dirac}{\mbox{\boldmath{$\partial$}}}}%
%BeginExpansion
\mbox{\boldmath{$\partial$}}%
%EndExpansion
\wedge$ and $\delta=-%
%TCIMACRO{\TeXButton{dirac}{\mbox{\boldmath{$\partial$}}}}%
%BeginExpansion
\mbox{\boldmath{$\partial$}}%
%EndExpansion
\lrcorner$ \ (Appendix A5) we can write%
\begin{equation}
d\star\mathcal{R}=-\theta^{5}(-%
%TCIMACRO{\TeXButton{dirac}{\mbox{\boldmath{$\partial$}}}}%
%BeginExpansion
\mbox{\boldmath{$\partial$}}%
%EndExpansion
\lrcorner\mathcal{R)=-\star(}%
%TCIMACRO{\TeXButton{dirac}{\mbox{\boldmath{$\partial$}}}}%
%BeginExpansion
\mbox{\boldmath{$\partial$}}%
%EndExpansion
\lrcorner\mathcal{R)}=-\star((\partial_{\mu}\mathbf{R}_{\nu}^{\mu})\theta
^{\nu}). \label{10.02}%
\end{equation}

Also,%
\begin{align}
\frac{1}{2}[%
%TCIMACRO{\TeXButton{omega}{\mbox{\boldmath{$\omega$}}}}%
%BeginExpansion
\mbox{\boldmath{$\omega$}}%
%EndExpansion
,\mathbf{\star}\mathcal{R}]  &  =\frac{1}{2}[%
%TCIMACRO{\TeXButton{omega}{\mbox{\boldmath{$\omega$}}}}%
%BeginExpansion
\mbox{\boldmath{$\omega$}}%
%EndExpansion
_{\mu},\mathbf{R}_{\alpha\beta}]\otimes\theta^{\mu}\wedge\mathbf{\star(}%
\theta^{\alpha}\wedge\theta^{\beta}\mathbf{)}\nonumber\\
&  =-\frac{1}{2}[%
%TCIMACRO{\TeXButton{omega}{\mbox{\boldmath{$\omega$}}}}%
%BeginExpansion
\mbox{\boldmath{$\omega$}}%
%EndExpansion
_{\mu},\mathbf{R}_{\alpha\beta}]\otimes\theta^{\mu}\wedge\theta^{5}%
\mathbf{(}\theta^{\alpha}\wedge\theta^{\beta}\mathbf{)}\nonumber\\
&  =-\frac{1}{4}[%
%TCIMACRO{\TeXButton{omega}{\mbox{\boldmath{$\omega$}}}}%
%BeginExpansion
\mbox{\boldmath{$\omega$}}%
%EndExpansion
_{\mu},\mathbf{R}_{\alpha\beta}]\otimes\{\theta^{\mu}\theta^{5}\mathbf{(}%
\theta^{\alpha}\wedge\theta^{\beta}\mathbf{)+}\theta^{5}\mathbf{\mathbf{(}%
}\theta^{\alpha}\mathbf{\wedge}\theta^{\beta}\mathbf{\mathbf{)}}\theta^{\mu
}\mathbf{\}}\nonumber\\
&  =\frac{\mathbf{\theta}^{5}}{4}[%
%TCIMACRO{\TeXButton{omega}{\mbox{\boldmath{$\omega$}}}}%
%BeginExpansion
\mbox{\boldmath{$\omega$}}%
%EndExpansion
_{\mu},\mathbf{R}_{\alpha\beta}]\otimes\{\theta^{\mu}\mathbf{(}\theta^{\alpha
}\wedge\theta^{\beta}\mathbf{)-\mathbf{(}}\theta^{\alpha}\wedge\theta^{\beta
})\theta^{\mu}\mathbf{\}}\nonumber\\
&  =\frac{\mathbf{\theta}^{5}}{2}[%
%TCIMACRO{\TeXButton{omega}{\mbox{\boldmath{$\omega$}}}}%
%BeginExpansion
\mbox{\boldmath{$\omega$}}%
%EndExpansion
_{\mu},\mathbf{R}_{\alpha\beta}]\otimes\{\theta^{\mu}\lrcorner\mathbf{(}%
\theta^{\alpha}\wedge\theta^{\beta}\mathbf{)}\nonumber\\
&  =-\star([%
%TCIMACRO{\TeXButton{omega}{\mbox{\boldmath{$\omega$}}}}%
%BeginExpansion
\mbox{\boldmath{$\omega$}}%
%EndExpansion
_{\mu},\mathbf{R}_{\beta}^{\mu}]\theta^{\beta}. \label{10.03}%
\end{align}

Using Eqs.(\ref{10.0}-\ref{10.03}) we get\footnote{Recall that $\mathbf{J}%
_{\nu}\in\sec\bigwedge\nolimits^{2}TM\hookrightarrow\sec\mathcal{C\ell
(}TM\mathcal{)}$.
\par
{}}
\begin{equation}
\mathcal{\partial}_{\mu}\mathbf{R}_{\nu}^{\mu}+[%
%TCIMACRO{\TeXButton{omega}{\mbox{\boldmath{$\omega$}}}}%
%BeginExpansion
\mbox{\boldmath{$\omega$}}%
%EndExpansion
_{\mu},\mathbf{R}_{\nu}^{\mu}]=\mathbf{D}_{e_{\mu}}\mathbf{R}_{\nu}^{\mu
}=\mathbf{J}_{\nu}. \label{10.1}%
\end{equation}
So, \ the gauge theory of gravitation has as field equations
\ the\ Eq.(\ref{10.1}), the nonhomogeneous field equations, and Eq.
(\ref{9.28}) the homogeneous field equations (which is Bianchi's identity). We
summarize that equations, as
\begin{equation}
\mathbf{D}_{e_{\mu}}\mathbf{R}_{\nu}^{\mu}=\mathbf{J}_{\nu}\text{,\hspace{1cm}
}\mathbf{D}_{e_{\rho}}\mathbf{R}_{\mu\nu}+\mathbf{D}_{e_{\mu}}\mathbf{R}%
_{\nu\rho}+\mathbf{D}_{e_{\nu}}\mathbf{R}_{\rho\mu}=0. \label{10.1bis}%
\end{equation}

Eqs.(\ref{10.1bis}) which looks like Maxwell equations, must, of course, be
compatible with Einstein's equations, which may be eventually used to
determine determines $\mathbf{R}_{\nu}^{\mu},%
%TCIMACRO{\TeXButton{omega}{\mbox{\boldmath{$\omega$}}}}%
%BeginExpansion
\mbox{\boldmath{$\omega$}}%
%EndExpansion
_{\mu}$ and $\mathbf{J}_{\nu}$.

\section{Another Set of Maxwell Like Nonhomogeneous Equations for Einstein
Theory}

We now show, e.g., how a special combination of the $\mathbf{R}_{\mathbf{b}%
}^{\mathbf{a}}$ are directly related with a combination of products of the
energy-momentum $1$-vectors $T_{\mathbf{a}}$\ and the tetrad fields
$\mathbf{e}_{\mathbf{a}}$ (see Eq.(\ref{10.3bis}) below) in Einstein theory.
In order to do that, we recall that Einstein's equations can be written in
components in an orthonormal basis as
\begin{equation}
R_{\mathbf{ab}}-\frac{1}{2}\eta_{\mathbf{ab}}R=T_{\mathbf{ab}}, \label{10.2}%
\end{equation}
where $R_{\mathbf{ab}}=R_{\mathbf{ba}}$ are the components of the Ricci tensor
($R_{\mathbf{ab}}=R_{\mathbf{a}\text{ }\mathbf{bc}}^{\text{ }\mathbf{c}}$),
$T_{\mathbf{ab}}$ are the components of the energy-momentum tensor of matter
fields and $R=\eta_{\mathbf{ab}}R^{\mathbf{ab}}$ is the curvature scalar. We
next introduce\footnote{Ricci $1$-form fields appear naturally when we
formulate Einstein's equations in temrs of tetrad fields. See Appendix B.} the
\textit{Ricci 1-vectors} and the \textit{energy-momentum 1-vectors} by
\begin{align}
R_{\mathbf{a}}  &  =R_{\mathbf{ab}}\mathbf{e}^{\mathbf{b}}\in\sec%
%TCIMACRO{\dbigwedge \nolimits^{1}}%
%BeginExpansion
{\displaystyle\bigwedge\nolimits^{1}}
%EndExpansion
TM\hookrightarrow\mathcal{C\ell}(TM),\label{10.3}\\
\hspace{0.15in}T_{\mathbf{a}}  &  =T_{\mathbf{ab}}\mathbf{e}^{\mathbf{b}}%
\in\sec%
%TCIMACRO{\dbigwedge \nolimits^{1}}%
%BeginExpansion
{\displaystyle\bigwedge\nolimits^{1}}
%EndExpansion
TM\hookrightarrow\mathcal{C\ell}(TM). \label{10.3bis}%
\end{align}

We have that
\begin{equation}
R_{\mathbf{a}}=-\mathbf{e}^{\mathbf{b}}\lrcorner\mathbf{R}_{\mathbf{ab}}.
\label{10.4}%
\end{equation}

Now, multiplying Eq.(\ref{10.2}) on the right by $e^{\mathbf{b}}$ we get
\begin{equation}
R_{\mathbf{a}}-\frac{1}{2}R\mathbf{e}_{\mathbf{a}}=T_{\mathbf{a}}.
\label{10.7}%
\end{equation}

Multiplying Eq.(\ref{10.7}) first on the right by $\mathbf{e}_{\mathbf{b}}$
and then on the left by $\mathbf{e}_{\mathbf{b}}$ and making the difference of
the resulting equations we get%
\begin{equation}
\left(  -\mathbf{e}^{\mathbf{c}}\lrcorner\mathbf{R}_{\mathbf{ac}}\right)
\mathbf{e}_{\mathbf{b}}-\mathbf{e}_{\mathbf{b}}\left(  -\mathbf{e}%
^{\mathbf{c}}\lrcorner\mathbf{R}_{\mathbf{ac}}\right)  -\frac{1}%
{2}R(\mathbf{e}_{\mathbf{a}}\mathbf{e}_{\mathbf{b}}-\mathbf{e}_{\mathbf{b}%
}\mathbf{e}_{\mathbf{a}})=(T_{\mathbf{a}}\mathbf{e}_{\mathbf{b}}%
-\mathbf{e}_{\mathbf{b}}T_{\mathbf{a}}). \label{10.8}%
\end{equation}

Defining%

\begin{align}
\mathcal{F}_{\mathbf{ab}}  &  =\left(  -\mathbf{e}^{\mathbf{c}}\lrcorner
\mathbf{R}_{\mathbf{ac}}\right)  \mathbf{e}_{\mathbf{b}}-\mathbf{e}%
_{\mathbf{b}}\left(  -\mathbf{e}^{\mathbf{c}}\lrcorner\mathbf{R}_{\mathbf{ac}%
}\right)  -\frac{1}{2}R(\mathbf{e}_{\mathbf{a}}\mathbf{e}_{\mathbf{b}%
}-\mathbf{e}_{\mathbf{b}}\mathbf{e}_{\mathbf{a}})\nonumber\\
&  =\frac{1}{2}(R_{\mathbf{ac}}\mathbf{e}^{\mathbf{c}}\mathbf{e}_{\mathbf{b}%
}+\mathbf{e}_{\mathbf{b}}\mathbf{e}^{\mathbf{c}}R_{\mathbf{ac}}-\mathbf{e}%
^{\mathbf{c}}R_{\mathbf{ac}}\mathbf{e}_{\mathbf{b}}-\mathbf{e}_{\mathbf{b}%
}R_{\mathbf{ac}}\mathbf{e}^{\mathbf{c}})-\frac{1}{2}R(\mathbf{e}_{\mathbf{a}%
}\mathbf{e}_{\mathbf{b}}-\mathbf{e}_{\mathbf{b}}\mathbf{e}_{\mathbf{a}})
\label{10.9}%
\end{align}
and
\begin{equation}
\mathcal{J}_{\mathbf{b}}=D_{\mathbf{e}_{\mathbf{a}}}(T^{\mathbf{a}}%
\mathbf{e}_{\mathbf{b}}-\mathbf{e}_{\mathbf{b}}T^{\mathbf{a}}), \label{10.10}%
\end{equation}
we have\footnote{Note that we could also produce another Maxwell like
equation, by using the extended covariant derivative operator in the
definition of the current, i.e., we can put $\mathcal{J}_{\mathbf{b}%
}=D_{\mathbf{e}_{\mathbf{a}}}(T^{\mathbf{a}}\mathbf{e}_{\mathbf{b}}%
-\mathbf{e}_{\mathbf{b}}T^{\mathbf{a}})$, and in that case we obtain
$\mathbf{D}_{\mathbf{e}_{\mathbf{a}}}\mathcal{F}_{\mathbf{b}}^{\mathbf{a}%
}=\mathcal{J}_{\mathbf{b}}$.}
\begin{equation}
D_{\mathbf{e}_{\mathbf{a}}}\mathcal{F}_{\mathbf{b}}^{\mathbf{a}}%
=\mathcal{J}_{\mathbf{b}} \label{10.11}%
\end{equation}

It is quite obvious that in a coordinate chart $\langle x^{\mu}\rangle$
covering an open set $U\subset M$ we can write%
\begin{equation}
D_{e_{\rho}}\mathcal{F}_{\beta}^{\rho}=\mathcal{J}_{\beta}, \label{10.11bis}%
\end{equation}
with $\mathcal{F}_{\beta}^{\rho}=g^{\rho\alpha}\mathcal{F}_{\alpha\beta}$%
\begin{align}
\mathcal{F}_{\alpha\beta}  &  =\left(  -e^{\gamma}\lrcorner\mathbf{R}%
_{\alpha\gamma}\right)  e_{\beta}-e_{\beta}\left(  -e^{\gamma}\lrcorner
\mathbf{R}_{\alpha\gamma}\right)  -\frac{1}{2}R(e_{\alpha}e_{\beta}-e_{\beta
}e_{\alpha})\label{10.12}\\
\mathcal{J}_{\beta}  &  =D_{e_{\rho}}(T^{\rho}e_{\beta}-e^{\rho}T_{\beta}).
\label{10.12bis}%
\end{align}

\begin{remark}
\label{nat of F and J}Eq.(\ref{10.11}) (or Eq.(\ref{10.11bis})) is a set
Maxwell like nonhomogeneous equations. It looks like the nonhomogeneous
classical Maxwell equations when that equations are written in components, but
Eq.(\ref{10.11bis}) is only a new way of writing the equation of the
nonhomogeneous field equations in the $Sl(2,\mathbb{C})$ like gauge theory
version of Einstein's theory, discussed in the previous section. In
particular, recall that any one of the six $\mathcal{F}_{\beta}^{\rho}\in\sec%
%TCIMACRO{\dbigwedge \nolimits^{2}}%
%BeginExpansion
{\displaystyle\bigwedge\nolimits^{2}}
%EndExpansion
TM\hookrightarrow\mathcal{C}\ell(TM)$. Or, in words, each one of the
$\mathcal{F}_{\beta}^{\rho}$ it is a bivector field, \textit{not} a set of
scalars which are components of a 2-form, as is the case in Maxwell theory.
Also, recall that according to Eq.(\ref{10.12bis}) each one of the four
$\mathcal{J}_{\beta}\in\sec\bigwedge\nolimits^{2}TM\hookrightarrow
\mathcal{C}\ell(TM)$.
\end{remark}

From Eq.(\ref{10.11}) \ it is not obvious that we must have $\mathcal{F}%
_{\mathbf{ab}}=0$ in vacuum, however that is exactly what happens if we take
into account Eq.(\ref{10.9}) which defines that object. Moreover,
$\mathcal{F}_{\mathbf{ab}}=0$\ does not imply that the curvature bivectors
$\mathbf{R}_{\mathbf{ab}}$ are null in vacuum. Indeed, in that case,
Eq.(\ref{10.9} ) implies only the identity (valid \textit{only} in vacuum)%

\begin{equation}
\left(  \mathbf{e}^{\mathbf{c}}\lrcorner\mathbf{R}_{\mathbf{ac}}\right)
\mathbf{e}_{\mathbf{b}}=\left(  \mathbf{e}^{\mathbf{c}}\lrcorner
\mathbf{R}_{\mathbf{bc}}\right)  \mathbf{e}_{\mathbf{a}}. \label{10.10''}%
\end{equation}

Moreover, recalling definition (Eq.(\ref{9.44})) we have%
\begin{equation}
\mathbf{R}_{\mathbf{ab}}=R_{\mathbf{abcd}}\mathbf{e}^{\mathbf{c}}%
\mathbf{e}^{\mathbf{d}}, \label{9.44'}%
\end{equation}
and we see that\ the $\mathbf{R}_{\mathbf{ab}}$are zero only if the Riemann
tensor is null which is not the case in any non trivial general relativistic model.

The important fact that we want to emphasize here is that although eventually
interesting, Eq.(\ref{10.11}) does not seem (according to our opinion) to
contain anything new in it. More precisely, all information given by that
equation is already contained in the original Einstein's equation, for indeed
it has been obtained from it by simple algebraic manipulations. We state
again: According to our view terms like%

\begin{align}
\mathcal{F}_{\mathbf{ab}}  &  =\frac{1}{2}(\mathbf{R}_{\mathbf{ac}}%
\mathbf{e}^{\mathbf{c}}\mathbf{e}_{\mathbf{b}}+\mathbf{e}_{\mathbf{b}%
}\mathbf{e}^{\mathbf{c}}\mathbf{R}_{\mathbf{ac}}-\mathbf{e}^{\mathbf{c}%
}\mathbf{R}_{\mathbf{ac}}\mathbf{e}_{\mathbf{b}}-\mathbf{e}_{\mathbf{b}%
}\mathbf{R}_{\mathbf{ac}}\mathbf{e}^{\mathbf{c}})-\frac{1}{2}R(\mathbf{e}%
_{\mathbf{a}}\mathbf{e}_{\mathbf{b}}-\mathbf{e}_{\mathbf{b}}\mathbf{e}%
_{\mathbf{a}}),\nonumber\\
\mathfrak{R}_{\mathbf{ab}}  &  =(T_{\mathbf{a}}\mathbf{e}_{\mathbf{b}%
}-\mathbf{e}_{\mathbf{b}}T_{\mathbf{a}})-\frac{1}{2}R(\mathbf{e}_{\mathbf{a}%
}\mathbf{e}_{\mathbf{b}}-\mathbf{e}_{\mathbf{b}}\mathbf{e}_{\mathbf{a}%
}),\nonumber\\
\mathbf{F}_{\mathbf{ab}}  &  =\frac{1}{2}R(\mathbf{e}_{\mathbf{a}}%
\mathbf{e}_{\mathbf{b}}-\mathbf{e}_{\mathbf{b}}\mathbf{e}_{\mathbf{a}}),
\label{10.13}%
\end{align}
are pure gravitational objects. We can see any relationship of any one of
these objects with the ones appearing in Maxwell theory. Of course, these
objects may eventually be used to formulate interesting equations, like
Eq.(\ref{10.11}) which are equivalent to Einstein's field equations, but this
fact does not seem to us to point to any new Physics.\footnote{Note that
$\mathbf{F}_{\mathbf{ab}}$ differs from a factor, namely $R$ from the
$\mathbf{F}_{\mathbf{ab}}^{\prime}$ given by Eq.(70).} Even more, from the
mathematical point of view, to find solutions to the new Eq.(\ref{10.11}) is
certainly a hard as to find solutions to the original Einstein equations.

\subsection{$Sl\left(  2,\mathbb{C}\right)  $ Gauge Theory and Sachs
Antisymmetric Equation}

We discuss in this subsection yet another algebraic exercise. First recall
that in section 2 we define the paravector fields,%

\[
\mathbf{q}_{\mathbf{a}}=\mathbf{e}_{\mathbf{a}}\mathbf{e}_{\mathbf{0}}=%
%TCIMACRO{\TeXButton{sigma}{\mbox{\boldmath{$\sigma$}}}}%
%BeginExpansion
\mbox{\boldmath{$\sigma$}}%
%EndExpansion
_{\mathbf{a}},\hspace{0.15in}\mathbf{\check{q}}_{\mathbf{a}}=(-%
%TCIMACRO{\TeXButton{sigma}{\mbox{\boldmath{$\sigma$}}}}%
%BeginExpansion
\mbox{\boldmath{$\sigma$}}%
%EndExpansion
_{\mathbf{0}},%
%TCIMACRO{\TeXButton{sigma}{\mbox{\boldmath{$\sigma$}}}}%
%BeginExpansion
\mbox{\boldmath{$\sigma$}}%
%EndExpansion
_{\mathbf{i}}),\hspace{0.15in}%
%TCIMACRO{\TeXButton{sigma}{\mbox{\boldmath{$\sigma$}}}}%
%BeginExpansion
\mbox{\boldmath{$\sigma$}}%
%EndExpansion
_{\mathbf{0}}=1.
\]

Recall that\footnote{In Sachs book he wrote: $[D_{e_{\rho}},D_{e_{\lambda}%
}]e_{\mu}=R_{\mu\text{ }\rho\lambda}^{\text{ }\alpha}e_{\alpha}=+R_{\alpha
\mu\rho\lambda}e^{\alpha}$. This produces some changes in signals in relation
to our formulas below. Our Eq.(\ref{9.41}) agrees with the conventions in
\cite{choquet}.}
\begin{align}
\lbrack D_{e_{\rho}},D_{e_{\lambda}}]e_{\mu}  &  =R_{\mu\text{ }\rho\lambda
}^{\text{ }\alpha}e_{\alpha}=-R_{\alpha\mu\rho\lambda}e^{\alpha}=R_{\mu
\alpha\rho\lambda}e^{\alpha},\nonumber\\
R_{\mu\text{ }\rho\lambda}^{\text{ }\alpha}  &  =\mathcal{R}(e_{\mu}%
,\theta^{\alpha},e_{\rho},e_{\lambda}). \label{9.41}%
\end{align}

Then a simple calculation shows that
\begin{align}
\lbrack D_{e_{\rho}},D_{e_{\lambda}}]e_{\mu}  &  =e_{\mu}\lrcorner
\mathbf{R}_{\rho\lambda}=-\mathbf{R}_{\rho\lambda}\llcorner e_{\mu
},\label{9.43a}\\
R_{\mu\alpha\rho\lambda}e^{\alpha}  &  =\frac{1}{2}(e_{\mu}\mathbf{R}%
_{\rho\lambda}-\mathbf{R}_{\rho\lambda}e_{\mu}). \label{9.43b}%
\end{align}

Multiplying Eq.(\ref{9.43b}) on the left by $\mathbf{e}_{\mathbf{0}}$ we get,
recalling that $%
%TCIMACRO{\TeXButton{omega}{\mbox{\boldmath{$\omega$}}}}%
%BeginExpansion
\mbox{\boldmath{$\omega$}}%
%EndExpansion
_{\mathbf{e}_{\mathbf{a}}}^{\dagger}=-\mathbf{e}^{\mathbf{0}}%
%TCIMACRO{\TeXButton{omega}{\mbox{\boldmath{$\omega$}}}}%
%BeginExpansion
\mbox{\boldmath{$\omega$}}%
%EndExpansion
_{\mathbf{e}_{\mathbf{a}}}\mathbf{e}^{\mathbf{0}}$ (Eq.(79) we get%
\begin{equation}
R_{\mu\alpha\rho\lambda}\mathbf{q}^{\alpha}=\frac{1}{2}(\mathbf{q}_{\mu
}\mathbf{R}_{\rho\lambda}^{\dagger}+\mathbf{R}_{\rho\lambda}\mathbf{q}_{\mu}).
\label{9.43c}%
\end{equation}
\ 

Now, to derive Sachs\footnote{Numeration is from Sachs' book \cite{s1}.}
Eq.(6.50a) all we need to do is to multiply Eq.(\ref{10.8}) on the right by
$\mathbf{e}^{\mathbf{0}}$ and perform some algebraic manipulations. We then
get (with \textit{our} normalization) for the equivalent of Einstein's
equations using the paravector fields and a coordinate chart $\langle x^{\mu
}\rangle$ covering an open set $U\subset M$, the following equation
\begin{equation}
\mathbf{R}_{\rho\lambda}\mathbf{q}^{\lambda}+\mathbf{q}^{\lambda}%
\mathbf{R}_{\rho\lambda}^{\dagger}+R\mathbf{q}_{\rho}=2\mathbf{T}_{\rho}.
\label{sachs1}%
\end{equation}
For the Hermitian conjugate we have
\begin{equation}
-\mathbf{R}_{\rho\lambda}^{\dagger}\mathbf{\check{q}}^{\lambda}-\mathbf{\check
{q}}^{\lambda}\mathbf{R}_{\rho\lambda}+R\mathbf{\check{q}}_{\rho
}=2\mathbf{\check{T}}_{\rho}. \label{sachs1'}%
\end{equation}
where as above $\mathbf{R}_{\rho\lambda}$ are the the curvature bivectors
given by Eq.(\ref{9.40}) and
\begin{equation}
\mathbf{T}_{\rho}=T_{\rho}^{\mu}\mathbf{q}_{\mu}\in\sec\bigwedge
\nolimits^{2}TM\hookrightarrow\mathcal{C\ell(}TM\mathcal{)}. \label{sachs2}%
\end{equation}

After that, we multiply Eq.(\ref{sachs1}) on the right by $\mathbf{\check{q}%
}_{\gamma}$ and Eq.(\ref{sachs1'}) on the left by $\mathbf{q}_{\gamma}$ ending
with two new equations. If we sum them, we get a \ `symmetric'
equation\footnote{Eq.(6.52) in Sachs' book \cite{s1}.} completely equivalent
to Einstein's equation (from where we started). If we make the difference of
the equations we get an antisymmetric equation. The antisymmetric equation can
be written, introducing%
\begin{align}
\mathbb{F}_{\rho\gamma}  &  =\frac{1}{2}(\mathbf{R}_{\rho\lambda}%
\mathbf{q}^{\lambda}\mathbf{\check{q}}_{\gamma}\mathbf{+q}_{\gamma
}\mathbf{\check{q}}^{\lambda}\mathbf{R}_{\rho\lambda}\mathbf{+q}^{\lambda
}\mathbf{R}_{\rho\lambda}^{\dagger}\mathbf{\check{q}}_{\gamma}\mathbf{+q}%
_{\gamma}\mathbf{R}_{\rho\lambda}^{\dagger}\mathbf{\check{q}}^{\lambda
})\label{sachs3}\\
&  +\frac{1}{2}R(\mathbf{q}_{\rho}\mathbf{\check{q}}_{\gamma}\mathbf{-q}%
_{\gamma}\mathbf{\check{q}}_{\rho})\nonumber
\end{align}
and%
\begin{equation}
\mathbb{J}_{\gamma}=D_{e_{\rho}}(\mathbf{T}^{\rho}\mathbf{\check{q}}_{\gamma
}-\mathbf{q}_{\gamma}\mathbf{\check{T}}^{\rho}), \label{sachs4}%
\end{equation}
as%
\begin{equation}
D_{e_{\rho}}\mathbb{F}_{\gamma}^{\rho}=\mathbb{J}_{\gamma}. \label{sachs5}%
\end{equation}
\ \ 

\begin{remark}
It is important to recall that each one of the six $\mathbb{F}_{\rho\gamma}$
and each one of the four $\mathbb{J}_{\gamma}$ are not a set of scalars, but
sections of $\mathcal{C}\ell^{(0)}\mathcal{(}TM\mathcal{)}$. Also, take notice
that Eq.(\ref{sachs5}), of course, is completely equivalent to our
Eq.(\ref{10.11}). Its matrix translation in $\mathbb{C}\ell^{(0)}(M)\simeq
S(M)\otimes_{\mathbb{C}}\bar{S}(M)$ gives Sachs equation (6.52-) in \cite{s1}
if we take into account his different Sachs different \ `normalization' of the
connection coefficients and the \textit{ad hoc} factor with dimension of
electric charge that he introduced. We cannot see at present any new
information encoded in that equations which could be translated in
\ interesting geometrical properties of the manifold, but of course,
eventually someone may find that they encode such a useful
information.\footnote{Anyway, it seems to us that until the written of the
present paper the true mathematical nature of Sachs equations have not been
understood, by people that read Sachs books and articles. To endorse our
statement, we quote that in Carmeli's review(\cite{carmelimr}) of Sachs book,
he did not realize that Sachs theory was indeed (as we showed above) a
description in the Pauli bundle of a $Sl(2,\mathbb{C)}$ gauge formulation of
Einstein's theory as described in his own book \cite{carmeli}. Had he
disclosed that fact (as we did) he probably had not written that Sachs'
approach was a possible unified field theory of gravitation and
electromagnetism.}
\end{remark}

Using \ the equations, \ $D_{\mathbf{e}_{\mathbf{a}}}\mathbf{e}_{0}=0$ and
$D_{e_{\rho}}^{\mathbf{S}}\mathbf{q}_{\mu}=0$ (respectively, \ Eq.(88) and
Eq.(108) in \cite{twospinor} ) and (\ref{9.28}) we may verify that
\begin{equation}
D_{e_{\rho}}^{\mathbf{S}}\mathbb{F}_{\mu\nu}+D_{e_{\mu}}^{\mathbf{S}%
}\mathbb{F}_{\nu\rho}+D_{e_{\nu}}^{\mathbf{S}}\mathbb{F}_{\rho\mu}=0,
\label{sachs6}%
\end{equation}
where $D_{e_{\rho}}^{\mathbf{S}}$ is Sachs \ `covariant' derivative that we
discussed in \cite{twospinor}. In \cite{s2} Sachs concludes that the last
equation implies that there are no magnetic monopoles in nature. Of course,
his conclusion would follow from Eq.(\ref{sachs6}) only if it happened that
$\mathbb{F}_{\gamma}^{\rho}$ were the components in a coordinate basis of a
$2$-form field $F\in\sec\bigwedge\nolimits^{2}T^{\ast}M$. However, this is not
the case, because as already noted above, this is not the mathematical nature
of the $\mathbb{F}_{\gamma}^{\rho}$ . Contrary to what we stated with relation
to Eq.(\ref{sachs5}) we cannot even say that Eq.(\ref{sachs6}) is really
interesting, because it uses a covariant derivative operator, which, as
discussed in \cite{twospinor} is not \ well justified, and in anyway
$D_{e_{\rho}}^{\mathbf{S}}\neq D_{e_{\rho}}$. We cannot see any relationship
of Eq.(\ref{sachs6}) with the legendary magnetic monopoles.

We thus conclude this section stating that Sachs claims in \cite{s1,s2,s3} of
having produced an unified field theory of electricity and electromagnetism
are not endorsed by our analysis.

\section{Energy-Momentum \textquotedblleft Conservation\textquotedblright\ in
General Relativity}

\subsection{Einstein's Equations in terms of Superpotentials $\star
S^{\mathbf{a}}$}

In this section we discuss some issues and statements concerning the problem
of the energy-momentum conservation in Einstein's theory, presented with
several different formalisms in the literature, which according to our view
are very confusing, or even wrong. \footnote{Mastering that there is indeed a
serious problem, will enable readers to appreciate some of our comments
concerning MEG.}. To start, recall that from Eq.(\ref{10.0}) it follows that
\begin{equation}
d(\star\mathcal{J}\mathbf{-}\frac{1}{2}[%
%TCIMACRO{\TeXButton{omega}{\mbox{\boldmath{$\omega$}}}}%
%BeginExpansion
\mbox{\boldmath{$\omega$}}%
%EndExpansion
,\mathbf{\star}\mathcal{R}])=0, \label{10.0BISS}%
\end{equation}
and we could think that this equation could be used to identify a
\textit{conservation} \textit{law} for the energy momentum of matter plus the
gravitational field, with $\frac{1}{2}[%
%TCIMACRO{\TeXButton{omega}{\mbox{\boldmath{$\omega$}}}}%
%BeginExpansion
\mbox{\boldmath{$\omega$}}%
%EndExpansion
,\mathbf{\star}\mathcal{R}]$ describing a mathematical object related to the
energy- momentum of the gravitational field. However, this is not the case,
because this term (due to the presence of $%
%TCIMACRO{\TeXButton{omega}{\mbox{\boldmath{$\omega$}}}}%
%BeginExpansion
\mbox{\boldmath{$\omega$}}%
%EndExpansion
$) is gauge dependent. The appearance of a gauge dependent term is \ a
recurrent fact in all known proposed\footnote{At least, the ones known by the
authors.
\par
{}} formulations of a \ `conservation law for energy-momentum' for Einstein
theory. We discuss now some statements found in the literature based on some
of that proposed \ `solutions' to the problem of energy-momentum conservation
in General Relativity and say why we think they are unsatisfactory. We also
mention a way with which the problem could be satisfactorily solved, but which
implies in a departure from the orthodox interpretation of Einstein's theory.

Now, to keep the mathematics as simple and transparent as possible instead of
working with Eq.(\ref{10.0BISS}), we work with a more simple (but equivalent)
formulation \cite{rq,thirring} of Einstein's equation where the gravitational
field is described by a set of $2$-forms $\star S^{\mathbf{a}}$,
$\mathbf{a}=0,1,2,3$ \ called superpotentials. This approach will permit to
identify very quickly certain objects that at first sight seems
\textit{appropriate \ energy-momentum }currents for the gravitational field in
Einstein's theory. The calculations that follows are done in the Clifford
algebra of multiforms fields $\mathcal{C\ell}\left(  T^{\ast}M\right)  $,
something that,,as the reader will testify, simplify considerably similar
calculations done with traditional methods.

We start again with Einstein's equations given by Eq.(\ref{10.2}), but this
time we multiply on the left by $\theta^{\mathbf{b}}\in\sec\bigwedge
\nolimits^{1}T^{\ast}M\hookrightarrow\mathcal{C\ell}\left(  T^{\ast}M\right)
$ getting an equation relating the \textit{Ricci }$\mathit{1}$\textit{-forms}
$\ \mathcal{R}^{\mathbf{a}}=R_{\mathbf{b}}^{\mathbf{a}}\theta^{\mathbf{b}}$
$\in\sec\bigwedge\nolimits^{1}T^{\ast}M\hookrightarrow\mathcal{C\ell}\left(
T^{\ast}M\right)  $ with the \textit{energy-momentum 1-forms} $\mathcal{T}%
^{\mathbf{a}}=T_{\mathbf{b}}^{\mathbf{a}}\theta^{\mathbf{b}}\in\sec
\bigwedge\nolimits^{1}T^{\ast}M\hookrightarrow\mathcal{C\ell}\left(  T^{\ast
}M\right)  $, i.e.,
\begin{equation}
\mathcal{G}^{\mathbf{a}}=\mathcal{R}^{\mathbf{a}}-\frac{1}{2}R\theta
^{\mathbf{a}}=\mathcal{T}^{\mathbf{a}}. \label{10.14}%
\end{equation}

We take the dual of this equation,%
\begin{equation}
\star\mathcal{G}^{\mathbf{a}}=\star\mathcal{T}^{\mathbf{a}}. \label{10.15}%
\end{equation}

Now, we observe that \cite{rq,thirring} we can write%
\begin{equation}
\star\mathcal{G}^{\mathbf{a}}=-d\star\mathcal{S}^{\mathbf{a}}-\star
\mathfrak{t}_{\mathbf{\ }}^{\mathbf{a}}, \label{10.16}%
\end{equation}
where \
\begin{align}
\mathcal{S}^{\mathbf{c}}  &  =-\frac{1}{2}%
%TCIMACRO{\TeXButton{omega}{\mbox{\boldmath{$\omega$}}}}%
%BeginExpansion
\mbox{\boldmath{$\omega$}}%
%EndExpansion
_{\mathbf{ab}}\wedge\star(\theta^{\mathbf{a}}\wedge\theta^{\mathbf{b}}%
\wedge\theta^{\mathbf{c}}),\nonumber\\
\star\mathfrak{t}_{\mathbf{\ }}^{\mathbf{c}}  &  =\frac{1}{2}%
%TCIMACRO{\TeXButton{omega}{\mbox{\boldmath{$\omega$}}}}%
%BeginExpansion
\mbox{\boldmath{$\omega$}}%
%EndExpansion
_{\mathbf{ab}}\wedge\lbrack%
%TCIMACRO{\TeXButton{omega}{\mbox{\boldmath{$\omega$}}}}%
%BeginExpansion
\mbox{\boldmath{$\omega$}}%
%EndExpansion
_{\mathbf{d}}^{\mathbf{c}}\star(\theta^{\mathbf{a}}\wedge\theta^{\mathbf{b}%
}\wedge\theta^{\mathbf{d}})+%
%TCIMACRO{\TeXButton{omega}{\mbox{\boldmath{$\omega$}}}}%
%BeginExpansion
\mbox{\boldmath{$\omega$}}%
%EndExpansion
_{\mathbf{d}}^{\mathbf{b}}\star(\theta^{\mathbf{a}}\wedge\theta^{\mathbf{d}%
}\wedge\theta^{\mathbf{c}})]. \label{10.17}%
\end{align}

The proof of Eq.(\ref{10.17}) follows at once from the fact that
\begin{equation}
\star\mathcal{G}^{\mathbf{d}}=\frac{1}{2}\mathcal{R}_{\mathbf{ab}}\wedge
\star(\theta^{\mathbf{a}}\wedge\theta^{\mathbf{b}}\wedge\theta^{\mathbf{d}}).
\label{10.18}%
\end{equation}

Indeed, recalling the identities in Eq.(\ref{Aidentities}) we can write%
\begin{align}
\frac{1}{2}\mathcal{R}_{\mathbf{ab}}\wedge\star(\theta^{\mathbf{a}}%
\wedge\theta^{\mathbf{b}}\wedge\theta^{\mathbf{d}})  &  =-\frac{1}{2}%
\star\lbrack\mathcal{R}_{\mathbf{ab}}\lrcorner(\theta^{\mathbf{a}}\wedge
\theta^{\mathbf{b}}\wedge\theta^{\mathbf{d}})]\nonumber\\
&  =-\frac{1}{2}R_{\mathbf{abcd}}\star\lbrack(\theta^{\mathbf{c}}\wedge
\theta^{\mathbf{d}})\lrcorner(\theta^{\mathbf{a}}\wedge\theta^{\mathbf{b}%
}\wedge\theta^{\mathbf{d}})]\nonumber\\
&  =-\star(\mathcal{R}^{\mathbf{d}}-\frac{1}{2}R\theta^{\mathbf{d}}).
\label{10.19}%
\end{align}

On the other hand we have,%
\begin{align}
&  2\star\mathcal{G}^{\mathbf{d}}=d%
%TCIMACRO{\TeXButton{omega}{\mbox{\boldmath{$\omega$}}}}%
%BeginExpansion
\mbox{\boldmath{$\omega$}}%
%EndExpansion
_{\mathbf{ab}}\wedge\star(\theta^{\mathbf{a}}\wedge\theta^{\mathbf{b}}%
\wedge\theta^{\mathbf{d}})+%
%TCIMACRO{\TeXButton{omega}{\mbox{\boldmath{$\omega$}}}}%
%BeginExpansion
\mbox{\boldmath{$\omega$}}%
%EndExpansion
_{\mathbf{ac}}\wedge%
%TCIMACRO{\TeXButton{omega}{\mbox{\boldmath{$\omega$}}}}%
%BeginExpansion
\mbox{\boldmath{$\omega$}}%
%EndExpansion
_{\mathbf{b}}^{\mathbf{c}}\wedge\star(\theta^{\mathbf{a}}\wedge\theta
^{\mathbf{b}}\wedge\theta^{\mathbf{d}})\nonumber\\
&  =d[%
%TCIMACRO{\TeXButton{omega}{\mbox{\boldmath{$\omega$}}}}%
%BeginExpansion
\mbox{\boldmath{$\omega$}}%
%EndExpansion
_{\mathbf{ab}}\wedge\star(\theta^{\mathbf{a}}\wedge\theta^{\mathbf{b}}%
\wedge\theta^{\mathbf{d}})]-%
%TCIMACRO{\TeXButton{omega}{\mbox{\boldmath{$\omega$}}}}%
%BeginExpansion
\mbox{\boldmath{$\omega$}}%
%EndExpansion
_{\mathbf{ab}}\wedge d\star(\theta^{\mathbf{a}}\wedge\theta^{\mathbf{b}}%
\wedge\theta^{\mathbf{d}})\nonumber\\
&  +%
%TCIMACRO{\TeXButton{omega}{\mbox{\boldmath{$\omega$}}}}%
%BeginExpansion
\mbox{\boldmath{$\omega$}}%
%EndExpansion
_{\mathbf{ac}}\wedge%
%TCIMACRO{\TeXButton{omega}{\mbox{\boldmath{$\omega$}}}}%
%BeginExpansion
\mbox{\boldmath{$\omega$}}%
%EndExpansion
_{\mathbf{b}}^{\mathbf{c}}\wedge\star(\theta^{\mathbf{a}}\wedge\theta
^{\mathbf{b}}\wedge\theta^{\mathbf{d}})\nonumber\\
&  =d[%
%TCIMACRO{\TeXButton{omega}{\mbox{\boldmath{$\omega$}}}}%
%BeginExpansion
\mbox{\boldmath{$\omega$}}%
%EndExpansion
_{\mathbf{ab}}\wedge\star(\theta^{\mathbf{a}}\wedge\theta^{\mathbf{b}}%
\wedge\theta^{\mathbf{d}})]-%
%TCIMACRO{\TeXButton{omega}{\mbox{\boldmath{$\omega$}}}}%
%BeginExpansion
\mbox{\boldmath{$\omega$}}%
%EndExpansion
_{\mathbf{ab}}\wedge%
%TCIMACRO{\TeXButton{omega}{\mbox{\boldmath{$\omega$}}}}%
%BeginExpansion
\mbox{\boldmath{$\omega$}}%
%EndExpansion
_{\mathbf{p}}^{\mathbf{a}}\star(\theta^{\mathbf{p}}\wedge\theta^{\mathbf{b}%
}\wedge\theta^{\mathbf{d}})\nonumber\\
&  -%
%TCIMACRO{\TeXButton{omega}{\mbox{\boldmath{$\omega$}}}}%
%BeginExpansion
\mbox{\boldmath{$\omega$}}%
%EndExpansion
_{\mathbf{ab}}\wedge%
%TCIMACRO{\TeXButton{omega}{\mbox{\boldmath{$\omega$}}}}%
%BeginExpansion
\mbox{\boldmath{$\omega$}}%
%EndExpansion
_{\mathbf{p}}^{\mathbf{b}}\star(\theta^{\mathbf{a}}\wedge\theta^{\mathbf{p}%
}\wedge\theta^{\mathbf{d}})-%
%TCIMACRO{\TeXButton{omega}{\mbox{\boldmath{$\omega$}}}}%
%BeginExpansion
\mbox{\boldmath{$\omega$}}%
%EndExpansion
_{\mathbf{ab}}\wedge%
%TCIMACRO{\TeXButton{omega}{\mbox{\boldmath{$\omega$}}}}%
%BeginExpansion
\mbox{\boldmath{$\omega$}}%
%EndExpansion
_{\mathbf{p}}^{\mathbf{d}}\star(\theta^{\mathbf{a}}\wedge\theta^{\mathbf{b}%
}\wedge\theta^{\mathbf{p}})]\nonumber\\
&  +%
%TCIMACRO{\TeXButton{omega}{\mbox{\boldmath{$\omega$}}}}%
%BeginExpansion
\mbox{\boldmath{$\omega$}}%
%EndExpansion
_{\mathbf{ac}}\wedge%
%TCIMACRO{\TeXButton{omega}{\mbox{\boldmath{$\omega$}}}}%
%BeginExpansion
\mbox{\boldmath{$\omega$}}%
%EndExpansion
_{\mathbf{b}}^{\mathbf{c}}\wedge\star(\theta^{\mathbf{a}}\wedge\theta
^{\mathbf{b}}\wedge\theta^{\mathbf{d}})\nonumber\\
&  =d[%
%TCIMACRO{\TeXButton{omega}{\mbox{\boldmath{$\omega$}}}}%
%BeginExpansion
\mbox{\boldmath{$\omega$}}%
%EndExpansion
_{\mathbf{ab}}\wedge\star(\theta^{\mathbf{a}}\wedge\theta^{\mathbf{b}}%
\wedge\theta^{\mathbf{d}})]-%
%TCIMACRO{\TeXButton{omega}{\mbox{\boldmath{$\omega$}}}}%
%BeginExpansion
\mbox{\boldmath{$\omega$}}%
%EndExpansion
_{\mathbf{ab}}\wedge\lbrack%
%TCIMACRO{\TeXButton{omega}{\mbox{\boldmath{$\omega$}}}}%
%BeginExpansion
\mbox{\boldmath{$\omega$}}%
%EndExpansion
_{\mathbf{p}}^{\mathbf{d}}\star(\theta^{\mathbf{a}}\wedge\theta^{\mathbf{b}%
}\wedge\theta^{\mathbf{p}})+%
%TCIMACRO{\TeXButton{omega}{\mbox{\boldmath{$\omega$}}}}%
%BeginExpansion
\mbox{\boldmath{$\omega$}}%
%EndExpansion
_{\mathbf{p}}^{\mathbf{b}}\star(\theta^{\mathbf{a}}\wedge\theta^{\mathbf{p}%
}\wedge\theta^{\mathbf{d}})]\nonumber\\
&  =-2(d\star\mathcal{S}^{\mathbf{d}}+\star\mathfrak{t}^{\mathbf{d}}).
\label{10.20}%
\end{align}

Now, we can then write Einstein's equation in a very interesting, but
\textit{dangerous} form, i.e.:%
\begin{equation}
-d\star\mathcal{S}^{\mathbf{a}}=\star\mathcal{T}^{\mathbf{a}}+\star
\mathfrak{t}^{\mathbf{a}}. \label{10.21}%
\end{equation}

In writing Einstein's equations in that way, we have associated to the
gravitational field a set of $2$-form fields $\star\mathcal{S}^{\mathbf{a}}$
called \textit{superpotentials} that have as sources the currents
$(\star\mathcal{T}^{\mathbf{a}}+\star\mathfrak{t}^{\mathbf{a}})$. However,
superpotentials are not uniquely defined since, e.g., superpotentials
\ $(\star\mathcal{S}^{\mathbf{a}}+\star\alpha^{\mathbf{a}})$, with
$\star\alpha^{\mathbf{a}}$ closed, i.e., $d\star\alpha^{\mathbf{a}}=0$ give
the same second member for Eq.(\ref{10.21}).

\subsection{Is There Any Energy-Momentum Conservation Law in GR?}

Why did we say that Eq.(\ref{10.21}) is a dangerous one?

The reason is that (as in the case of Eq.(\ref{10.0BISS})) we can be led to
think that we have discovered a conservation law for the energy momentum of
matter plus gravitational field, since from Eq.(\ref{10.21}) it follows that
\begin{equation}
d(\star\mathcal{T}^{\mathbf{a}}+\star\mathfrak{t}^{\mathbf{a}})=0.
\label{10.22}%
\end{equation}
This thought however is only an example of wishful thinking, because the
$\star\mathfrak{t}^{\mathbf{a}}$ \ depends on the connection (see
Eq.(\ref{10.17})) and thus are gauge dependent. They do not have the same
tensor transformation law as the $\star\mathcal{T}^{\mathbf{a}}$. So, Stokes
theorem cannot be used to derive from Eq.(\ref{10.22}) conserved quantities
that are independent of the gauge, which is clear. However, and this is less
known, Stokes theorem, also cannot be used to derive conclusions that are
independent of the local coordinate chart used to perform calculations
\cite{boro}. In fact, the currents $\star\mathfrak{t}^{\mathbf{a}}$ are
nothing more than the old pseudo energy momentum tensor of Einstein in a new
dress. Non recognition of this fact can lead to many misunderstandings. We
present some of them in what follows, in order to to call our readers'
attention of potential errors of inference that can be done when we use
sophisticated mathematical formalisms without a perfect domain of their contents.

\qquad\textbf{(i)}\ First, it is easy to see that from Eq.(\ref{10.15}) it
follows that \cite{mtw}
\begin{equation}
\mathbf{D}^{c}\mathbf{\star}\mathfrak{G}=\mathbf{D}^{c}\star\mathfrak{T}=0,
\label{10.22.0}%
\end{equation}
where $\mathbf{\star}\mathfrak{G}=\mathbf{e}_{\mathbf{a}}\otimes
\star\mathcal{G}^{\mathbf{a}}$ $\in\sec TM\otimes\sec\bigwedge\nolimits^{3}%
T^{\ast}M$ and $\star\mathfrak{T}=\mathbf{e}_{\mathbf{a}}\otimes
\star\mathcal{T}^{\mathbf{a}}\in\sec TM\otimes\sec\bigwedge\nolimits^{3}%
T^{\ast}M$ . Now, in \cite{mtw} it is written (without proof) a \ `Stokes
theorem' \medskip%
\begin{equation}%
\begin{tabular}
[c]{|c|}\hline
$%
%TCIMACRO{\dint \limits_{{\footnotesize 4}\text{-cube}}}%
%BeginExpansion
{\displaystyle\int\limits_{{\footnotesize 4}\text{-cube}}}
%EndExpansion
\mathbf{D}^{c}\mathbf{\star}\mathfrak{T}\mathbf{=}%
%TCIMACRO{\dint \limits_{\substack{{\footnotesize 3}\text{ boundary}\\\text{ of
%this }{\footnotesize 4}\text{-cube}}}}%
%BeginExpansion
{\displaystyle\int\limits_{\substack{{\footnotesize 3}\text{ boundary}\\\text{
of this }{\footnotesize 4}\text{-cube}}}}
%EndExpansion
\mathbf{\star}\mathfrak{T}$\\\hline
\end{tabular}
\ \ \ \label{mtw}%
\end{equation}

\begin{center}
\medskip
\end{center}

We searched in the literature for a proof of Eq.(\ref{mtw}) which appears also
in many other texts and scientific papers, as e.g., in \cite{dalton,vatorr1}
and could find none, which we can consider as valid. The reason is simply. If
expressed in details, e.g., the first member of Eq.(\ref{mtw}) reads
\begin{equation}%
%TCIMACRO{\dint \limits_{{\footnotesize 4}\text{-cube}}}%
%BeginExpansion
{\displaystyle\int\limits_{{\footnotesize 4}\text{-cube}}}
%EndExpansion
\mathbf{e}_{\mathbf{a}}\otimes(d\star\mathcal{T}^{\mathbf{a}}+\omega
_{\mathbf{b}}^{\mathbf{a}}\wedge\mathcal{T}^{\mathbf{b}}). \label{10.22.01}%
\end{equation}
and it is necessary to explain what is the meaning (if any) of the integral.
Since is integrand is a sum of tensor fields, this integral says that we are
\textit{summing} tensors belonging to the tensor spaces of different spacetime
points. As, well known, this cannot be done in general, unless there is a way
for identification of the tensor spaces at different spacetime points. This
requires, of course, the introduction of additional structure on the spacetime
representing a given gravitational field, and such extra structure is lacking
in Einstein theory. We unfortunately, must conclude that Eq.(\ref{mtw}) do not
express any conservation law, for it lacks as yet, a precise mathematical
meaning.\footnote{Of course, if some could give a mathematical meaning to
Eq.(\ref{mtw}), we will be glad to be informed of that fact.}

\ \ In Einstein theory possible superpotentias are, of course, the
$\star\mathcal{S}^{\mathbf{a}}$ that we found above (Eq.(\ref{10.17})), with
\begin{equation}
\star\mathcal{S}_{\mathbf{c}}=[-\frac{1}{2}\omega_{\mathbf{ab}}\lrcorner
(\theta^{\mathbf{a}}\wedge\theta^{\mathbf{b}}\wedge\theta_{\mathbf{c}}%
)]\theta^{\mathbf{5}}. \label{10.22.1.1}%
\end{equation}

Then, if we integrate Eq.(\ref{10.21}) over a \ `certain finite $3$%
-dimensional volume', say a ball $B$, and use Stokes theorem we have%
\begin{equation}
P^{\mathbf{a}}=%
%TCIMACRO{\dint \limits_{B}}%
%BeginExpansion
{\displaystyle\int\limits_{B}}
%EndExpansion
\star\left(  \mathcal{T}^{\mathbf{a}}+\mathfrak{t}^{\mathbf{a}}\right)  =-%
%TCIMACRO{\dint \limits_{\partial B}}%
%BeginExpansion
{\displaystyle\int\limits_{\partial B}}
%EndExpansion
\star\mathcal{S}^{\mathbf{a}}. \label{10.22.2}%
\end{equation}

In particular the energy or (\textit{inertial mass}) of the gravitational
field plus matter generating the field is defined by
\begin{equation}
P^{\mathbf{0}}=E=m_{i}=-\lim_{R\rightarrow\infty}%
%TCIMACRO{\dint \limits_{\partial B}}%
%BeginExpansion
{\displaystyle\int\limits_{\partial B}}
%EndExpansion
\star\mathcal{S}^{\mathbf{0}} \label{10.22.3'}%
\end{equation}

\textbf{(ii) }Now, a frequent misunderstanding is the following. Suppose that
in a \textit{given} gravitational theory\ there exists an energy-momentum
conservation law for matter plus the gravitational field expressed in the form
of Eq.(\ref{10.22}), where $\mathcal{T}^{\mathbf{a}}$ \ are the
energy-momentum 1-forms of matter and $\mathfrak{t}^{\mathbf{a}}$ are
\textit{true}\footnote{This means that the $t^{\mathbf{a}}$ are not pseudo
1-forms, as in Einstein's theory.} energy-momentum 1-forms of the
gravitational field. \ This means that \ the $3$-forms $(\mathbf{\star
}\mathcal{T}^{\mathbf{a}}+\star\mathfrak{t}^{\mathbf{a}})$ are closed, i.e.,
\ they satisfy Eq.(\ref{10.22}). Is this enough to warrant that the energy of
a closed universe is zero? Well, that would be the case if starting from
\ Eq.(\ref{10.22}) we could jump to an equation like Eq.(\ref{10.21}) and then
to Eq.(\ref{10.22.3'}) (as done, e.g., in \cite{thirring2}). But that sequence
of inferences in general cannot be done, for indeed, as it is well known,it is
not the case that closed three forms are always exact. Take a closed universe
with topology, say $\mathbb{R\times}S^{3}$. In this case $B=$ $S^{3}$ and we
have $\partial B=$ $\partial S^{3}=\varnothing$. Now, as it is well known
(see, e.g., \cite{nakahara}), the third de Rham cohomology group of
$\mathbb{R\times}S^{3}$ is $H^{3}\left(  \mathbb{R\times}S^{3}\right)
=H^{3}\left(  S^{3}\right)  =\mathbb{R}$. Since this group is non trivial it
follows that in such manifold closed forms are not exact. Then from
Eq.(\ref{10.22}) it did not follow the validity of an equation analogous to
Eq.(\ref{10.21}). So, in that case an equation like Eq.(\ref{10.22.2}) cannot
even be written.

Despite that commentary, keep in mind that in Einstein's theory the energy of
a closed universe\footnote{Note that if we suppose that the universe contains
spinor fields, then it must be a spin manifold, i.e., it is parallezible
according to Geroch's theorem \cite{g1}.} if it is given by Eq.(\ref{10.22.3'}%
) is indeed zero, since in that theory the $3$-forms $(\mathbf{\star
}\mathcal{T}^{\mathbf{a}}+\star\mathfrak{t}^{\mathbf{a}})$ are indeed exact
(see Eq.(\ref{10.21})). This means that accepting $\mathfrak{t}^{\mathbf{a}}$
as the energy-momentum $1$-form fields of the gravitational field, it follows
that gravitational energy \textit{must} be negative in a closed universe.

\textbf{(iii)} But, is the above formalism a consistent one? Given a
coordinate chart $\langle x^{\mu}\rangle$ of the maximal atlas of $M$ , with
some algebra we can show that for a gravitational model represented by a
diagonal asymptotic flat metric\footnote{A metric is said to be asymptotically
flat in given coordinates, if $g_{\mu\nu}=n_{\mu\nu}(1+\mathrm{O}\left(
r^{-k}\right)  )$, with $k=2$ or $k=1$ depending on the author. See, eg.,
\cite{schoenyau1, schoenyau2,wald}.}, the inertial mass $E=m_{i}$ is given by
\begin{equation}
m_{i}=\lim_{R\rightarrow\infty}\frac{-1}{16\pi}%
%TCIMACRO{\dint \limits_{\partial B}}%
%BeginExpansion
{\displaystyle\int\limits_{\partial B}}
%EndExpansion
\frac{\partial}{\partial x^{\beta}}(g_{11}g_{22}g_{33}g^{\alpha\beta}%
)d\sigma_{\alpha}, \label{10.22.4}%
\end{equation}
where $\partial B=S^{2}(R)$ is a $2$-sphere of radius $R$,
$(-n_{\mathbf{\alpha}})$ is the outward unit normal and $d\sigma_{\alpha
}=-R^{2}n_{\alpha}dA$. \ If we apply Eq.(\ref{10.22.4}) to calculate, e.g.,
the energy of the Schwarzschild space time\footnote{For a Scharzschild
spacetime we have $g=\left(  1-\frac{2m}{r}\right)  dt\otimes dt-\left(
1-\frac{2m}{r}\right)  ^{-1}dr\otimes dr-r^{2}(d\theta\otimes d\theta+\sin
^{2}\theta d\varphi\otimes d\varphi)$.} generate by a gravitational mass $m$,
we expect to have one unique and unambiguous result, namely $m_{i}=m$.

However, as showed in details, e.g., in \cite{boro} the calculation of $E$
depends on the spatial coordinate system naturally adapted to the reference
frame $Z=\frac{1}{\sqrt{\left(  1-\frac{2m}{r}\right)  }}\frac{\partial
}{\partial t}$ , even if these coordinates produce asymptotically flat
metrics. Then, even if in one given chart we may obtain $m_{i}=m$ there are
others where $m_{i}\neq m$!

Moreover, note also that, as showed above, for a closed universe, Einstein's
theory implies on general grounds (once we accept that the $\mathfrak{t}^{a}$
describes the energy-momentum distribution of the gravitational field) that
$m_{i}=0$. This result, it is important to quote, does not contradict the so
called "positive mass theorems" of,e.g., references
\cite{schoenyau1,schoenyau2,witten}, because that theorems refers to the total
energy of an isolated system. A system of that kind is supposed to be modelled
by a Lorentzian spacetime having a spacelike, asymptotically Euclidean
hypersurface.\footnote{The proof also uses as hypothesis the so called energy
dominance condition \cite{hawellis}} However, we want to emphasize here, that
although the energy results positive, its value is not unique, since depends
on the asymptotically flat coordinates chosen to perform the calculations, as
it is clear from the example of the Schwarzschild \ field, as we already
commented above and detailed in \cite{boro}.

In view of what has been presented above, it is our view that all discourses
(based on Einstein's equivalence principle) concerning the use of
pseudo-energy momentum tensors as \textit{reasonable} descriptions of energy
and momentum of gravitational fields in Einstein's theory are not convincing.

The fact is: there are \textit{in general} no conservation laws of
energy-momentum in General Relativity in general. And, at this point it is
better to quote page 98 of Sachs\&Wu\footnote{Note, please, that in this
reference Sachs refers to R. K. Sachs and not to M. Sachs.} \cite{sw}:

{\footnotesize \ " As mentioned in section 3.8, conservation laws have a great
predictive power. It is a shame to lose the special relativistic total energy
conservation law (Section 3.10.2) in general relativity. Many of the attempts
to resurrect it are quite interesting; many are simply garbage."}

We quote also Anderson \cite{anderson}:

{\footnotesize " In an interaction that involves the gravitational field a
system can loose energy without this energy being transmitted to the
gravitational field."}

In General Relativity, we already said, every gravitational field is modelled
(module diffeomorphisms) by a Lorentzian spacetime. In the particular case,
when this spacetime structure admits a \textit{timelike} Killing vector, we
can formulate a law of energy conservation. If the spacetime admits three
linearly independent \textit{spacelike} Killing vectors, we have a law of
conservation of momentum. The crucial fact to have in mind here is that a
general Lorentzian spacetime, does not admits such Killing vectors in general.
As one example, we quote that the popular Friedmann-Robertson-Walker expanding
universes models do not admit timelike Killing vectors, in general.

At present, the authors know only one possibility of resurrecting a
\textit{trustworthy} conservation law of energy-momentum valid in all
circumstances in a theory of the gravitational field that \textit{resembles}
General Relativity (in the sense of keeping Einstein's equation). It consists
in reinterpreting that theory as a field theory in flat Minkowski spacetime.
Theories of this kind have been proposed in the past by, e.g., Feynman
\cite{feynman}, Schwinger \cite{schwinger},Thirring \cite{thirring0} and
Weinberg \cite{weinberg1,weinberg2} and have been extensively studied by
Logunov and collaborators \cite{logunov1,logunov2}. Another presentation of a
theory of that kind, is one where the gravitational field is represented by a
distortion field in Minkowski spacetime. A first attempt to such a theory
using Clifford bundles has been given in \cite{rq}. Another presentation has
been given in \cite{ladogu}, but that work, which contains many interesting
ideas, unfortunately contains also some equivocated statements that make (in
our opinion) the theory, as originally presented by that authors invalid. This
has been discussed with details in \cite{femoro}.

Before closing this section we observe that recently people think to have find
a valid way of having a genuine energy-momentum conservation law in general
relativity, by using the so-called \textit{teleparallel} version of that
theory \cite{deandrade}. If that is really the case will be analyzed in a
sequel paper \cite{roqui}, where we discuss conservation laws in a general
Riemann-Cartan spacetime, using Clifford bundle methods.

One of our intentions in writing this section was to leave the reader aware of
the \textit{shameful} fact of non energy-momentum conservation in General
Relativity \ when we comment in the next section some papers by Evans\&AIAS
where they try to explain the functioning of MEG, a \ `motionless electric
generator' that according to those authors pumps energy from the vacuum.

\subsection{\textquotedblleft Explanation\textquotedblright\ of MEG according
to AIAS}

Our comments on AIAS papers dealing with MEG are the following:

(\textbf{i)} AIAS claim\footnote{See the list of their papers related to the
subject in the bibliography.} that the $\mathbf{B}_{3}$ electromagnetic field
of their new "$O(3)$ electrodynamics" is to be identified with $\mathbf{F}%
_{\mathbf{12}}$ (giving by Eq.(\ref{10.13})).

Well, this is a nonsequitur because we already showed above that
$\mathbf{F}_{\mathbf{12}}$ (Eq.(\ref{10.13})) has nothing to do with
electromagnetic fields, it is only a combination of the curvature bivectors,
which is a pure gravitational object. \ 

(\textbf{ii}) With that identification AIAS claims that it is the energy of
the "electromagnetic" field $\mathbf{F}_{\mathbf{12}}$ that makes MEG to work.
In that way MEG must be understood as motionless electromagnetic generator
that (according to AIAS) pumps energy from the\ `vacuum' defined by the
$\mathbf{B}_{3}$ field.

Well, Eq.(\ref{10.13}) shows that $\mathbf{F}_{\mathbf{12}}=0$ on the vacuum.
It follows that if MEG really works, then it is pumping energy from another
source, or it is violating the law of energy-momentum conservation. So, it is
unbelievable how Physics journals have published AIAS papers on MEG using
arguments as the one just discussed, that are completely wrong.

(\textbf{iii) }We would like to leave it clear here that it is our my opinion
that MEG does not work, even if the USPTO granted a patent for that invention,
what we considered a very sad and dangerous fact. We already elaborated on
this point in the introduction and more discussion on the subject of MEG can
be found\footnote{The reader must be aware that there are many nonsequitur
posts in this yahoo group, but there are also many serious papers written by
serious and competent people.} at http://groups.yahoo.com/group/free\_energy/.

(\textbf{iv}) And what to say about the new electrodynamics of the AIAS group
and its $\mathbf{B}_{3}$ field?

Well, in \cite{carrod,warxaias} we analyzed in deep all known presentations of
the "new $O(3)$ electrodynamics" of the AIAS group. It has been proved beyond
any doubt that almost all AIAS papers are simply a pot pourri of non sequitur
Mathematics and Physics. That is not only our opinion, and the reader is
invited (if he become interested on that issue) to read a review of
\cite{carrod} in \cite{angles}.

Recently (\cite{e0}-\cite{e3}) Evans is claiming to have produced an unified
theory and succeeded in publishing his odd ideas in ISI indexed Physical
journals. In the next section we discuss his `unified' theory, showing that it
is again, as it is the case of the old Evans\&AIAS papers, simply a compendium
of nonsense Mathematics and Physics.

(\textbf{v}) And if we are wrong concerning our opinion that MEG does not work?

Well, in that (improbable) case that MEG works, someone can claim that its
functioning vindicates the General Theory of Relativity, since as proved in
the last section in that theory there is no trustworthy law of energy-momentum
conservation. That would be really amazing...

\section{Einstein Field Equations for the Tetrad Fields $\theta^{\mathbf{a}}$}

In the main text we gave a Clifford bundle formulation of the field equations
of general relativity in a form that resembles a $Sl(2,\mathbb{C)}$ gauge
theory and also a formulation in terms of a set of $2$-form fields
$\star\mathcal{S}^{\mathbf{a}}$. Here we want to \textit{recall }yet another
face\textit{ }of Einstein's equations, i.e., we show how to write the field
equations directly for the tetrad fields $\theta^{\mathbf{a}}$ in such a way
that the obtained equations are equivalent to Einstein's field equations. This
is done in order to compare the correct equations for that objects which some
other equations proposed for these objects that appeared recently in the
literature (and which will be discussed below). Before proceeding, we mention
that, of course, we could write analogous (and equivalent) equations for the
dual tetrads $\mathbf{e}_{\mathbf{a}}$.

As shown in details in papers \ \cite{rq,qr} \ the correct wave like equations
satisfied by the $\theta^{\mathbf{a}}$ are\footnote{Of course, there are
analogous equations for the $\mathbf{e}_{\mathbf{a}}$ \cite{hestenes}, where
in that case, the Dirac operator must be defined (in an obvious way) as acting
on sections of the Clifford bundle of multivectors, that has been introduced
in section 3.}:
\begin{equation}
-({%
%TCIMACRO{\TeXButton{partial}{\mbox{\boldmath$\partial$}}}%
%BeginExpansion
\mbox{\boldmath$\partial$}%
%EndExpansion
}\cdot{%
%TCIMACRO{\TeXButton{partial}{\mbox{\boldmath$\partial$}}}%
%BeginExpansion
\mbox{\boldmath$\partial$}%
%EndExpansion
})\theta^{\mathbf{a}}+{%
%TCIMACRO{\TeXButton{partial}{\mbox{\boldmath$\partial$}}}%
%BeginExpansion
\mbox{\boldmath$\partial$}%
%EndExpansion
}\wedge({%
%TCIMACRO{\TeXButton{partial}{\mbox{\boldmath$\partial$}}}%
%BeginExpansion
\mbox{\boldmath$\partial$}%
%EndExpansion
}\cdot\theta^{\mathbf{a}})+{%
%TCIMACRO{\TeXButton{partial}{\mbox{\boldmath$\partial$}}}%
%BeginExpansion
\mbox{\boldmath$\partial$}%
%EndExpansion
}\lrcorner({%
%TCIMACRO{\TeXButton{partial}{\mbox{\boldmath$\partial$}}}%
%BeginExpansion
\mbox{\boldmath$\partial$}%
%EndExpansion
}\wedge\theta^{\mathbf{a}})=\mathcal{T}^{\mathbf{a}}-\frac{1}{2}%
T\theta^{\mathbf{a}}. \label{11.1}%
\end{equation}

In Eq.(\ref{11.1}), $\mathcal{T}^{\mathbf{a}}=T_{\mathbf{b}}^{\mathbf{a}%
}\theta^{\mathbf{b}}\in\sec\bigwedge\nolimits^{1}T^{\ast}M\hookrightarrow
\sec\mathcal{C}\ell(T^{\ast}M)$ are the energy momentum $1$-form fields and
$T=T_{\mathbf{a}}^{\mathbf{a}}=-R=-R_{\mathbf{a}}^{\mathbf{a}}$, with
$T_{\mathbf{ab}}$ the energy momentum tensor of matter. When $\theta
^{\mathbf{a}}$ is an exact differential, and in this case we write
$\theta^{\mathbf{a}}\mapsto$ $\theta^{\mu}=dx^{\mu}$ and if the coordinate
functions are harmonic, i.e., $\delta\theta^{\mu}=-{%
%TCIMACRO{\TeXButton{partial}{\mbox{\boldmath$\partial$}}}%
%BeginExpansion
\mbox{\boldmath$\partial$}%
%EndExpansion
}\theta^{\mu}=0$, Eq.(\ref{11.1}) becomes%
\begin{equation}
\square\theta^{\mu}+\frac{1}{2}R\theta^{\mu}=-\mathcal{T}^{\mu},
\label{11.1BIS}%
\end{equation}
where we have used Eq. (A.16),
\begin{equation}
({%
%TCIMACRO{\TeXButton{partial}{\mbox{\boldmath$\partial$}}}%
%BeginExpansion
\mbox{\boldmath$\partial$}%
%EndExpansion
}\cdot{%
%TCIMACRO{\TeXButton{partial}{\mbox{\boldmath$\partial$}}}%
%BeginExpansion
\mbox{\boldmath$\partial$}%
%EndExpansion
})=\square\label{11.BISS}%
\end{equation}
i.e., ${%
%TCIMACRO{\TeXButton{partial}{\mbox{\boldmath$\partial$}}}%
%BeginExpansion
\mbox{\boldmath$\partial$}%
%EndExpansion
}\cdot{%
%TCIMACRO{\TeXButton{partial}{\mbox{\boldmath$\partial$}}}%
%BeginExpansion
\mbox{\boldmath$\partial$}%
%EndExpansion
}$ is the (covariant) D'Alembertian operator.

In Eq.(\ref{11.1}) ${%
%TCIMACRO{\TeXButton{partial}{\mbox{\boldmath$\partial$}}}%
%BeginExpansion
\mbox{\boldmath$\partial$}%
%EndExpansion
}=\theta^{\mathbf{a}}D_{\mathbf{e}_{a}}={%
%TCIMACRO{\TeXButton{partial}{\mbox{\boldmath$\partial$}}}%
%BeginExpansion
\mbox{\boldmath$\partial$}%
%EndExpansion
}\wedge+$ ${%
%TCIMACRO{\TeXButton{partial}{\mbox{\boldmath$\partial$}}}%
%BeginExpansion
\mbox{\boldmath$\partial$}%
%EndExpansion
}\lrcorner$ $=d-\delta$ is the Dirac (like) operator acting on sections of the
Clifford bundle $\mathcal{C}\ell(T^{\ast}M)$ defined in the previous Appendix.

With these formulas we can write%
\begin{align}
{%
%TCIMACRO{\TeXButton{partial}{\mbox{\boldmath$\partial$}}}%
%BeginExpansion
\mbox{\boldmath$\partial$}%
%EndExpansion
}^{2}  &  ={%
%TCIMACRO{\TeXButton{partial}{\mbox{\boldmath$\partial$}}}%
%BeginExpansion
\mbox{\boldmath$\partial$}%
%EndExpansion
}\cdot{%
%TCIMACRO{\TeXButton{partial}{\mbox{\boldmath$\partial$}}}%
%BeginExpansion
\mbox{\boldmath$\partial$}%
%EndExpansion
}+{%
%TCIMACRO{\TeXButton{partial}{\mbox{\boldmath$\partial$}}}%
%BeginExpansion
\mbox{\boldmath$\partial$}%
%EndExpansion
}\wedge{%
%TCIMACRO{\TeXButton{partial}{\mbox{\boldmath$\partial$}}}%
%BeginExpansion
\mbox{\boldmath$\partial$}%
%EndExpansion
},\nonumber\\
{%
%TCIMACRO{\TeXButton{partial}{\mbox{\boldmath$\partial$}}}%
%BeginExpansion
\mbox{\boldmath$\partial$}%
%EndExpansion
}\wedge{%
%TCIMACRO{\TeXButton{partial}{\mbox{\boldmath$\partial$}}}%
%BeginExpansion
\mbox{\boldmath$\partial$}%
%EndExpansion
}  &  =-{%
%TCIMACRO{\TeXButton{partial}{\mbox{\boldmath$\partial$}}}%
%BeginExpansion
\mbox{\boldmath$\partial$}%
%EndExpansion
}\cdot{%
%TCIMACRO{\TeXButton{partial}{\mbox{\boldmath$\partial$}}}%
%BeginExpansion
\mbox{\boldmath$\partial$}%
%EndExpansion
}+{%
%TCIMACRO{\TeXButton{partial}{\mbox{\boldmath$\partial$}}}%
%BeginExpansion
\mbox{\boldmath$\partial$}%
%EndExpansion
}\wedge{%
%TCIMACRO{\TeXButton{partial}{\mbox{\boldmath$\partial$}}}%
%BeginExpansion
\mbox{\boldmath$\partial$}%
%EndExpansion
}\lrcorner+{%
%TCIMACRO{\TeXButton{partial}{\mbox{\boldmath$\partial$}}}%
%BeginExpansion
\mbox{\boldmath$\partial$}%
%EndExpansion
}\lrcorner{%
%TCIMACRO{\TeXButton{partial}{\mbox{\boldmath$\partial$}}}%
%BeginExpansion
\mbox{\boldmath$\partial$}%
%EndExpansion
}\wedge,\hspace{0.15in} \label{11.2}%
\end{align}
with%
\begin{align}
{%
%TCIMACRO{\TeXButton{partial}{\mbox{\boldmath$\partial$}}}%
%BeginExpansion
\mbox{\boldmath$\partial$}%
%EndExpansion
}\cdot{%
%TCIMACRO{\TeXButton{partial}{\mbox{\boldmath$\partial$}}}%
%BeginExpansion
\mbox{\boldmath$\partial$}%
%EndExpansion
}  &  =\eta^{\mathbf{ab}}(D_{\mathbf{e}_{\mathbf{a}}}D_{\mathbf{e}%
_{\mathbf{b}}}-\omega_{\mathbf{ab}}^{\mathbf{c}}D_{\mathbf{e}_{\mathbf{c}}%
}),\nonumber\\
{%
%TCIMACRO{\TeXButton{partial}{\mbox{\boldmath$\partial$}}}%
%BeginExpansion
\mbox{\boldmath$\partial$}%
%EndExpansion
}\wedge{%
%TCIMACRO{\TeXButton{partial}{\mbox{\boldmath$\partial$}}}%
%BeginExpansion
\mbox{\boldmath$\partial$}%
%EndExpansion
}  &  =\theta^{\mathbf{a}}\wedge\theta^{\mathbf{b}}(D_{\mathbf{e}_{\mathbf{a}%
}}D_{\mathbf{e}_{\mathbf{b}}}-\omega_{\mathbf{ab}}^{\mathbf{c}}D_{\mathbf{e}%
_{\mathbf{c}}}). \label{11.3}%
\end{align}

\bigskip Note that $D_{e_{a}}\theta^{\mathbf{b}}=-\omega_{\mathbf{ac}%
}^{\mathbf{b}}\theta^{\mathbf{c}}$ and a \ somewhat long, but simple
calculation \footnote{The calculation is done in detail in \cite{rq,qr}.}
shows that%
\begin{equation}
({%
%TCIMACRO{\TeXButton{partial}{\mbox{\boldmath$\partial$}}}%
%BeginExpansion
\mbox{\boldmath$\partial$}%
%EndExpansion
}\wedge{%
%TCIMACRO{\TeXButton{partial}{\mbox{\boldmath$\partial$}}}%
%BeginExpansion
\mbox{\boldmath$\partial$}%
%EndExpansion
})\theta^{\mathbf{a}}=\mathcal{R}^{\mathbf{a}}, \label{11.5}%
\end{equation}
where, as already defined, $\mathcal{R}^{\mathbf{a}}=R_{\mathbf{b}%
}^{\mathbf{a}}\theta^{\mathbf{b}}$ are the Ricci 1-forms. We also observe
(that for the best of our knowledge) ${%
%TCIMACRO{\TeXButton{partial}{\mbox{\boldmath$\partial$}}}%
%BeginExpansion
\mbox{\boldmath$\partial$}%
%EndExpansion
}\wedge{%
%TCIMACRO{\TeXButton{partial}{\mbox{\boldmath$\partial$}}}%
%BeginExpansion
\mbox{\boldmath$\partial$}%
%EndExpansion
}$ that has been named the Ricci operator in \cite{qr} has no analogue \ in
classical differential geometry.

Note that Eq. (\ref{11.1}) can be written after some algebra as
\begin{equation}
\mathcal{R}^{\mu}-\frac{1}{2}R\theta^{\mu}=\mathcal{T}^{\mu}, \label{11.6}%
\end{equation}
with $\mathcal{R}^{\mu}=R_{\nu}^{\mu}dx^{\nu}$ and $\mathcal{T}^{\mu}=T_{\nu
}^{\mu}dx^{\nu}$, $\theta^{\mu}=dx^{\mu}$ in a coordinate chart of the maximal
atlas of $M$ covering an open set $U\subset M$.

We are now prepared to make some crucial comments concerning some recent
papers (\cite{cleevans},\cite{e1}-\cite{e4}).

(\textbf{i}) In (\cite{cleevans},\cite{e1}-\cite{e4}) authors claims that the
$\mathbf{e}_{\mathbf{a}}$, $\mathbf{a}=0,1,2,3$ satisfy the equations\medskip

\begin{center}%
\begin{tabular}
[c]{|l|}\hline
$(\square+T)\mathbf{e}_{\mathbf{a}}=0.$\\\hline
\end{tabular}
\medskip
\end{center}

They thought to have produced a valid derivation for that equations. We will
not comment on that derivation here. Enough is to say that if that equation
was true it would imply that $(\square+T)\theta^{\mathbf{a}}=0$. This is not
the case. Indeed, as a careful reader may verify, the true equation satisfied
by any one of the $\theta^{\mathbf{a}}$ is Eq.(\ref{11.1}).

(\textbf{ii}) We quote that authors of \cite{e1,e2,e3} explicitly wrote
several times that the "electromagnetic potential"\footnote{In \cite{e1,e2,e3}
authors do identify their "electromagnetic potential" with the \ bivector
valued connection 1-form ${%
%TCIMACRO{\TeXButton{omega}{\mbox{\boldmath$\omega$}}}%
%BeginExpansion
\mbox{\boldmath$\omega$}%
%EndExpansion
}$ that we introduced in section above. As we explained with details this
cannot be done because that quantity is related to gravitation, not
electromagnetism.} $\mathbf{A}$ in their theory (a 1-form with values in a
vector space) satisfies the following wave equation,\medskip

\begin{center}%
\begin{tabular}
[c]{|c|}\hline
$(\square+T)\mathbf{A}=0.$\\\hline
\end{tabular}
\medskip
\end{center}

Now, this equation cannot be correct even for the usual $U(1)$ gauge potential
of classical electrodynamics \footnote{Which must be one of the gauge
components of the gauge field.} $A\in\sec%
%TCIMACRO{\dbigwedge \nolimits^{1}}%
%BeginExpansion
{\displaystyle\bigwedge\nolimits^{1}}
%EndExpansion
T^{\ast}M\subset\sec\mathcal{C\ell(}T^{\ast}M)$. Indeed, in vacuum Maxwell
equation reads (see Eq.(\ref{1.7}))%
\begin{equation}
{%
%TCIMACRO{\TeXButton{partial}{\mbox{\boldmath$\partial$}}}%
%BeginExpansion
\mbox{\boldmath$\partial$}%
%EndExpansion
}F=0, \label{11.9}%
\end{equation}
where $F={%
%TCIMACRO{\TeXButton{partial}{\mbox{\boldmath$\partial$}}}%
%BeginExpansion
\mbox{\boldmath$\partial$}%
%EndExpansion
}A={%
%TCIMACRO{\TeXButton{partial}{\mbox{\boldmath$\partial$}}}%
%BeginExpansion
\mbox{\boldmath$\partial$}%
%EndExpansion
}\wedge A=dA$, \textit{if} we work in the Lorenz gauge ${%
%TCIMACRO{\TeXButton{partial}{\mbox{\boldmath$\partial$}}}%
%BeginExpansion
\mbox{\boldmath$\partial$}%
%EndExpansion
}\cdot A={%
%TCIMACRO{\TeXButton{partial}{\mbox{\boldmath$\partial$}}}%
%BeginExpansion
\mbox{\boldmath$\partial$}%
%EndExpansion
}\lrcorner A=-\delta A=0$. \ Now, since we have according to Eq.(\ref{1.6bis})
that ${%
%TCIMACRO{\TeXButton{partial}{\mbox{\boldmath$\partial$}}}%
%BeginExpansion
\mbox{\boldmath$\partial$}%
%EndExpansion
}^{2}=-(d\delta+\delta d),$we get
\begin{equation}
{%
%TCIMACRO{\TeXButton{partial}{\mbox{\boldmath$\partial$}}}%
%BeginExpansion
\mbox{\boldmath$\partial$}%
%EndExpansion
}^{2}A=0. \label{11.11}%
\end{equation}

A simple calculation then shows that in the coordinate basis introduced above
we have,
\begin{equation}
({%
%TCIMACRO{\TeXButton{partial}{\mbox{\boldmath$\partial$}}}%
%BeginExpansion
\mbox{\boldmath$\partial$}%
%EndExpansion
}^{2}A)_{\alpha}=g^{\mu\nu}D_{\mu}D_{\nu}A_{\alpha}+R_{\alpha}^{\nu}A_{\nu}
\label{11.12}%
\end{equation}
and we see that Eq.(\ref{11.11}) reads in components%
\begin{equation}
D_{\alpha}D^{\alpha}A_{\mu}+R_{\mu}^{\nu}A_{\nu}=0. \label{11.13}%
\end{equation}

Eq.(\ref{11.13}) can be found, e.g., in Eddington's book \ \cite{eddington} on
page 175. Take also notice that in Einstein theory in vacuum the term $R_{\mu
}^{\nu}A_{\nu}=0$.

Finally we make a single comment on reference \cite{cleevans}, because this
paper is related to Sachs \ `unified' theory in the sense that authors try to
identify Sachs `electromagnetic' field (discussed in the main text) with a
supposedly existing longitudinal electromagnetic field predict by their
theory. Well, on \cite{cleevans} we can read at the beginning of section 1.1:

\textquotedblleft The antisymmetrized form of special relativity [1] has
spacetime metric given by the enlarged structure%
\begin{equation}
\eta^{\mu\nu}=\frac{1}{2}\left(  \sigma^{\mu}\sigma^{\nu\ast}+\sigma^{\nu
}\sigma^{\mu\ast}\right)  , \tag{1.1.}%
\end{equation}
where $\sigma^{\mu}$ are the Pauli matrices satisfying a clifford (sic)
algebra
\[
\{\sigma^{\mu},\sigma^{\nu}\}=2\delta^{\mu\nu},
\]
which are represented by
\begin{equation}
\sigma^{0}=\left(
\begin{array}
[c]{cc}%
1 & 0\\
0 & 1
\end{array}
\right)  ,\sigma^{1}=\left(
\begin{array}
[c]{cc}%
0 & 1\\
1 & 0
\end{array}
\right)  ,\sigma^{2}=\left(
\begin{array}
[c]{cc}%
0 & -i\\
i & 0
\end{array}
\right)  ,\sigma^{3}=\left(
\begin{array}
[c]{cc}%
1 & 0\\
0 & -1
\end{array}
\right)  . \tag{1.2}%
\end{equation}
The $\ast$ operator denotes quaternion conjugation, which translates to a
spatial parity transformation.\textquotedblright

Well, we comment as follows: the $\ast$ is not really defined anywhere in
\cite{cleevans}. If it refers to a spatial parity operation, we infer that
$\sigma^{0\ast}=\sigma^{0}$and \ $\sigma^{i\ast}=-\sigma^{i}$. Also,
$\eta^{\mu\nu}$ is not defined, but Eq.(3.5) of \cite{cleevans} make us to
infers that $\eta^{\mu\nu}=$diag$(1,-1,-1,1)$. In that case Eq.(1.1) above is
true (if the first member is understood as $\eta^{\mu\nu}\sigma^{0}$) but the
equation $\{\sigma^{\mu},\sigma^{\nu}\}=2\delta^{\mu\nu}$ is false. Enough is
to see that \ $\{\sigma^{0},\sigma^{i}\}=2\sigma^{i}$ $\neq2\delta^{0i}$. \ 

\section{Conclusions}

In this paper we introduced the concept of Clifford valued differential forms,
which are sections of $\mathcal{C\ell}(TM)\otimes%
%TCIMACRO{\dbigwedge }%
%BeginExpansion
{\displaystyle\bigwedge}
%EndExpansion
T^{\ast}M$. We showed how this theory can be used to produce a very elegant
description the theory of linear connections, where a given linear connection
is represented by a bivector valued 1-form. Crucial to the program was the
introduction the notion of the exterior covariant derivative of sections of
$\mathcal{C\ell}(TM)\otimes%
%TCIMACRO{\dbigwedge }%
%BeginExpansion
{\displaystyle\bigwedge}
%EndExpansion
T^{\ast}M$. Our \textit{natural} definitions parallel in a noticeable way the
formalism of the theory of connections in a principal bundle and the covariant
derivative operators acting on associate bundles to that principal bundle. We
identified Cartan curvature 2-forms and \ \textit{curvature bivectors}. The
curvature 2-forms satisfy Cartan's second structure equation and the curvature
bivectors satisfy equations in analogy with equations of gauge theories. This
immediately suggest to write Einstein's theory in that formalism, something
that has already been done and extensively studied in the past. However, we
did not enter into the details of that theory in this paper. We only discussed
the relation between the nonhomogeneous $Sl(2,\mathbb{C})$ \ gauge equation
satisfied by the curvature bivector and the problem of the \ energy-momentum
\ `conservation' in General Relativity, and also between that theory and M.
Sachs \ `unified' field theory as described in \cite{s1,s2}.

To make a complete analysis of M. Sachs `unified' field theory we also
recalled the concept of covariant derivatives of spinor fields, when these
objects are represented as sections of real spinor bundles
(\cite{lami,moro,28}) and study how this theory has as matrix representative
the standard spinor fields (dotted and undotted) already introduced long ago,
see, e.g., \cite{carmeli, penrose, penrindler, pirani}. What was new in our
approach is that we identify a possible profound physical meaning concerning
some of the rules used in the standard formulation of the (matrix) formulation
of spinor fields, e.g., why the covariant derivative of the Pauli matrices
must be null. Those rules implies in constraints for the geometry of the
spacetime manifold. A possible realization of that constraints is one where
the fields defining a global tetrad must be such that $\mathbf{e}_{\mathbf{0}%
}$ is a geodesic field and the $\mathbf{e}_{\mathbf{i}}$ $_{\text{ }}$are
Fermi transported (i.e., are not rotating relative to the "fixed stars") along
each integral line of $\mathbf{e}_{\mathbf{0}}$. For the best of our knowledge
this important fact is here disclosed for the first time.

We use our formalism to discuss several issues in presentations of
gravitational theory and other theories. In particular, we scrutinized Sachs
"unified" the theory as discussed recently in \cite{s2,s3} and as originally
introduced in \cite{s1}. \ It is really difficult to believe that after that
more than 40 years Sachs succeeded in publishing his doubtable results without
anyone denouncing his errors. The case is worth to have in mind when we
realized that Sachs has more than 900 citations in the Science Citation Index.
Some one may say: who cares? Well, we cared, for reasons mainly described in
the introduction, and here we showed that there are some crucial mathematical
errors in that theory. To start, \cite{s1,s2,s3} identified erroneously his
basic variables $q_{\mu}$ as being (matrix representations) of quaternion
fields. Well, they are not. The real mathematical structure of these objects
is that they are matrix representations of particular sections of the even
Clifford bundle of multivectors $\mathcal{C\ell}(TM)$ as we proved in section
2. Next we show that the identification of a `new' antisymmetric field
$\mathbb{F}_{\alpha\beta}$ (Eq.(\ref{sachs3})) in his theory is indeed nothing
more than the identification of some combinations of the curvature
bivectors\footnote{The curvature bivectors are physically and mathematically
equivalent to the Cartan curvature 2-forms, since they carry the same
information. This statement is obvious from our study in section 4.}, an
object that appears naturally when we try to formulate Einstein's
gravitational theory as a $Sl(2,\mathbb{C)}$ gauge theory. In that way, any
tentative of identifying $\mathbb{F}_{\alpha\beta}$ with any kind of
electromagnetic field as did by Sachs in \cite{s1,s2} is clearly wrong. We
also present the wave like equations solved by the (co)tetrad
fields\footnote{The set $\{\theta^{\mathbf{a}}\}$ is the dual basis of
$\{e_{\mathbf{a}}\}$.} $\theta^{\mathbf{a}}$. Equipped with the correct
mathematical formulation of some sophisticated notions of modern Physics
theories we identified \textit{fatal mathematical flaws} in several papers by
Evans\&AIAS\footnote{Recall that Evans is as quoted as Sachs, according to the
Sceience Citation Index...} that use Sachs `unified' theory. \ In a series of
papers, quoted in the bibliography Evans\&AIAS claims that MEG works with the
energy of the $\mathbf{B}_{3}$ field that they identified with the field
$\mathbf{F}_{\mathbf{12}}$ (given by Eq.(\ref{10.13})) that appears in Sachs
theory. They thought, following Sachs, that that field represents an
electromagnetic field. It is amazing how referees of that papers could accept
that argument, for in vacuum $\mathbf{F}_{\mathbf{12}}=0$ (see Eq.(\ref{10.13}%
)). Also, as already said $\mathbf{F}_{\mathbf{12}}$ and also $\mathbb{F}%
_{\mathbf{12}}$ re not electromagnetic fields. However, since there are no
conservation laws of energy-momentum in general relativity, if MEG
works\footnote{We stated above our opinion that despite MEG is a patented
device it does not work.}, maybe it is only demonstrating this aspect of
General Relativity, that may authors on the subject try (hard) to hide under
the carpet.

\begin{acknowledgement}
Authors are grateful to Ricardo A. Mosna for very useful observations.
\end{acknowledgement}

\appendix

\section{Clifford Bundles $\mathcal{C\ell}(T^{\ast}M)$ and $\mathcal{C\ell
}(TM)$}

Let $\mathcal{L}=(M,g,D,\tau_{g},\uparrow)$ be a Lorentzian spacetime. This
means that $(M,g,\tau_{g},\uparrow)$ is a four dimensional Lorentzian
manifold, time oriented by $\uparrow$ and spacetime oriented by $\tau_{g}$ ,
with $M\simeq\mathbb{R}^{4}$ and $g\in\sec(T^{\ast}M\times T^{\ast}M)$ being a
Lorentzian metric of signature (1,3). $T^{\ast}M$ [$TM$] is the cotangent
[tangent] bundle. $T^{\ast}M=\cup_{x\in M}T_{x}^{\ast}M$, $TM=\cup_{x\in
M}T_{x}M$, and $T_{x}M\simeq T_{x}^{\ast}M\simeq\mathbb{R}^{1,3}$, where
$\mathbb{R}^{1,3}$ is the Minkowski vector space \cite{sw}.. $D$ is the
Levi-Civita connection of $g$, $i.e$\textit{.\/}, $Dg=0$, $\mathcal{R}(D)=0$.
Also $\Theta(D)=0$, $\mathcal{R}$ and $\Theta$ being respectively the torsion
and curvature tensors. Now, the Clifford bundle of differential forms
$\mathcal{C\ell}(T^{\ast}M)$ is the bundle of algebras\footnote{We can show
using the definitions of section 5 that $\mathcal{C}\ell(T^{\ast}M)$ is a
vector bundle associated with the \emph{\ orthonormal frame bundle}, i.e.,
$\mathcal{C}\!\ell(M)$ $=P_{SO_{+(1,3)}}\times_{ad}Cl_{1,3}$.\ Details about
this construction can be found, e.g., in \cite{moro}.} $\mathcal{C\ell
}(T^{\ast}M)=\cup_{x\in M}\mathcal{C\ell}(T_{x}^{\ast}M)$ , where $\forall
x\in M,\mathcal{C\ell}(T_{x}^{\ast}M)=\mathbb{R}_{1,3}$, the so-called
\emph{spacetime} \emph{algebra }\cite{lounesto}. Locally as a linear space
over the real field $R$, $\mathcal{C\ell}(T_{x}^{\ast}M)$ is isomorphic to the
Cartan algebra $%
%TCIMACRO{\dbigwedge }%
%BeginExpansion
{\displaystyle\bigwedge}
%EndExpansion
(T_{x}^{\ast}M)$ of the cotangent space and $%
%TCIMACRO{\dbigwedge }%
%BeginExpansion
{\displaystyle\bigwedge}
%EndExpansion
T_{x}^{\ast}M=\sum_{k=0}^{4}%
%TCIMACRO{\dbigwedge }%
%BeginExpansion
{\displaystyle\bigwedge}
%EndExpansion
{}^{k}T_{x}^{\ast}M$, where $%
%TCIMACRO{\dbigwedge \nolimits^{k}}%
%BeginExpansion
{\displaystyle\bigwedge\nolimits^{k}}
%EndExpansion
T_{x}^{\ast}M$ is the $\binom{4}{k}$-dimensional space of $k$-forms. The
Cartan bundle $%
%TCIMACRO{\dbigwedge }%
%BeginExpansion
{\displaystyle\bigwedge}
%EndExpansion
T^{\ast}M=\cup_{x\in M}%
%TCIMACRO{\dbigwedge }%
%BeginExpansion
{\displaystyle\bigwedge}
%EndExpansion
T_{x}^{\ast}M$ can then be thought \cite{lami} as \textquotedblleft
imbedded\textquotedblright\ in $\mathcal{C\ell}(T^{\ast}M)$. In this way
sections of $\mathcal{C\ell}(T^{\ast}M)$ can be represented as a sum of
nonhomogeneous differential forms. Let $\{\mathbf{e}_{\mathbf{a}}\}\in\sec
TM,(\mathbf{a}=0,1,2,3)$ be an orthonormal basis $g(\mathbf{e}_{\mathbf{a}%
},\mathbf{e}_{\mathbf{b}})=\eta_{\mathbf{ab}}=\mathrm{diag}(1,-1,-1,-1)$ and
let $\{\theta^{\mathbf{a}}\}\in\sec%
%TCIMACRO{\dbigwedge \nolimits^{1}}%
%BeginExpansion
{\displaystyle\bigwedge\nolimits^{1}}
%EndExpansion
T^{\ast}M\hookrightarrow\sec\mathcal{C\ell}(T^{\ast}M)$ be the dual basis.
Moreover, we denote by $g^{-1}$ the metric in the cotangent bundle.

An analogous construction can be done for the tangent space. The corresponding
Clifford bundle is denoted $\mathcal{C\ell}(TM)$ and their sections are called
multivector fields. All formulas presented below for $\mathcal{C\ell}(T^{\ast
}M)$ have a corresponding in $\mathcal{C\ell}(TM)$ and this fact has been used
in the text.

\subsection{Clifford product, scalar contraction and exterior products}

The fundamental \emph{Clifford product} (in what follows to be denoted by
juxtaposition of symbols) is generated by $\theta^{\mathbf{a}}\theta
^{\mathbf{b}}+\theta^{\mathbf{b}}\theta^{\mathbf{a}}=2\eta^{\mathbf{ab}}$ and
if $\mathcal{C}\in\sec\mathcal{C\ell}(T^{\ast}M)$ we have%

\begin{equation}
C=s+v_{\mathbf{a}}\theta^{\mathbf{a}}+\frac{1}{2!}b_{\mathbf{cd}}%
\theta^{\mathbf{c}}\theta^{\mathbf{d}}+\frac{1}{3!}a_{\mathbf{abc}}%
\theta^{\mathbf{a}}\theta^{\mathbf{b}}\theta^{\mathbf{c}}+p\theta^{\mathbf{5}%
}\;, \label{a.1}%
\end{equation}
where $\theta^{\mathbf{5}}=\theta^{0}\theta^{\mathbf{1}}\theta^{\mathbf{2}%
}\theta^{\mathbf{3}}$ is the volume element and $s$, $v_{\mathbf{a}}$,
$b_{\mathbf{cd}}$, $a_{\mathbf{abc}}$, $p\in\sec%
%TCIMACRO{\dbigwedge \nolimits^{0}}%
%BeginExpansion
{\displaystyle\bigwedge\nolimits^{0}}
%EndExpansion
T^{\ast}M\subset\sec\mathcal{C\ell}(T^{\ast}M)$.

Let $A_{r},\in\sec%
%TCIMACRO{\dbigwedge \nolimits^{r}}%
%BeginExpansion
{\displaystyle\bigwedge\nolimits^{r}}
%EndExpansion
T^{\ast}M\hookrightarrow\sec\mathcal{C\ell}(T^{\ast}M),B_{s}\in\sec%
%TCIMACRO{\dbigwedge \nolimits^{s}}%
%BeginExpansion
{\displaystyle\bigwedge\nolimits^{s}}
%EndExpansion
T^{\ast}M\hookrightarrow\sec\mathcal{C\ell}(T^{\ast}M)$. For $r=s=1$, we
define the \emph{scalar product} as follows:

For $a,b\in\sec%
%TCIMACRO{\dbigwedge \nolimits^{1}}%
%BeginExpansion
{\displaystyle\bigwedge\nolimits^{1}}
%EndExpansion
T^{\ast}M\hookrightarrow\sec\mathcal{C\ell}(T^{\ast}M),$%
\begin{equation}
a\cdot b=\frac{1}{2}(ab+ba)=g^{-1}(a,b). \label{a.2}%
\end{equation}
We also define the \emph{exterior product} ($\forall r,s=0,1,2,3)$ by
\begin{align}
A_{r}\wedge B_{s}  &  =\langle A_{r}B_{s}\rangle_{r+s},\nonumber\\
A_{r}\wedge B_{s}  &  =(-1)^{rs}B_{s}\wedge A_{r} \label{a.3}%
\end{align}
where $\langle\rangle_{k}$ is the component in the subspace $%
%TCIMACRO{\dbigwedge \nolimits^{k}}%
%BeginExpansion
{\displaystyle\bigwedge\nolimits^{k}}
%EndExpansion
T^{\ast}M$ of the Clifford field. The exterior product is extended by
linearity to all sections of $\mathcal{C}\ell(T^{\ast}M).$

For $A_{r}=a_{1}\wedge...\wedge a_{r},B_{r}=b_{1}\wedge...\wedge b_{r}$, the
\textit{scalar product} is defined as
\begin{align}
A_{r}\cdot B_{r}  &  =(a_{1}\wedge...\wedge a_{r})\cdot(b_{1}\wedge...\wedge
b_{r})\nonumber\\
&  =\det\left[
\begin{array}
[c]{ccc}%
a_{1}\cdot b_{1} & \ldots & a_{1}\cdot b_{k}\\
\ldots & \ldots & \ldots\\
a_{k}\cdot b_{1} & \ldots & a_{k}\cdot b_{k}%
\end{array}
\right]  . \label{a.4}%
\end{align}

We agree that if $r=s=0$, the scalar product is simple the ordinary product in
the real field.

Also, if $r\neq s$ then $A_{r}\cdot B_{s}=0$.

For $r\leq s,A_{r}=a_{1}\wedge...\wedge a_{r},B_{s}=b_{1}\wedge...\wedge
b_{s\text{ }}$we define the \textit{left contraction} by
\begin{equation}
\lrcorner:(A_{r},B_{s})\mapsto A_{r}\lrcorner B_{s}=%
%TCIMACRO{\dsum \limits_{i_{1}<...<i_{r}}}%
%BeginExpansion
{\displaystyle\sum\limits_{i_{1}<...<i_{r}}}
%EndExpansion
\epsilon_{1......s}^{i_{1}.....i_{s}}(a_{1}\wedge...\wedge a_{r}%
)\cdot(b_{i_{1}}\wedge...\wedge b_{i_{r}})^{\sim}b_{i_{r}+1}\wedge...\wedge
b_{i_{s}}, \label{a.5}%
\end{equation}
\ where $\sim$ denotes the reverse mapping (\emph{reversion})
\begin{equation}
\sim:\sec%
%TCIMACRO{\dbigwedge \nolimits^{p}}%
%BeginExpansion
{\displaystyle\bigwedge\nolimits^{p}}
%EndExpansion
T^{\ast}M\ni a_{1}\wedge...\wedge a_{p}\mapsto a_{p}\wedge...\wedge a_{1},
\label{a.6}%
\end{equation}
and extended by linearity to all sections of $\mathcal{C\ell}(T^{\ast}M)$. We
agree that for $\alpha,\beta\in\sec%
%TCIMACRO{\dbigwedge \nolimits^{0}}%
%BeginExpansion
{\displaystyle\bigwedge\nolimits^{0}}
%EndExpansion
T^{\ast}M$ the contraction is the ordinary (pointwise) product in the real
field and that if $\alpha\in\sec%
%TCIMACRO{\dbigwedge \nolimits^{0}}%
%BeginExpansion
{\displaystyle\bigwedge\nolimits^{0}}
%EndExpansion
T^{\ast}M$, $A_{r},\in\sec%
%TCIMACRO{\dbigwedge \nolimits^{r}}%
%BeginExpansion
{\displaystyle\bigwedge\nolimits^{r}}
%EndExpansion
T^{\ast}M,B_{s}\in\sec%
%TCIMACRO{\dbigwedge \nolimits^{s}}%
%BeginExpansion
{\displaystyle\bigwedge\nolimits^{s}}
%EndExpansion
T^{\ast}M$ then $(\alpha A_{r})\lrcorner B_{s}=A_{r}\lrcorner(\alpha B_{s})$.
Left contraction is extended by linearity to all pairs of elements of sections
of $\mathcal{C\ell}(T^{\ast}M)$, i.e., for $A,B\in\sec\mathcal{C\ell}(T^{\ast
}M)$%

\begin{equation}
A\lrcorner B=\sum_{r,s}\langle A\rangle_{r}\lrcorner\langle B\rangle_{s},r\leq
s. \label{a.7}%
\end{equation}

It is also necessary to introduce in $\mathcal{C\ell}(T^{\ast}M)$ the operator
of \emph{right contraction} denoted by $\llcorner$. The definition is obtained
from the one presenting the left contraction with the imposition that $r\geq
s$ and taking into account that now if $A_{r},\in\sec%
%TCIMACRO{\dbigwedge \nolimits^{r}}%
%BeginExpansion
{\displaystyle\bigwedge\nolimits^{r}}
%EndExpansion
T^{\ast}M,B_{s}\in\sec%
%TCIMACRO{\dbigwedge \nolimits^{s}}%
%BeginExpansion
{\displaystyle\bigwedge\nolimits^{s}}
%EndExpansion
T^{\ast}M$ then $A_{r}\llcorner(\alpha B_{s})=(\alpha A_{r})\llcorner B_{s}$.

\subsection{Some useful formulas}

The main formulas used in the Clifford calculus in the main text can be
obtained from the following ones, where $a\in\sec%
%TCIMACRO{\dbigwedge \nolimits^{1}}%
%BeginExpansion
{\displaystyle\bigwedge\nolimits^{1}}
%EndExpansion
T^{\ast}M$ and $A_{r},\in\sec%
%TCIMACRO{\dbigwedge \nolimits^{r}}%
%BeginExpansion
{\displaystyle\bigwedge\nolimits^{r}}
%EndExpansion
T^{\ast}M,B_{s}\in\sec%
%TCIMACRO{\dbigwedge \nolimits^{s}}%
%BeginExpansion
{\displaystyle\bigwedge\nolimits^{s}}
%EndExpansion
T^{\ast}M$:
\begin{align}
aB_{s}  &  =a\lrcorner B_{s}+a\wedge B_{s},B_{s}a=B_{s}\llcorner a+B_{s}\wedge
a,\label{a.8}\\
a\lrcorner B_{s}  &  =\frac{1}{2}(aB_{s}-(-)^{s}B_{s}a),\nonumber\\
A_{r}\lrcorner B_{s}  &  =(-)^{r(s-1)}B_{s}\llcorner A_{r},\nonumber\\
a\wedge B_{s}  &  =\frac{1}{2}(aB_{s}+(-)^{s}B_{s}a),\nonumber\\
A_{r}B_{s}  &  =\langle A_{r}B_{s}\rangle_{|r-s|}+\langle A_{r}\lrcorner
B_{s}\rangle_{|r-s-2|}+...+\langle A_{r}B_{s}\rangle_{|r+s|}\nonumber\\
&  =\sum\limits_{k=0}^{m}\langle A_{r}B_{s}\rangle_{|r-s|+2k},\text{ }%
m=\frac{1}{2}(r+s-|r-s|). \label{1.201}%
\end{align}

\subsection{Hodge star operator}

Let $\star$ be the usual Hodge star operator $\star:%
%TCIMACRO{\dbigwedge \nolimits^{k}}%
%BeginExpansion
{\displaystyle\bigwedge\nolimits^{k}}
%EndExpansion
T^{\ast}M\rightarrow%
%TCIMACRO{\dbigwedge \nolimits^{4-k}}%
%BeginExpansion
{\displaystyle\bigwedge\nolimits^{4-k}}
%EndExpansion
T^{\ast}M$. If $B\in\sec%
%TCIMACRO{\dbigwedge \nolimits^{k}}%
%BeginExpansion
{\displaystyle\bigwedge\nolimits^{k}}
%EndExpansion
T^{\ast}M$, $A\in\sec%
%TCIMACRO{\dbigwedge \nolimits^{4-k}}%
%BeginExpansion
{\displaystyle\bigwedge\nolimits^{4-k}}
%EndExpansion
T^{\ast}M$ and $\tau\in\sec%
%TCIMACRO{\dbigwedge \nolimits^{4}}%
%BeginExpansion
{\displaystyle\bigwedge\nolimits^{4}}
%EndExpansion
T^{\ast}M$ is the volume form, then $\star B$ is defined by
\[
A\wedge\star B=(A\cdot B)\tau.
\]

Then we can show that if $A_{p}\in\sec%
%TCIMACRO{\dbigwedge \nolimits^{p}}%
%BeginExpansion
{\displaystyle\bigwedge\nolimits^{p}}
%EndExpansion
T^{\ast}M\hookrightarrow\sec\mathcal{C\!\ell}(T\ast M)$ we have
\begin{equation}
\star A_{p}=\widetilde{A_{p}}\theta^{\mathbf{5}}. \label{a.hodge}%
\end{equation}
This equation is enough to prove very easily the following identities (which
are used in the main text):%
\begin{align}
A_{r}\wedge\star B_{s}  &  =B_{s}\wedge\star A_{r};\hspace{0.15in}%
r=s,\nonumber\\
A_{r}\lrcorner\star B_{s}  &  =B_{s}\lrcorner\star A_{r};\hspace
{0.15in}r+s=4,\nonumber\\
A_{r}\wedge\star B_{s}  &  =(-1)^{r(s-1)}\star(\tilde{A}_{r}\lrcorner
B_{s});\hspace{0.15in}r\leq s,\nonumber\\
A_{r}\lrcorner\star B_{s}  &  =(-1)^{rs}\star(\tilde{A}_{r}\wedge
B_{s});\hspace{0.15in}r+s\leq4 \label{Aidentities}%
\end{align}

Let $d$ and $\delta$ be respectively the differential and Hodge codifferential
operators acting on sections of $%
%TCIMACRO{\dbigwedge }%
%BeginExpansion
{\displaystyle\bigwedge}
%EndExpansion
T^{\ast}M$. If $\omega_{p}\in\sec%
%TCIMACRO{\dbigwedge \nolimits^{p}}%
%BeginExpansion
{\displaystyle\bigwedge\nolimits^{p}}
%EndExpansion
T^{\ast}M\hookrightarrow\sec\mathcal{C\ell}(T^{\ast}M)$, then $\delta
\omega_{p}=(-)^{p}\star^{-1}d\star\omega_{p}$, with $\star^{-1}\star
=\mathrm{identity}$. When applied to a $p$-form we have%
\[
\star^{-1}=(-1)^{p(4-p)+1}\star\hspace{0.15in}.
\]

\subsection{Action of $D_{\mathbf{e}_{\mathbf{a}}}$ on Sections of
$\mathcal{C\!\ell}(TM)$ and $\mathcal{C\!\ell}(T^{\ast}M)$}

Let $D_{\mathbf{e}_{\mathbf{a}}}$ be the Levi-Civita covariant derivative
operator acting on sections of the tensor bundle. It can be easily shown (see,
e.g., \cite{cru}) that $D_{\mathbf{e}_{\mathbf{a}}}$ is also a covariant
derivative operator on the Clifford bundles $\mathcal{C\!\ell}(TM)$ and
$\mathcal{C\!\ell}(T^{\ast}M).$

Now, if \ $A_{p}\in\sec%
%TCIMACRO{\dbigwedge \nolimits^{p}}%
%BeginExpansion
{\displaystyle\bigwedge\nolimits^{p}}
%EndExpansion
T^{\ast}M\hookrightarrow\sec\mathcal{C\!\ell}(M)$ \ we can show, very easily
by explicitly performing the calculations\footnote{A derivation of this
formula from the genral theory of connections can be found in \cite{moro}.}
that%
\begin{equation}
D_{\mathbf{e}_{\mathbf{a}}}A_{p}=\partial_{\mathbf{e}_{\mathbf{a}}}A_{p}%
+\frac{1}{2}[\omega_{\mathbf{e}_{\mathbf{a}}},A_{p}], \label{der1}%
\end{equation}
where the $\omega_{\mathbf{e}_{\mathbf{a}}}\in\sec%
%TCIMACRO{\dbigwedge \nolimits^{2}}%
%BeginExpansion
{\displaystyle\bigwedge\nolimits^{2}}
%EndExpansion
T^{\ast}M\hookrightarrow\sec\mathcal{C\!\ell}(M)$ may be called
\textit{Clifford} \textit{connection 2-forms. }They\textit{ }are given
by:\textit{ }%
\begin{equation}
\omega_{\mathbf{e}_{\mathbf{a}}}=\frac{1}{2}\omega_{\mathbf{a}}^{\mathbf{bc}%
}\theta_{\mathbf{b}}\theta_{\mathbf{c}}=\frac{1}{2}\omega_{\mathbf{a}%
}^{\mathbf{bc}}\theta_{\mathbf{b}}\wedge\theta_{\mathbf{c}}, \label{der2}%
\end{equation}
where (in standard notation)%
\begin{equation}
D_{\mathbf{e}_{\mathbf{a}}}\theta_{\mathbf{b}}=\omega_{\mathbf{ab}%
}^{\mathbf{c}}\theta_{\mathbf{c}},\hspace{0.15in}D_{\mathbf{e}_{\mathbf{a}}%
}\theta^{\mathbf{b}}=-\omega_{\mathbf{ac}}^{\mathbf{b}}\theta^{\mathbf{c}%
},\hspace{0.15cm}\omega_{\mathbf{a}}^{\mathbf{bc}}=-\omega_{\mathbf{a}%
}^{\mathbf{cb}} \label{der3}%
\end{equation}

An analogous formula to Eq.(\ref{der1}) is valid for the covariant derivative
of sections of $\mathcal{C\!\ell}(TM)$ and they are used in several places in
the main text.

\subsection{Dirac Operator, Differential and Codifferential}

The Dirac operator acting on sections of $\mathcal{C\!\ell}(T^{\ast}M)$ is the
invariant first order differential operator
\begin{equation}
{%
%TCIMACRO{\TeXButton{dirac}{\mbox{\boldmath$\partial$}}}%
%BeginExpansion
\mbox{\boldmath$\partial$}%
%EndExpansion
}=\theta^{\mathbf{a}}D_{\mathbf{e}_{\mathbf{a}}}, \label{1.5}%
\end{equation}
and we can show(see, e.g., \cite{rq}) the very important result:
\begin{equation}
{%
%TCIMACRO{\TeXButton{dirac}{\mbox{\boldmath$\partial$}}}%
%BeginExpansion
\mbox{\boldmath$\partial$}%
%EndExpansion
}={%
%TCIMACRO{\TeXButton{dirac}{\mbox{\boldmath$\partial$}}}%
%BeginExpansion
\mbox{\boldmath$\partial$}%
%EndExpansion
}\wedge\,+\,{%
%TCIMACRO{\TeXButton{dirac}{\mbox{\boldmath$\partial$}}}%
%BeginExpansion
\mbox{\boldmath$\partial$}%
%EndExpansion
}\lrcorner=d-\delta. \label{1.6}%
\end{equation}

The square of the Dirac operator ${%
%TCIMACRO{\TeXButton{dirac}{\mbox{\boldmath$\partial$}}}%
%BeginExpansion
\mbox{\boldmath$\partial$}%
%EndExpansion
}^{2}$ is called the \textit{Hodge Laplacian}. It is not to be confused with
the covariant D'Alembertian which is given by $\square={%
%TCIMACRO{\TeXButton{dirac}{\mbox{\boldmath$\partial$}}}%
%BeginExpansion
\mbox{\boldmath$\partial$}%
%EndExpansion
\cdot%
%TCIMACRO{\TeXButton{dirac}{\mbox{\boldmath$\partial$}}}%
%BeginExpansion
\mbox{\boldmath$\partial$}%
%EndExpansion
}$. The following identities are used in the text%
\begin{align}
dd  &  =\delta\delta=0,\nonumber\\
d{%
%TCIMACRO{\TeXButton{dirac}{\mbox{\boldmath$\partial$}}}%
%BeginExpansion
\mbox{\boldmath$\partial$}%
%EndExpansion
}^{2}  &  ={%
%TCIMACRO{\TeXButton{dirac}{\mbox{\boldmath$\partial$}}}%
%BeginExpansion
\mbox{\boldmath$\partial$}%
%EndExpansion
}^{2}d;\hspace{0.15in}\delta{%
%TCIMACRO{\TeXButton{dirac}{\mbox{\boldmath$\partial$}}}%
%BeginExpansion
\mbox{\boldmath$\partial$}%
%EndExpansion
}^{2}={%
%TCIMACRO{\TeXButton{dirac}{\mbox{\boldmath$\partial$}}}%
%BeginExpansion
\mbox{\boldmath$\partial$}%
%EndExpansion
}^{2}\delta,\nonumber\\
\delta\star &  =(-1)^{p+1}\star d;\hspace{0.15in}\star\delta=(-1)^{p}\star
d,\nonumber\\
d\delta\star &  =\star d\delta;\hspace{0.15in}\star d\delta=\delta
d\star;\hspace{0.15in}\star{%
%TCIMACRO{\TeXButton{dirac}{\mbox{\boldmath$\partial$}}}%
%BeginExpansion
\mbox{\boldmath$\partial$}%
%EndExpansion
}^{2}={%
%TCIMACRO{\TeXButton{dirac}{\mbox{\boldmath$\partial$}}}%
%BeginExpansion
\mbox{\boldmath$\partial$}%
%EndExpansion
}^{2}\star\label{A.identities1}%
\end{align}

\subsection{Maxwell Equation}

Maxwell equations in the Clifford bundle of differential forms resume in one
single equation. Indeed, if $F\in\sec%
%TCIMACRO{\dbigwedge \nolimits^{2}}%
%BeginExpansion
{\displaystyle\bigwedge\nolimits^{2}}
%EndExpansion
T^{\ast}M\subset\sec\mathcal{C\ell}(T^{\ast}M)$ is the electromagnetic field
and $J_{e}\in\sec%
%TCIMACRO{\dbigwedge \nolimits^{1}}%
%BeginExpansion
{\displaystyle\bigwedge\nolimits^{1}}
%EndExpansion
T^{\ast}M\subset\sec\mathcal{C\ell}(T^{\ast}M)$ is the electromagnetic
current, we have Maxwell equation\footnote{Then, there is no misprint in the
title of this section.}:
\begin{equation}
{%
%TCIMACRO{\TeXButton{dirac}{\mbox{\boldmath$\partial$}}}%
%BeginExpansion
\mbox{\boldmath$\partial$}%
%EndExpansion
}F=J_{e}. \label{1.7}%
\end{equation}

Eq.(\ref{1.7}) is equivalent to the pair of equations%
\begin{align}
dF  &  =0,\label{1.8a}\\
\delta F  &  =-J_{e}. \label{1.8b}%
\end{align}

Eq.(\ref{1.8a}) is called the homogenous equation and Eq.(\ref{1.8b}) is
called the nonhomogeneous equation. Note that it can be written also as:%
\begin{equation}
d\star F=-\star J_{e}. \label{1.9}%
\end{equation}

\end{document}